\documentclass[fleqn,usenatbib]{mnras}

\usepackage{txfonts}
\usepackage[authoryear]{natbib}
\usepackage{lscape}                                
\usepackage{pdflscape}                           
\usepackage{color}
\usepackage{graphicx}
\usepackage{longtable}
\usepackage{epstopdf}
\usepackage{ulem}
\usepackage{soul}

\newcommand{\elecd}{$n_{\rm e}$}
\newcommand{\te}{$T_{\rm e}$}
\newcommand{\hb}{H$\beta$}

\newcommand{\fci}{[C~{\sc i}]}

\newcommand{\foi}{[O~{\sc i}]}
\newcommand{\foii}{[O~{\sc ii}]}
\newcommand{\foiii}{[O~{\sc iii}]}

\newcommand{\fsii}{[S~{\sc ii}]}
\newcommand{\fsiii}{[S~{\sc iii}]}
\newcommand{\fsiv}{[S~{\sc iv}]}
\newcommand{\fnitroi}{[N~{\sc i}]}
\newcommand{\fnii}{[N~{\sc ii}]}
\newcommand{\sfmgi}{Mg~{\sc i}]}

\newcommand{\mgii}{Mg~{\sc ii}}
\newcommand{\fariii}{[Ar~{\sc iii}]}
\newcommand{\fariv}{[Ar~{\sc iv}]}
\newcommand{\farv}{[Ar~{\sc v}]}
\newcommand{\fcaii}{[Ca~{\sc ii}]}
\newcommand{\fclii}{[Cl~{\sc ii}]}
\newcommand{\fcliii}{[Cl~{\sc iii}]}
\newcommand{\fcliv}{[Cl~{\sc iv}]}
\newcommand{\fcrii}{[Cr~{\sc ii}]}

\newcommand{\fcriv}{[Cr~{\sc iv}]}
\newcommand{\fneiii}{[Ne~{\sc iii}]}
\newcommand{\fneiv}{[Ne~{\sc iv}]}
\newcommand{\fnev}{[Ne~{\sc v}]}

\newcommand{\fkriv}{[Kr~{\sc iv}]}

\newcommand{\fkiv}{[K~{\sc iv}]}
\newcommand{\fkv}{[K~{\sc v}]}

\newcommand{\fniqii}{[Ni~{\sc ii}]}
\newcommand{\fniqiii}{[Ni~{\sc iii}]}

\newcommand{\fniqv}{[Ni~{\sc v}]}

\newcommand{\fcav}{[Ca~{\sc v}]}

\newcommand{\fmnv}{[Mn~{\sc v}]}

\newcommand{\ffeii}{[Fe~{\sc ii}]}
\newcommand{\ffeiii}{[Fe~{\sc iii}]}
\newcommand{\ffeiv}{[Fe~{\sc iv}]}
\newcommand{\ffev}{[Fe~{\sc v}]}
\newcommand{\ffevi}{[Fe~{\sc vi}]}
\newcommand{\ffevii}{[Fe~{\sc vii}]}
\newcommand{\fpii}{[P~{\sc ii}]}

\newcommand{\oiii}{O~{\sc iii}}
\newcommand{\oiv}{O~{\sc iv}}

\newcommand{\nitroi}{N~{\sc i}}
\newcommand{\nii}{N~{\sc ii}}
\newcommand{\niii}{N~{\sc iii}}

\newcommand{\silii}{Si~{\sc ii}}

\newcommand{\oi}{O~{\sc i}}
\newcommand{\oii}{O~{\sc ii}}
\newcommand{\ci}{C~{\sc i}}
\newcommand{\cii}{C~{\sc ii}}
\newcommand{\ciii}{C~{\sc iii}}
\newcommand{\civ}{C~{\sc iv}}

\newcommand{\nei}{Ne~{\sc i}}
\newcommand{\neii}{Ne~{\sc ii}}

\newcommand{\nai}{Na~{\sc i}}
\newcommand{\hi}{H\,{\sc i}}
\newcommand{\hii}{H~{\sc ii}}

\newcommand{\hei}{He~{\sc i}}
\newcommand{\heii}{He~{\sc ii}}
\newcommand{\mc}{\multicolumn}
\newcommand{\nodata}{---}

\newcommand{\gloria}[1]{\color{cyan}#1\color{black}}

\title[DC PNe C/O ratios]{C/O ratios in planetary nebulae with dual-dust chemistry from faint optical recombination lines}

\author[J. Garc\'{\i}a-Rojas et al.]{
J. Garc\'{\i}a-Rojas$^{1,2}$\thanks{E-mail: jogarcia@iac.es}, 
G. Delgado-Inglada$^3$,
D.~A. Garc\'{\i}a-Hern\'andez$^{1,2}$, 
F. Dell'Agli$^{1,2}$,
\newauthor
M. Lugaro$^{4,6}$, 
A.~I. Karakas$^{5,6}$,
M. Rodr\'{\i}guez$^{7}$
\\
$^{1}$Instituto de Astrof\'\i sica de Canarias, E-38205 La Laguna, Tenerife, Spain \\
$^{2}$Universidad de La Laguna, Dpto. Astrof\'isica,  E-38206 
           La Laguna, Tenerife, Spain\\
$^{3}$ Instituto de Astronom\'ia, Universidad Nacional Aut\'onoma de M\'exico,
Apdo. Postal 70264, Ciudad de M\'exico, 04510, Mexico\\
$^{4}$ Konkoly Observatory, Research Centre for Astronomy and Earth Sciences, Hungarian Academy of Sciences, Konkoly Thege Mikl\'os \'ut 15-17, 1121,\\ Budapest, Hungary\\
$^{5}$ Research School of Astronomy and Astrophysics, the Australian National University, Canberra, ACT 2611, Australia\\
$^{6}$ Monash Centre for Astrophysics, School of Physics and Astronomy, Monash University, VIC 3800, Australia\\
$^{7}$ Instituto Nacional de Astrof\'{\i}sica, \'Optica y Electr\'onica, Apdo. Postal 51 y 216, 7200 Puebla, Mexico\\
}

\date{Accepted XXX. Received YYY; in original form ZZZ}

\pubyear{2015}

\begin{document}
\label{firstpage}
\pagerange{\pageref{firstpage}--\pageref{lastpage}}
\maketitle
%
\begin{abstract}
We present deep high-resolution (R$\sim$15,000) and high-quality UVES optical spectrophotometry of nine planetary nebulae with dual-dust chemistry. We compute physical conditions from several diagnostics. Ionic abundances for a large number of ions of N, O, Ne, S, Cl, Ar, K, Fe and Kr are derived from collisionally excited lines. Elemental abundances are computed using state-of-the-art ionization correction factors. We derive accurate C/O ratios from optical recombination lines. We have re-analyzed additional high-quality spectra of 14 PNe from the literature following the same methodology. 
Comparison with asymptotic giant branch models reveal that about half of the total sample objects are consistent with being descendants of low-mass progenitor stars (M $<$ 1.5 M$_{\odot}$). Given the observed N/O, C/O, and He/H ratios, we cannot discard that some of the objects come from more massive progenitor stars (M $>$ 3--4 M$_{\odot}$) that have suffered a mild HBB. None of the objects seem to be descendant of very massive progenitors. We propose that in most of the planetary nebulae studied here, the PAHs have been formed through the dissociation of the CO molecule. The hypothesis of a last thermal pulse that turns O-rich PNe into C-rich PNe is discarded, except in three objects, that show C/O$>$1. We also discuss the possibility of a He pre-enrichment to explain the most He-enriched objects. We cannot discard another scenarios like extra mixing, stellar rotation or binary interactions to explain the chemical abundances behaviour observed in our sample.
\end{abstract}




\section{Introduction}\label{Intro}

Planetary nebulae (PNe) can be easily observed at very large distances and the chemical composition of the gas and other properties can be derived. Some abundances (e.g., Ar/H, Cl/H) may remain practically unchanged by stellar evolution, reflecting the primordial composition of the interstellar matter where the progenitor stars were born. Other abundance ratios (e.g., $^{12}$C/$^{13}$C, N/O or C/O), however, are strongly modified during the previous asymptotic giant branch (AGB) phase. During the thermally-pulsing phase on the AGB, the products of the He burning shell (e.g., $^{12}$C) are brought to the stellar surface via third dredge-up (TDU) episodes, which consist in the penetration of the stellar convective envelope into regions where nuclear burning of helium has taken place, and can convert originally O-rich stars (C/O$<$1) in C-rich stars (C/O$>$1) \citep{herwig05, karakaslattanzio14}. The activation of proton capture nucleosynthesis at the base of the convective envelope (hot bottom burning, HBB) in the more massive AGB stars (M$>$3--4 M$_{\odot}$, at solar metallicity) prevents the formation of C-rich atmospheres and enriches the stellar surface in specific isotopes such as $^{13}$C and $^{14}$N \citep{boothroydsackmann92, mazzitellietal99, venturaetal15}. Thus, low-mass ($\sim$1.5$-$3 M$_{\odot}$) stars are expected to be C-rich at the end of the AGB phase while more massive HBB stars may remain O-rich during all their AGB evolution. At supersolar metallicity, however, only stars in a small mass range (3$-$4 M$_{\odot}$) can become C-rich \citep{karakas14,marigoetal17}.

The low effective temperature and the high density of the circumstellar envelope of AGB stars
favour the production of dust. The kind of dust formed depends on the surface chemistry: carbon stars form mainly solid carbon and silicon carbide, while oxygen-rich stars are responsible for the condensation of silicates \citep[e.g.][]{nannietal13, dellaglietal17}. 
It has been believed for a long time that post-AGB stars and PNe belong to only
one of the two chemical branches (C-rich or O-rich) mentioned above. PNe with
rare Wolf-Rayet [WC] central stars \citep{gornytylenda00} were the first to show
simultaneously the presence of  carbon-based (e.g., polycyclic aromatic
hydrocarbons; PAHs) and oxygen-based dust (e.g., amorphous/crystalline silicates)
\citep[][]{watersetal98}. However, {\it Spitzer/IRS} observations of Galactic
bulge and disc PNe have shown that such dual-dust chemistry (DC) phenomenon is
clearly not restricted to PNe with [WC] central stars and is seen also in other PNe with
normal central stars \citep{pereacalderonetal09, garciahdezgorny14}.

The origin of the simultaneous presence of PAHs and silicates in DC PNe is still
a mystery. \citet{pereacalderonetal09} discussed several mechanisms to explain it in the Galactic bulge. Their most plausible scenario is a final
thermal-pulse on the AGB (or just after), which may turn an O-rich outflow into C-rich. The crystallization of amorphous silicates may be due to the enhanced
mass loss, while the PAHs may form from the newly released C-rich material \citep{matsuuraetal04}. \citet{guzmanramirezetal11} and \citet{guzmanramirezetal14} proposed
instead a chemical model to explain PAHs presence in the IR spectra of O-rich Galactic Bulge PNe; in their scenario, first suggested by \citet{matsuuraetal04} and \citet{cernicharo04}, hydrocarbon chains can form within O-rich gas through gas-phase chemical reactions and they concluded that the 
formation of PAHs in DC bulge PNe is best explained through hydrocarbon chemistry in an UV-irradiated, dense torus, which could be associated with the action of a central binary system. Both
proposals are based on the assumption that DC PNe in the bulge are of low-mass
($\sim$1$-$2 M$_{\odot}$) and intrinsically O-rich. 

Recent {\it Spitzer} observations of a large sample ($\sim$150) of compact
($\leq$5$''$) Galactic disc and bulge PNe have shown that DC PNe represent an
important fraction ($\sim$20\%) of PNe in the disc, while they dominate
($\sim$50\%) the PNe sample in the bulge \citep{stanghellinietal12}.
Low-resolution (R$\sim$1000) optical spectroscopy of a significant portion of
this sample of compact Galactic disc and bulge PNe with {\it Spitzer} spectra
allowed \citet{garciahdezgorny14} to derive the nebular abundances of He, N, O, Ar, Ne, S, and Cl. Remarkably, DC PNe both with
crystalline and amorphous silicate dust features (DC$_{cr}$ and DC$_{am+cr}$) in
the bulge and disc display median He and N (and N/O) abundances higher than
those observed in PNe with only O-rich or C-rich dust \citep[see Figs. 5 and 6 in][]{garciahdezgorny14};
DC PNe are also located closer to the Galactic disc than the
other types of PNe. Curiously, some DC PNe are C-rich
\citep[C/O$>$1;][]{garciarojasetal13}, suggesting that the underlying assumptions of
the Guzm\'an-Ram\'{\i}rez et al. analysis may not apply in all cases. The chemical
abundances observed in DC PNe (bulge and disc) are indeed consistent with them
being the descendants of solar metallicity and relatively massive ($\sim$3$-$5
M$_{\odot}$) AGB stars, with the most massive experiencing HBB 
\citep[see also][]{garciarojasetal13}. This is at odds with the generally accepted idea
that the Galactic bulge is mostly composed by old low-mass stars but it is in
line with the recent findings of a lack of C-rich AGB stars in the bulge of M31 
\citep{boyeretal13}. However, another
possibility could be that DC PNe evolve from suprasolar metallicity and less
massive stars (say $\sim$1.5$-$2.5 M$_{\odot}$), which do not become C-rich
depending on the number of TDU episodes experienced \citep{karakas14}. 
Hence, accurate C/O ratios of a larger sample of DC PNe would help to understand this puzzle.

The determination of C abundances in PNe is somewhat difficult because, traditionally, C abundances have been derived from UV C III] $\lambda$1909 \AA\ and C II] $\lambda$2326 \AA\ collisionally excited lines (hereinafter CELs), whose intensity is strongly affected by interstellar reddening and should be observed from space. Moreover, DC PNe are extremely faint objects in the UV, making this approach unreliable. Moreover, there are not detections of these lines reported in the literature for any of the objects in our sample. Alternatively, the use of large telescopes and the new CCDs with improved efficiency in the blue has made it possible to measure the faint optical recombination line (hereinafter, ORL) of {\cii} $\lambda$4267 \AA\, whose abundance is almost independent of the assumed electron temperature. The abundances obtained from this and other C ORLs are always higher than the ones computed from UV C CELs; this behaviour is the so-called ``abundance discrepancy problem'' and it has been present for the last 70 years. However, we avoid this problem by measuring C abundances from {\cii} ORLs along with the faint ORLs of multiplet 1 of {\oii} at $\lambda$$\sim$4650 \AA\ and making use of state-of-the-art ionization correction factors for PNe \citep[][hereinafter D-I14]{delgadoingladaetal14} to take into account the unobserved ionization stages of C and O. This approach allows one to derive reliable C/O nebular ratios since C/O ratios computed in this way have been shown to be consistent to those computed from CELs \citep{wangliu07,delgadoingladarodriguez14}. We will discuss this further in  Sect.~\ref{sec:ionic_ab}.

An example is the work by \citet{garciarojasetal13}, which reported accurate C/O nebular ratios based on optical recombination lines 
in a few (8) DC PNe with early [WC]-type central stars; 5 DC PNe in their sample have O-rich (C/O $<$ 1) nebulae but 3 of them have C-rich nebulae. Interestingly,
the DC PNe with O-rich nebulae are those with the typical DC {\it Spitzer}
spectrum, showing very weak PAH bands and crystalline/amorphous silicates, while
the C-rich ones display very unusual {\it Spitzer} spectra with strong PAH bands
(and a strong unidentified broad emission around $\sim$24 $\mu$m) and very weak
crystalline silicate features. 

The aim of this paper is to exploit these observational advances to expand the sample of dual chemistry planetary nebulae for which C/O ratios are known. We present elemental abundances (He, C, N, O, Ne, S, Cl, Ar, K, Fe, and Kr, when possible) from new observations of nine dual chemistry PNe and for another six PNe already studied by \citet[][hereinafter D-I15]{delgadoingladaetal15} to expand the sample of eight DC PNe previously analysed by \citet{garciarojasetal12} and \citet{garciarojasetal13}. Furthermore, we perform a detailed comparison of the nebular abundances observed with two different sets of AGB models to infer new information on the origin of the dual chemistry phenomenon in PNe. 

This paper is organized as follows: in Sect.~\ref{obsred} we describe the observations and data reduction; in Sect.~\ref{sec:lineint} we discuss the methodology for line flux measurements and line identification as well as the extinction correction; in Sect.~\ref{sec:chemistry} we compute physical conditions and chemical abundances of the objects. In Sect.~\ref{sec:models} we describe the nucleosynthesis models we use; the comparison between observations and nucleosynthesis models is made in Sect.~\ref{sec:results}. Finally, our results are discussed in Sect.~\ref{sec:discuss} and our conclusions are presented in Sect.~\ref{sec:conclu}.

\section{Observations and Data Reduction}\label{obsred}

We selected our sample (9 objects) from the list of Galactic disc and bulge DC PNe of \citet{stanghellinietal12} following several criteria: a) relatively high surface brightness in the optical; b) previous detection of the temperature sensitive {\foiii} $\lambda$4363 emission line with smaller telescopes; c) high ionization degree to be sure that most of the oxygen is in the form of O$^{2+}$; and d) different types of {\it Spitzer} spectra (dust types; see Table 1) and central stars. We note that all the objects included in the \citet{stanghellinietal12} sample are compact \citep[with optical diameters $\leq$ 5 $''$, except H\,1-50 and M\,1-60 which have optical diameters $\sim$10 $''$;][]{ackeretal92} objects and, hence, our sample is biased to (presumably) young and relatively low-mass PNe; indeed, in the \citet{stanghellinietal12} PNe sample there is a complete lack of very massive N-rich PNe with O-rich dust (the expected outcome of the most massive AGB stars experiencing strong HBB). As we will see in this paper, most of our sources seem to be consistent with very low-mass progenitors; see Sect.~ref{sec:discuss}.

The spectra of all sample sources were taken with the Ultraviolet-Visual Echelle Spectrograph \citep[UVES, ][]{dodoricoetal00}, attached to the 8.2m Kueyen (UT2) Very Large Telescope at Cerro Paranal Observatory (Chile). The observations were carried out in visitor mode. All except one of the objects (He\,2-96) were observed on 2015 May 26 under clear/dark conditions and with a seeing below 1.3$''$. He\,2-96 was observed on 2015 May 24 under clear/dark conditions and with a seeing between 1$''$ and 2.2$''$.

We used two standard settings, DIC1 (346+580) and DIC2 (437+860), in both the red and blue arms of the telescope, covering nearly the full optical range between 3100--10420 \AA. In the setting DIC1 (346+580) the dichroic splits the light beam in two wavelengths ranges: 3100--3885 \AA\ in the blue arm and 4785--6805 \AA\ in the red arm; in the setting DIC2 (437+860) the dichroic configuration changes to split the light beam in the wavelength ranges: 3750--4995 in the blue arm and 6700--10420 \AA\ in the red arm. The wavelength regions 5773--5833  \AA\ and 8540--8650  \AA\ were not observed because of the gap between the two CCDs used in the red arm. Additionally, there are small gaps at the reddest wavelengths that were not observed because the redmost orders do not fit completely within the CCD. The journal of observations is shown in Table~\ref{tab:tobs}. We used the Atmospheric Dispersion Corrector (ADC) to prevent atmospheric dispersion effects. Additionally, most of the objects are compact and were observed with the slit in the parallactic angle. The airmasses at which the objects were observed are presented in Table~\ref{tab:tobs}.
We labelled the four spectral ranges as B1 for DIC1-346, B2 for DIC2-437, R1 for DIC1-580 and R2: for DIC2-860 (see Table~\ref{tab:tobs}). We obtained 3 exposures between 150 and 200s in DIC1 configuration and between 300 and 600 s in DIC2 configuration. These exposures were taken consecutively following the sequence DIC1 (346+580) $\rightarrow$ DIC2 (437+860). Additional single short exposures of 30 s in each configuration and object were taken to prevent the saturation of the brightest emission lines.
The slit length was fixed to 10$''$ in the two bluest spectral ranges (B1 and B2) and 12$''$ in the two reddest ones (R1 and R2), obtaining an adequate interorder separation. The slit width was set to 1.5$''$, which gives an effective spectral resolution of $\Delta\lambda/\lambda$$\sim$ 15,000 ($\sim$17.5 km s$^{-1}$), which is needed to deblend the {\oii}+{\niii}+{\ciii} complex in some of our objects. In Table~\ref{tab:tobs} we also present the extracted area for each object.

\setcounter{table}{0}
\begin{table*}
\centering
\begin{small}
\caption{Journal of observations.}
\label{tab:tobs}
\begin{tabular}{cccccccccccc}
\noalign{\hrule} \noalign{\vskip3pt}
PN G & Object& Date & Setting & Exp. time &  Airmass & Extracted  & Galactic  &  $R_G$  & $z$ & {\it Spitzer}  & Spectral \\
        & 	    &          &             &  (s) &   &  area & Comp.  & (kpc)$^{\rm a}$  & (pc)$^{\rm b}$ & dust-type$^{\rm c}$ & type$^{\rm d}$ \\
\noalign{\vskip3pt} \noalign{\hrule} \noalign{\vskip3pt}
359.7-02.6& H\,1-40	& 2015/05/26 	& B1, R1 	& 30, 3$\times$200 	&1.1 --1.2	& 3.9$''$$\times$1.5$''$ &  bulge &	0.73  & $-$342 & DC$_{\rm am+cr}$  & none 	\\
		&              	& 			& B2, R2	& 30, 3$\times$600 	&		& 			&	&	&		 &	& \\
358.7-05.2& H\,1-50	& 2015/05/26 	& B1, R1 	& 30, 3$\times$150 	&1.1--1.2	& 3.9$''$$\times$1.5$''$ &  bulge &  2.64 & $-$983 & DC$_{\rm cr}$   & none	\\
		&              	& 			& B2, R2	& 30, 3$\times$300 	&		& 			&	&	&		&	&  \\
296.3-03.0& He\,2-73& 2015/05/26 	& B1, R1 	& 30, 3$\times$150 	&1.3 --1.4	& 4.7$''$$\times$1.5$''$ &  disc  & 7.87   & $-$365 & DC$_{\rm cr}$  & none		\\
		&              	& 			& B2, R2	& 30, 3$\times$600 	&		& 			&	&	&		&	&	\\
309.0+00.8& He\,2-96& 2015/05/24 	& B1, R1 	& 30, 3$\times$200 	&1.2 --1.3 & 4.2$''$$\times$1.5$''$ &  disc  & 1.80 & 28 &  DC$_{\rm am+cr}$ & none		\\
		&              	& 			& B2, R2	& 30, 3$\times$600 	&		& 			&	&	&		&	& \\
327.8-06.1& He\,2-158& 2015/05/26 	& B1, R1 	& 30, 3$\times$200 	&1.3 --1.5 & 3.7$''$$\times$1.5$''$ &  disc  & 13.90	&  $-$2151 &  DC$_{\rm cr}$  &	none	\\
		&              	& 			& B2, R2	& 30, 3$\times$550 	&		& 			&	&	&		&	&	\\
006.4+02.0& M\,1-31	& 2015/05/26 	& B1, R1 	& 30, 3$\times$200 	&1.2 --1.4 & 4.2$''$$\times$1.5$''$ &  bulge &2.94	& 181 &  DC$_{\rm cr}$  & {\it wels}	\\
		&              	& 			& B2, R2	& 30, 3$\times$600 	&		& 			&	&	&		&	& \\
013.1+04.1&M\,1-33	& 2015/05/26 	& B1, R1 	& 30, 3$\times$200 	&1.0 --1.4 & 5.2$''$$\times$1.5$''$ &  disc  & 1.87	& 533 & DC$_{\rm am+cr}$ & unknown	\\
		&              	& 			& B2, R2	& 30, 3$\times$600 	&		& 			&	&	&		&	& \\
019.7-04.5& M\,1-60	& 2015/05/26 	& B1, R1 	& 30, 3$\times$200 	&1.0 --1.1 & 4.7$''$$\times$1.5$''$ &  disc  & 3.34	& $-$751 & DC$_{\rm cr}$ & [WC4]		\\
		&              	& 			& B2, R2	& 30, 3$\times$600 	&		& 			&	&	&		&	& \\
006.0-03.6& M\,2-31	& 2015/05/26 	& B1, R1 	& 30, 3$\times$200 	&1.1 --1.3 & 4.7$''$$\times$1.5$''$ &  bulge &  1.80	&  $-$403 & DC$_{\rm cr}$  &	[WC4]	\\
		&              	& 			& B2, R2	& 30, 3$\times$600 	&		& 			&	&	&		&	& \\
\noalign{\vskip3pt} \noalign{\hrule} \noalign{\vskip3pt}
\end{tabular}
\end{small}
\begin{description}
\item[$^{\rm a}$] $R_G$ = [$R_{\odot}^2$ + [$cos(b) \times D$]$^2 -2 \times R_{\odot} \times D \times cos(l) \times cos(b)$]$^{1/2}$, where $R_G$ and $D$ are the Galactocentric and heliocentric distances, respectively, and $b$ and $l$ are Galactic latitude and longitude, respectively. Galactocentric distances are obtained using statistical heliocentric distances from \citet{stanghellinihaywood10}, except for He\,2-96 for which we adopted the statistical distance computed by \citet{tajitsutamura98} from IRAS fluxes. The Sun is assumed to be at $R_{\odot}$ = 8.0 kpc.
\item[$^{\rm b}$] $z$ is the height above the Galactic plane in pc. $z$ = $D \times cos(b)$, where $D$ and $b$ are the heliocentric distance and the Galactic latitude, respectively.
\item[$^{\rm c}$] DC$_{\rm cr}$: Double chemistry with crystalline silicates, DC$_{\rm am+cr}$: Double chemistry with amorphous+crystalline silicates \citep[see][]{garciahdezgorny14}.
\item[$^{\rm d}$] Central Star classification from \citet{ackerneiner03}. None --- Neither [WC] nor {\it wels}.
\end{description}
\end{table*}

The raw frames were reduced using the public ESO UVES pipeline \citep{ballesteretal00} under the GASGANO graphic user interface, following the standard procedure of bias subtraction, aperture
extraction, background subtraction, flat-fielding and wavelength calibration. The final products of the pipeline were 2D wavelength calibrated spectra; our own Python scripts were used thereafter to collapse the spectra in the spatial direction and obtain our final 1D-spectra. 
The standard stars EG\,274, Feige\,110 and HR\,5501 \citep{hamuyetal92, hamuyetal94} were observed to perform the flux
calibration and were also fully reduced with the pipeline. The flux calibration and radial velocity correction were performed using the standard procedures with IRAF\footnote{IRAF is distributed by National Optical Astronomy Observatory, which is operated by Association of Universities for Research in Astronomy, under cooperative agreement with the National Science Foundation.} \citep{tody93}.

\subsection{Description of the sample}

\begin{figure*}
\includegraphics[width=\textwidth]{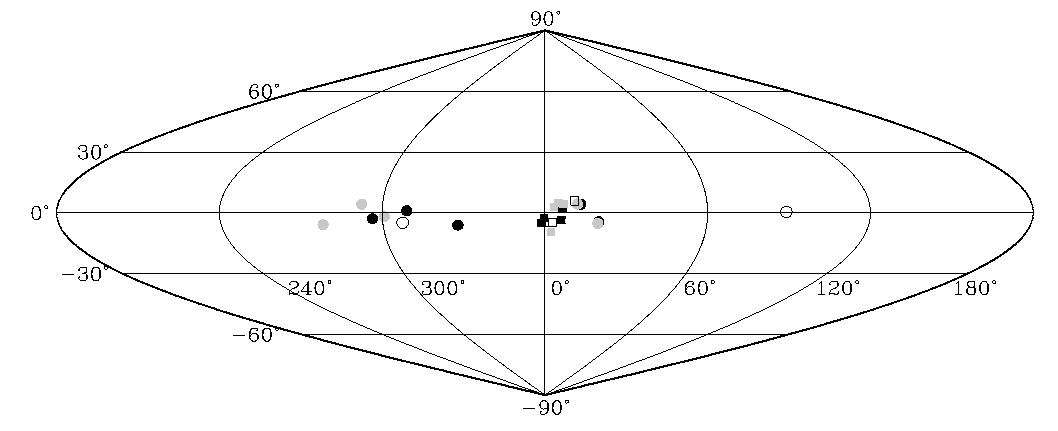}
\caption{Distribution of the PNe of our sample in the Galactic coordinates. Circles represent objects of the Galactic disk and squares, objects belonging to the Galactic bulge. Filled black symbols are new observed objects whose properties are summarized in Table~\ref{tab:tobs}. Gray and white symbols are objects from the literature and are also analysed in this work. They are described in Sect.~\ref{sec:lit_data}.} 
\label{fig:pos_gal}
\end{figure*}

In Fig.~\ref{fig:pos_gal} we show the location of our sample objects in the Galactic coordinates plane. We have included our new observed objects (black dots) as well as the literature data described in Sect.~\ref{sec:lit_data} (grey and white symbols). The symbols differentiate between bulge (squares) and disc (circles) PNe. This choice will be maintained throughout the paper. We estimated Galactocentric distances, $R_G$, for most of the PNe using the statistical distances computed by \citet{stanghellinihaywood10}, except for He\,2-96, for which we assumed the distance computed by \citet{tajitsutamura98}. We adopted a Galactocentric distance for the Sun of 8 kpc. Additionally, we also computed the height over the Galactic plane, $z$, using the estimated heliocentric distances and the Galactic latitude for each PNe, trying to identify possible PNe belonging to the thick disc. Assuming that the distance scale of the thin disc is about 350 pc, three objects have high probabilities of being thick disc PNe (He\,2-158, M\,1-33 and M\,1-60). We will briefly discuss this topic in Sect.~\ref{sec:Cl_O}. The computed $R_G$ and $z$ are shown in Table~\ref{tab:tobs}. We checked that the selection of the distance scale does not change significantly the distribution of the objects along the Galactic disc by comparing with the Galactocentric distances obtained using the distance scale by \citet{zhang95}; although individual Galactocentric distances can change, the global behaviour remains the same. Finally, there is no evidence of binary central stars in our sample.

\section{Line fluxes, identifications, and extinction correction}\label{sec:lineint}

We used the {\it splot} routine of the IRAF package to measure the line fluxes. The total flux of each line was measured by integrating between two given limits, over a local continuum estimated by eye. In the cases of line blending we used a multiple Gaussian fitting. Owing to the small area covered by our slit, we could not properly extract a sky spectrum. However, taking into account the wider profiles of the emission lines with respect to the telluric lines, it was easy to distinguish telluric emission features from nebular emission lines. We checked carefully line blendings (especially with telluric emission lines) in our 2D wavelength calibrated spectra. The cases in which nebular emission lines are severely blended with sky emission features were not considered and are not included in our line identification tables (see Tables~\ref{tab:lineid1} to~\ref{tab:lineid9}  in the Appendix). Finally, several lines are strongly affected by atmospheric features in absorption, by internal reflections of the optics or by bleeding of the emission from the most brightest lines in adjacent orders, rendering their intensities unreliable. In some cases, where we consider we could deblend the line from the non-nebular emission feature, we decided to report the line flux anyway, and included a label in the line identification table as a note of caution. We have to emphasize that we did not correct for telluric emission/absorptions in the reddest part of the spectra. Although we have labeled in Tables ~\ref{tab:lineid1} to~\ref{tab:lineid9} lines clearly affected by telluric emission/absorptions, we can not discard that other lines at the reddest wavelenghts of our spectra could be slightly affected by narrow telluric absorptions, and these fluxes should be considered with caution.

The four different spectral ranges covered in the spectra have overlapping regions at the edges. To produce a homogeneous data set of line flux ratios, we have followed the same procedure described in \citet{garciarojasetal15}. 
Some lines that were saturated in the long exposures were measured in the short ones and rescaled to the {\hb} flux in a similar way. 

The final intensity of a given line in the overlapping regions is the average of the values obtained in both spectra. In general there was an excellent agreement between line fluxes measured in overlapping spectra. The differences for each line do not show any systematic trend and, for the brightest lines, are always lower than 10\%. The differences found for the faintest lines in the overlapping regions can be larger (up to 50\%), but the final adopted errors for each line take into account these uncertainties. Therefore, the final adopted uncertainties are always larger than the differences found between both ranges.
However, our final results are not substantially affected by these effects because we have not considered fluxes of faint lines of overlapping regions in our analysis. 

\section{Physical conditions and abundance determinations}\label{sec:chemistry}

\subsection{Electron densities and temperatures}\label{sec:physcond}

The high-quality of our spectra allowed us to use multiple emission-line ratios to derive physical conditions ({\elecd} and {\te}) in our objects. The computations of physical conditions and chemical abundances were carried out with {\sc Pyneb} $v$1.0.26 \citep{luridianaetal15}. To derive the electron density, {\elecd}, and the electron temperature, {\te}, we cross-converged the temperature and density derived from two sensitive diagnostic line ratios. The method {\it getCrossTemDen} included in {\sc Pyneb} allows the simultaneous determination of {\elecd} and {\te}. This routine makes an initial guess of the {\te} and then, it calculates iteratively the values of {\elecd} and {\te} until convergence. We have used the same state-of-the-art atomic data set as in \citet{garciarojasetal15} which are summarized in Table~\ref{atomic_cels}. Errors in the diagnostics were computed via Monte Carlo simulations.  We generate 500 random values for each line intensity using a Gaussian distribution centered in the observed line intensity with a sigma equal to the associated uncertainty. For higher number of Monte Carlo simulations, the errors in the computed quantities remain constant.
The computed electron temperatures and densities are presented in Table~\ref{tab:phy_cond}.

\setcounter{table}{1}
\begin{table}
\centering
\caption{Atomic data set used for collisionally excited lines.}
\label{atomic_cels}
\begin{tabular}{lcc}
\hline
& Transition  & Collision \\
Ion & Probabilities & Strengths \\
\hline
N$^+$ & \citet{froesefischertachiev04} & \citet{tayal11} \\
O$^+$ & \citet{froesefischertachiev04} & \citet{kisieliusetal09} \\
O$^{2+}$ &  \citet{froesefischertachiev04} &  \citet{storeyetal14} \\
                &  \citet{storeyzeippen00} &  \\
Ne$^{2+}$ & \citet{galavisetal97} & \citet{mclaughlinbell00} \\
Ne$^{3+}$ & \citet{butlerzeippen89} & \citet{giles81} \\
          &  \citet{bhatiakastner88} & \\
Ne$^{4+}$ & \citet{galavisetal97} & \citet{danceetal13} \\
          &  \citet{bhatiadoschek93} & \\
S$^+$ & \citet{podobedovaetal09} & \citet{tayalzatsarinny10} \\
S$^{2+}$ &  \citet{podobedovaetal09} & \citet{tayalgupta99} \\
Cl$^{+}$ & \citet{mendozazeippen83} & \citet{tayal04} \\
Cl$^{2+}$ & \citet{mendoza83} & \citet{butlerzeippen89} \\
Cl$^{3+}$ & \citet{kaufmansugar86} & \citet{galavisetal95} \\
         &  \cite{mendozazeippen82b} & \\
         &  \cite{ellismartinson84} & \\
Ar$^{2+}$ & \citet{mendoza83} & \citet{galavisetal95} \\
          &  \citet{kaufmansugar86} & \\
Ar$^{3+}$ & \citet{mendozazeippen82a} & \citet{ramsbottombell97} \\
Ar$^{4+}$ & \citet{mendozazeippen82b} & \citet{galavisetal95} \\
          &  \citet{kaufmansugar86} & \\
          &  \citet{lajohnluke93} & \\
K$^{3+}$ & \citet{kaufmansugar86} & \citet{galavisetal95} \\
          &  \citet{mendoza83} & \\
K$^{4+}$ & \citet{kaufmansugar86} & \citet{butleretal88} \\
          &  \citet{mendoza83} & \\
Fe$^{2+}$ & \citet{quinet96} & \citet{zhang96} \\
          &  \citet{johanssonetal00} & \\
Fe$^{3+}$ & \citet{froesefischeretal08} & \citet{zhangpradhan97} \\
Fe$^{4+}$ & \citet{naharetal00} & \citet{ballanceetal07} \\
Fe$^{5+}$ & \citet{chenpradhan00} & \citet{chenpradhan99} \\
Fe$^{6+}$ & \citet{witthoeftbadnell08} & \citet{witthoeftbadnell08} \\
Kr$^{3+}$ & \citet{biemonthansen86a} & \citet{schoning97} \\
\hline
\end{tabular}
\end{table}

\begin{landscape}
\begin{figure}
\includegraphics[width=20.5cm]{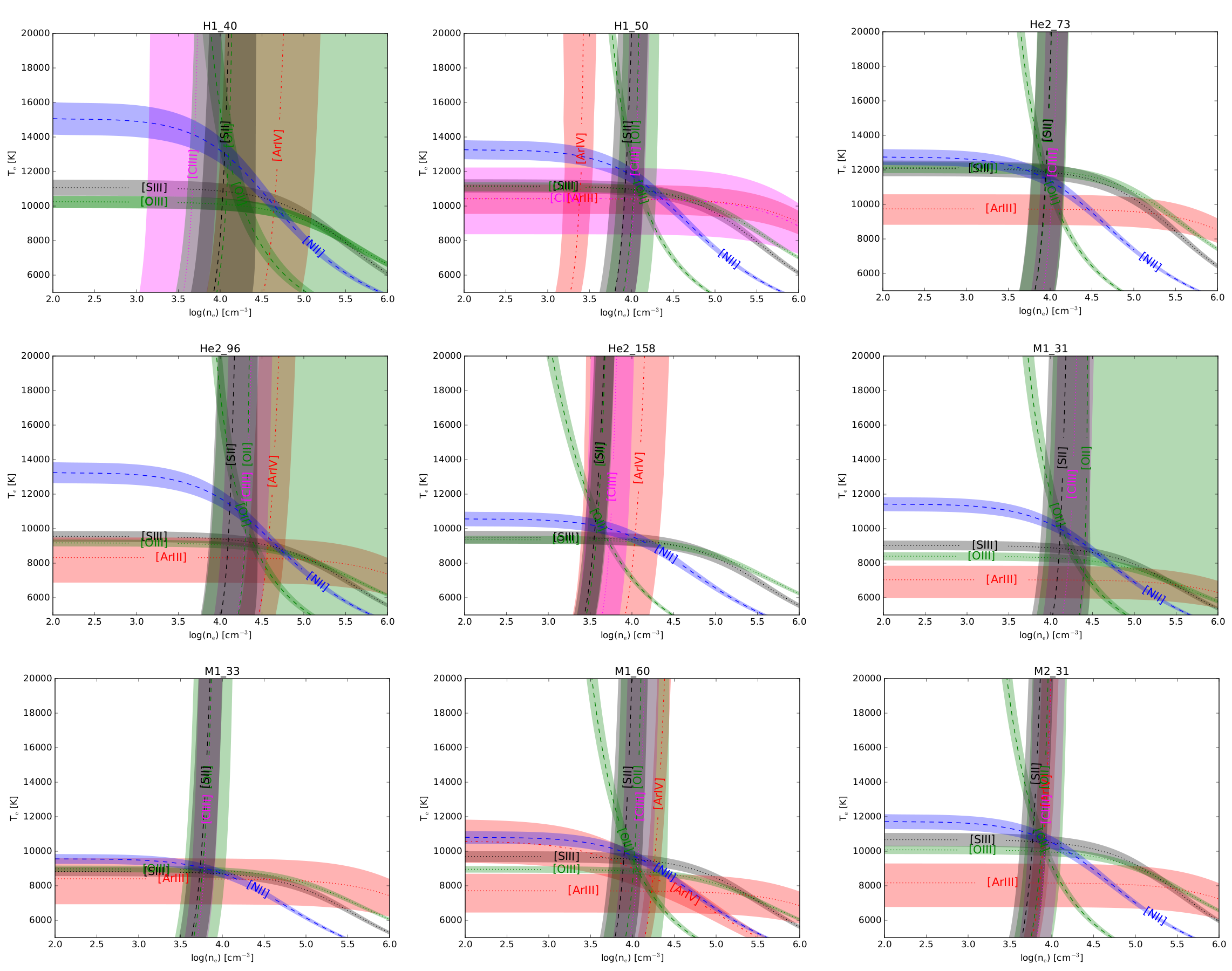}
\caption{{\te}--{\elecd} diagnostic plots. Colors correspond to species: grey for S, blue for N, green for O, magenta for Cl, and red for Ar. Different lines indicate ions: dashed for once ionized ions ({\fnii}, {\foii} and {\fsii}), dotted for two ionized ions ({\foiii}, {\fariii}, and {\fcliii}), and dotted-dashed for three times ionized ions ({\fariv}). \gloria{The width of the coloured bands is related to the uncertainties in the observed line ratios.}} 
\label{diag_tene}
\end{figure}
\end{landscape}

We assumed a two-zone ionization scheme. For {\elecd}, we obtained very similar results from the different diagnostics; therefore, we adopted the average of {\elecd}({\foii}), {\elecd}({\fsii}), {\elecd}({\fcliii}) and {\elecd}({\fariv}) as representative of the whole nebula for each PN. We adopted the average of electron temperatures obtained from {\fnii} and {\foii} lines as representative of the low-ionization zone (IP $<$ 17 eV); we refer to this value as {\te}(low).
Similarly, the average of electron temperatures obtained from lines of {\foiii}, {\fariii}, {\fsiii}, {\fcliv} and {\fariv} (when available) was assumed as representative of the high-ionization zone (17 eV $<$ IP; {\te}(high), see Table~\ref{tab:phy_cond}). In Fig.~\ref{diag_tene} we show the diagnostic diagrams computed for our sample objects using {\sc Pyneb}. This Figure shows the behaviour of the available electron density and temperature diagnostics (also shown in Table~\ref{tab:phy_cond}) in the $T_e$ vs. {\elecd} plane. Shadowed regions are estimates of the propagated uncertainties in the diagnostics given the observed uncertainties of the line fluxes used in each diagnostic; however, they do not show the effect of the Monte Carlo simulations used to compute the definitive uncertainties and should be taken as orientative only. The final values for {\elecd} and {\te} as well as their computed uncertainties using a Monte Carlo method are those shown in Table~\ref{tab:phy_cond}.

The measured fluxes of the auroral {\fnii} $\lambda$5755 line and the trans-auroral {\foii} $\lambda$$\lambda$7320+30 lines have a contribution from recombination. Taking into account these contributions can affect temperature determinations in the low ionization zone and therefore, affect low-ionized species abundance ratios, such as N$^+$/H$^+$, O$^+$/H$^+$, Cl$^+$/H$^+$, and Fe$^{2+}$/H$^+$. An estimate of these contributions can be computed using equations 1 and 2 by \citet{liuetal00} led to corrections between 1.7\% and 14.6\% in the case of the auroral {\fnii} $\lambda$5755 line and between 2.8\% and 14.8\% for the trans-auroral {\foii} $\lambda$$\lambda$7320+30 lines. These values translate into lower temperature determinations: from 80 K to 550 K lower for {\te}({\fnii}, and from 300 K to 850 K lower for {\te}({\foii}. These corrections are relatively low and the new computed physical conditions and abundances are always within the uncertainties of the previously computed ones. 
Therefore we have not considered such corrections in our computations.
The recombination contribution to the auroral $\lambda$4363 line is negligible for the two objects with the highest excitation in our sample: H\,1-50 and He\,2-73. We used equation 3 of \citet{liuetal00}, where the O$^{3+}$/H$^+$ ratio was estimated assuming O$^{3+}$/H$^+$=(He/He$^+$)$^{2/3}$$\times$(O$^+$/H$^+$+O$^{2+}$/H$^+$). Using the values derived for He$^+$/H$^+$ and He$^{2+}$/H$^+$ from optical recombination lines and O$^+$/H$^+$ and O$^{2+}$ /H$^+$ from collisionally excited lines (see Table~\ref{tab:ionic_ab}), this contribution amounts to $<$1\%, which has almost no effect on the determination of {\te}({\foiii}) and therefore, was not considered.

\setcounter{table}{2}
\begin{landscape}
\begin{table}
\caption{Plasma Diagnostic.}
\label{tab:phy_cond}
\begin{tabular}{l@{\hspace{2.0mm}}l@{\hspace{2.0mm}}c@{\hspace{2.0mm}}c@{\hspace{2.0mm}}c@{\hspace{2.0mm}}c@{\hspace{2.0mm}}c@{\hspace{2.0mm}}c@{\hspace{2.0mm}}c@{\hspace{2.0mm}}c@{\hspace{2.0mm}}c@{\hspace{2.0mm}}}
\noalign{\smallskip} \noalign{\smallskip} \noalign{\hrule} \noalign{\smallskip}
           Parameter &                                  Line ratio &            H\,1-40 &              H\,1-50&           He\,2-73 &            He\,2-96 &            He\,2-158 	&            M\,1-31 &           M\,1-33 &           M\,1-60 &              M\,2-31\\
\noalign{\smallskip} \noalign{\hrule} \noalign{\smallskip}
{\elecd} (cm$^{-3}$) &                        {\foii} $\lambda$3726/$\lambda$3729 &  12300$^{+12250}_{-6100}$   &  10950$^{+7250}_{-4350}$   &  9100$^{+5000}_{-3200}$   &  19950$^{+20500}_{-10000}$  &  3850$^{+950}_{-800}$   &  25300$^{+21800}_{-11700}$   &  5900$^{+3450}_{-2200}$   & 10550$^{+9550}_{-5000}$   &  7650$^{+4450}_{-2800}$ \\[3pt]
                     &                                    {\fsii} $\lambda$6731/$\lambda$6716 &  11150$^{+5700}_{-3800}$   &  8900$^{+3700}_{-2600}$   &  8900$^{+3400}_{-2450}$   &  13000$^{+9050}_{-5350}$   & 3650$^{+900}_{-750}$   &  13200$^{+8700}_{-5250}$   &  5650$^{+2500}_{-1750}$   & 8150$^{+3350}_{-1650}$   &  6000$^{+2100}_{-1500}$ \\[3pt]
                     &                                  {\fcliii} $\lambda$5538/$\lambda$5518 &  {\nodata}$^{\rm a}$  &  10800$^{+3500}_{-2600}$   & 10550$^{+4000}_{-2900}$   &  19350$^{+19250}_{-9550}$   & 5500 $^{+4250}_{-2400}$  &  16300$^{+12050}_{-6950}$   &  6000$^{+2100}_{-1550}$   &  11300$^{+6600}_{-4150}$   &  8100$^{+4500}_{-2850}$  \\[3pt]
                     &                                   {\fariv} $\lambda$4740/$\lambda$4711  & {\nodata}  &  {\nodata}   &  {\nodata}  &  37000$^{+30600}_{-15600}$   &  10750$^{+20300}_{-7050}$  & {\nodata}  & {\nodata}   &  17800$^{+3600}_{-3000}$   &  7750$^{+2000}_{-1600}$    \\[3pt]
                     &                                 adopted  & {\bf 11650$^{\bf +4400}_{\bf -3200}$}  &  {\bf 10650$^{\bf +1300}_{\bf -1150}$}  &  {\bf 9700$^{\bf +1700}_{\bf -1450}$}  &  {\bf 19250$^{\bf +3900}_{\bf -3400}$}  & {\bf 3550$^{\bf +450}_{\bf -400}$}  &  {\bf 16750$^{\bf +3850}_{\bf -3150}$}   & {\bf 5900$^{\bf +850}_{\bf -750}$}   & {\bf 13500$^{\bf +1800}_{\bf -1600}$}   & {\bf 7250$^{\bf +950}_{\bf -850}$}   \\[3pt]
                     &                                             &		      & 		 &		    &		       &		  \\
     $T_{\rm e}$ (K) &                          {\fnii} $\lambda$5755/$\lambda$6548  & 13350$\pm$1500   &  12250$\pm$1000   &  11700$\pm$700   &  11250$\pm$950   &  10250$\pm$300  &   9750$\pm$1100   &  9600$\pm$300   &  10200$\pm$500   &  11250$\pm$500     \\[3pt]
                     &       			    {\foii} $\lambda\lambda$3726+29/$\lambda\lambda$7320+30 & 13450$\pm$2700   &  13750$\pm$2200   &  12150$\pm$1900$^{\rm b}$   &  {\nodata}$^{\rm b}$   &  {\nodata}$^{\rm b}$  &  9250$\pm$2700   &  {\nodata}$^{\rm b}$   &  10950$\pm$1900   &  12000$\pm$1800   \\[3pt]
                     &                                   Low  & {\bf 13400$\pm$1100}   &  {\bf 12250$\pm$650}   &  {\bf 11750$\pm$500}  &  {\bf 11250$\pm$950}   &  {\bf 10250$\pm$300}    &     {\bf 9700$\pm$800}   &  {\bf 9600$\pm$300}   &  {\bf 10200$\pm$400}   &  {\bf 11250$\pm$400}    \\[3pt]
                     &                                             &		      & 		 &		    &		       &		&  	   	     &		        &		   & 		  \\
                     &                                   {\foiii} $\lambda$4363/$\lambda$4959   & 10140$\pm$350   &  11000$\pm$250   &  12000$\pm$300   &  8950$\pm$300   &  9300$\pm$200   &    8250$\pm$250   &  8950$\pm$150   &  8850$\pm$200   &  10000$\pm$250  \\[3pt]
                     &                                   {\fsiii} $\lambda$6312/$\lambda$9069  & 10950$\pm$450   &  10950$\pm$350   &  11850$\pm$450$^{\rm c}$   &  9250$\pm$450   &  9400$\pm$400  &   8800$\pm$350   &  8750$\pm$250   &  9500$\pm$300   &  10500$\pm$400     \\[3pt]
                     &                                  {\fariii} $\lambda$5192/$\lambda$7135  & 9700$\pm$750   &  10400$\pm$800   &  9700$\pm$900   &  8300$\pm$1250   &  {\nodata} &      7000$\pm$950  & 8400$\pm$1300   &  7700$\pm$1150   &  8150$\pm$1200    \\[3pt]
                     &                                  {\fcliv} $\lambda$5323/$\lambda$7531  & {\nodata}   &  10400$\pm$1900   &  {\nodata}   &  {\nodata}  &  {\nodata}  &  {\nodata}   &  {\nodata}  &  {\nodata} & {\nodata}  \\[3pt]
                     &                                  {\fariv} $\lambda\lambda$7136+7751/$\lambda\lambda$4711+40  & {\nodata}   &  {\nodata}     &  {\nodata}   &  {\nodata}  &  {\nodata}  &  {\nodata}   &  {\nodata}  &    8500$\pm$850   & {\nodata}  \\[3pt]
                       &                                   High  & {\bf 10250$\pm$250}   &  {\bf 10850$\pm$150}   &  {\bf 11600$\pm$150}   & {\bf 9100 $\pm$200}  & {\bf 9350$\pm$100}  &  {\bf 8500$\pm$150}   &  {\bf 8850$\pm$100}   &  {\bf 9100$\pm$100}   &  {\bf 10150$\pm$150}    \\[3pt]
\noalign{\smallskip} \noalign{\hrule} \noalign{\smallskip}
\end{tabular}
\begin{description}
\item[$^{\rm a}$] {\fcliii} $\lambda$5517 line affected by ghost and not considered.
\item[$^{\rm b}$] {\foii} $\lambda$$\lambda$7320+30 lines affected by telluric emission lines.
\item[$^{\rm c}$] {\fsiii} $\lambda$9069 line affected by telluric absorption lines. $\lambda$9531 line used instead.
\end{description}
\end{table}
\end{landscape}

\subsection{Ionic abundances from collisionally excited lines}\label{sec:ionabun}

We compute the ionic abundances from multiple ionic CELs of heavy elements. For each ion, we use the same CELs, when available, as those shown in Table~8 of \citet{garciarojasetal15}. We did not use auroral or trans-auroral lines to compute abundances owing to their large dependence on the assumed {\te}. 
We computed abundances using {\sc Pyneb} \citep{luridianaetal15} and the atomic data shown in Table~\ref{atomic_cels}. Errors in the line fluxes and the
physical conditions were propagated via Monte Carlo simulations. For all the objects we used a two-zone scheme of the nebula, adopting a unique {\elecd} value in the two zones, {\te}(low) for ions with IP$<$17 eV (i.e. N$^+$, O$^+$, S$^+$ and Fe$^{2+}$), and {\te}(high) for ions with IP$>$17 eV (i.e. O$^{2+}$, Ne$^{2+}$, Ne$^{3+}$, Ne$^{4+}$, S$^{2+}$, Cl$^{2+}$, Cl$^{3+}$, Ar$^{2+}$, Ar$^{3+}$, Ar$^{4+}$, K$^{3+}$, K$^{4+}$, Fe$^{3+}$, Fe$^{4+}$, Fe$^{5+}$, Fe$^{6+}$ and Kr$^{3+}$).  Ionic abundances are presented in Table~\ref{tab:ionic_ab}.

\setcounter{table}{3}
\begin{table*}
\begin{small}
\caption{Ionic abundances.}
\label{tab:ionic_ab}
\begin{center}
\begin{tabular}{lccccccccc}
\noalign{\smallskip} \noalign{\smallskip} \noalign{\hrule} \noalign{\smallskip}
 & \multicolumn{9}{c}{12 + log(X$^{+i}$/H$^+$)} \\
\noalign{\smallskip} \noalign{\hrule} \noalign{\smallskip}
                 Ion &        H\,1-40 			&        H\,1-50 			&      He\,2-73 			&      He\,2-96 			&     He\,2-158    	&        M\,1-31 			&     M\,1-33 			&     M\,1-60 			 &       M\,2-31     \\
\noalign{\smallskip} \noalign{\hrule} \noalign{\smallskip}
             He$^+$  & 11.10$\pm$0.02 		& 10.98$\pm$0.01 		& 10.96$\pm$0.01 		& 11.08$\pm$0.01 		& 11.04$\pm$0.01 	& 11.18$\pm$0.02 		& 11.12$\pm$0.01 		& 11.12$\pm$0.01 		& 11.05$\pm$0.05 		\\[3pt]
             He$^{2+}$& \nodata 				& 10.03$\pm$0.02 		& 10.30$\pm$0.02 		& \nodata				& \nodata		 	& \nodata 				& 9.15$\pm$0.03 		& 8.93$\pm$0.04 		& 8.56$\pm$0.06 		\\[3pt]
              N$^+$  & 6.78$^{+0.16}_{-0.09}$ 	& 6.89$\pm$0.08 		& 7.26$\pm$0.07 		& 7.07$^{+0.19}_{-0.09}$ 	& 7.32$\pm$0.05 	& 7.56$^{+0.22}_{-0.12}$ 	& 7.50$\pm$0.06 		& 7.58$^{+0.10}_{-0.06}$ 	& 7.04$\pm$0.06 		 \\[3pt]
              O$^+$  & 7.17$^{+0.39}_{-0.21}$ 	& 7.34$^{+0.16}_{-0.12}$ 	& 7.54$^{+0.15}_{-0.22}$ 	& 7.63$^{+0.37}_{-0.19}$ 	& 7.89$\pm$0.09 	& 7.88: 				& 7.70$\pm$0.10 		& 7.72$^{+0.17}_{-0.11}$ 	& 7.43$^{+0.11}_{-0.09}$ 	\\[3pt]
           O$^{2+}$ & 8.48$\pm$0.06 		& 8.63$\pm$0.04 		& 8.51$\pm$0.04 		& 8.63$\pm$0.05 		& 8.35$\pm$0.05 	& 8.70$\pm$0.05 		& 8.73$\pm$0.04 		& 8.71$\pm$0.04 		& 8.57$\pm$0.04 			\\[3pt]
          Ne$^{2+}$& 8.01$\pm$0.07		& 8.11$\pm$0.04 		& 8.03$\pm$0.04 		& 8.14$\pm$0.06 		& 7.76$\pm$0.05 	& 8.28$\pm$0.06 		& 8.34$\pm$0.04 		& 8.31$\pm$0.05 		& 8.08$\pm$0.04 		\\[3pt]
          Ne$^{3+}$ &   \nodata    			& 8.01$\pm$0.11   		& 8.04$\pm$0.09		&   \nodata    			&   \nodata     		&   \nodata    			&   \nodata    			&   \nodata    			&   \nodata    				 \\[3pt]
          Ne$^{4+}$ &   \nodata    			& 6.35$\pm$0.06   		& 6.83$\pm$0.06 		&   \nodata    			&   \nodata      		&   \nodata    			&   \nodata    			&   \nodata    			&   \nodata    				 \\[3pt]
                S$^+$ & 5.22$^{+0.30}_{-0.17}$ 	& 5.63$^{+0.12}_{-0.10}$ 	& 5.80$^{+0.12}_{-0.09}$ 	& 5.72$^{+0.28}_{-0.16}$ 	& 5.74$\pm$0.06 	& 6.04$^{+0.30}_{-0.20}$ 	& 5.97$^{+0.09}_{-0.07}$ 	& 6.07$^{+0.12}_{-0.09}$ 	& 5.74$^{+0.09}_{-0.06}$ 	\\[3pt]
           S$^{2+}$ & 6.58$\pm$0.04 		& 6.50$\pm$0.03 		& 6.55$\pm$0.04 		& 6.85$\pm$0.04 		& 6.54$\pm$0.03 	& 6.87$\pm$0.04 		& 6.88$\pm$0.03 		& 6.88$\pm$0.04 		& 6.63$\pm$0.03 			\\[3pt]
              Cl$^+$ & 3.39$^{+0.17}_{-0.22}$ 	& 3.75$\pm$0.10 		& 3.91$\pm$0.06 		& 3.78$^{+0.11}_{-0.08}$ 	& 3.85$^{+0.16}_{-0.27}$ & 4.07$^{+0.13}_{-0.08}$ 	& 4.11$\pm$0.06 		& 4.09$^{+0.08}_{-0.06}$ 	& 3.67$\pm$0.08 			\\[3pt]
          Cl$^{2+}$ & 4.88$^{+0.16}_{-0.11}$ 	& 4.83$\pm$0.04 		& 4.83$\pm$0.05 		& 5.06$^{+0.10}_{-0.08}$ 	& 4.80$\pm$0.07 	& 5.23$\pm$0.08 		& 5.15$\pm$0.04 		& 5.21$\pm$0.05 		& 4.92$\pm$0.05 			\\[3pt]
          Cl$^{3+}$  & 4.40$\pm$0.07 		& 4.95$\pm$0.03 		& 4.76$\pm$0.03 		& 4.34$\pm$0.04 		& 3.74$^{+0.14}_{-0.19}$ & 4.40$\pm$0.04 		& 4.68$\pm$0.03 		& 4.77$\pm$0.03 		& 4.67$\pm$0.03 			\\[3pt]
          Ar$^{2+}$ & 6.16$\pm$0.04 		& 6.10$\pm$0.03 		& 6.24$\pm$0.03 		& 6.39$\pm$0.04 		& 6.03$\pm$0.03 	& 6.54$\pm$0.04 		& 6.50$\pm$0.03		& 6.55$\pm$0.03 		& 6.21$\pm$0.03 			\\[3pt]
          Ar$^{3+}$  & 4.79$^{+0.23}_{-0.29}$ 	& 6.04$^{+0.07}_{-0.05}$ 	& 5.99$^{+0.07}_{-0.05}$ 	& 5.08$^{+0.09}_{-0.07}$ 	& 4.67$\pm$0.10 	& 5.55$\pm$0.14 		& 5.83$\pm$0.05 		& 5.82$\pm$0.05 		& 5.72$\pm$0.04 		\\[3pt]
          Ar$^{4+}$  &   \nodata    			& 4.91$\pm$0.04 		& 5.18$\pm$0.04 		&   \nodata    			&   \nodata        		&   \nodata    			&   \nodata    			&   \nodata    			&   \nodata    			\\[3pt]
          K$^{3+}$  &  3.59$^{+0.18}_{-0.25}$ 	& 4.30$\pm$0.06 		& 4.04$^{+0.05}_{-0.08}$ 	&   \nodata    			&   \nodata        		&   \nodata    			& 3.92$^{+0.14}_{-0.17}$ 	& 4.02$^{+0.11}_{-0.17}$ 	& 3.98$^{+0.09}_{-0.11}$ 	\\[3pt]
          K$^{4+}$  &   \nodata    			& 4.02$^{+0.13}_{-0.16}$ 	&   \nodata    			&   \nodata    			&   \nodata        		&   \nodata    			&   \nodata    			&   \nodata    			&   \nodata    			\\[3pt]
          Fe$^{2+}$  & 5.71$^{+0.15}_{-0.10}$ 	& 4.40$\pm$0.12 		& 4.65$\pm$0.11	 	& 4.19$^{+0.21}_{-0.16}$ 	& 5.14$\pm$0.09        & 5.01$^{+0.29}_{-0.18}$ 	& 4.92$\pm$0.09 		& 4.69$\pm$0.13	 	& 4.49$^{+0.15}_{-0.20}$ 	\\[3pt]
          Fe$^{3+}$  &   \nodata    			&   \nodata    			&   \nodata    			&   \nodata    			&   \nodata        		&   \nodata    			&   \nodata    			&   \nodata    			&   \nodata    				\\[3pt]
          Fe$^{4+}$  &   \nodata    			& 4.87$\pm$0.13 		& 5.76$\pm$0.08		&   \nodata    			&   \nodata        		&   \nodata    			&   \nodata    			&   \nodata    			&   \nodata    				\\[3pt]
          Fe$^{5+}$  &   \nodata    			& 4.41$^{+0.17}_{-0.24}$ 	& 5.20$^{+0.08}_{-0.10}$ 	&   \nodata    			&   \nodata        		&   \nodata    			&   \nodata    			&   \nodata    			&   \nodata    				\\[3pt]
          Fe$^{6+}$  &   \nodata    			& 5.08:$^{\rm a}$ 	&   \nodata    			&   \nodata    			&   \nodata        		&   \nodata    			&   \nodata    			&   \nodata    			&   \nodata    				\\[3pt]
          Kr$^{3+}$  &   \nodata    			& \nodata 				& 3.47$^{+0.11}_{-0.14}$ 	&   \nodata    			&   \nodata        		&   \nodata    			&   \nodata    			&   \nodata    			&   \nodata    				\\[3pt]
\noalign{\smallskip} \noalign{\hrule} \noalign{\smallskip}
\end{tabular}
\begin{description}
\item[$^{\rm a}$] {\ffevii} $\lambda$5158.41 line identification is dubious (see Table~\ref{tab:lineid2}).
\end{description}
\end{center}
\end{small}
\end{table*}

\subsection{Ionic abundances from recombination lines}\label{sec:ionic_ab}

A large number of {\hei} emission lines were detected in the spectra of all the PNe in our sample. These lines arise mainly from recombination, but some of them can be affected by collisional excitation and self-absorption effects. 
We use the effective recombination coefficients by \citet{porteretal12, porteretal13} for He$^+$. Both collisional contribution effects and the optical depth in the triplet lines are included in the computations. We determine the He$^+$/H$^+$ ratio from the three brightest {\hei} emission lines: $\lambda\lambda$4471, 5876, and 6678 using {\sc Pyneb}.
We measure several {\heii} emission lines in the spectra of H\,1-50 and He\,2-73, and only the bright {\heii} $\lambda$4686 line in M\,1-33, M\,1-60 and  M\,2-31. Hence, we only use the {\heii} $\lambda$4686 line to compute the He$^{2+}$/H$^+$ ratio. The computation adopts the recombination coefficients computed by \citet{storeyhummer95}. The adopted He$^+$/H$^+$ and He$^{2+}$/H$^+$ ratios are presented in Table~\ref{tab:ionic_ab}.

We detect several ORLs of different ions of C, N, O and Ne in our spectra. We only compute abundances from {\oii} and {\cii} recombination lines as they are crucial for our analysis. Ionic abundances using {\nii}, {\niii}, and {\neii} ORLs can be computed following the prescriptions given in \citet{garciarojasetal15}. {\neii} ORLs are only detected in H\,1-50, M\,1-33, and M\,1-60 but they are extremely faint (uncertainties are always higher than 40\%) and, therefore, abundances obtained from these lines should be treated with caution. 

\begin{figure*}
\includegraphics[width=\textwidth]{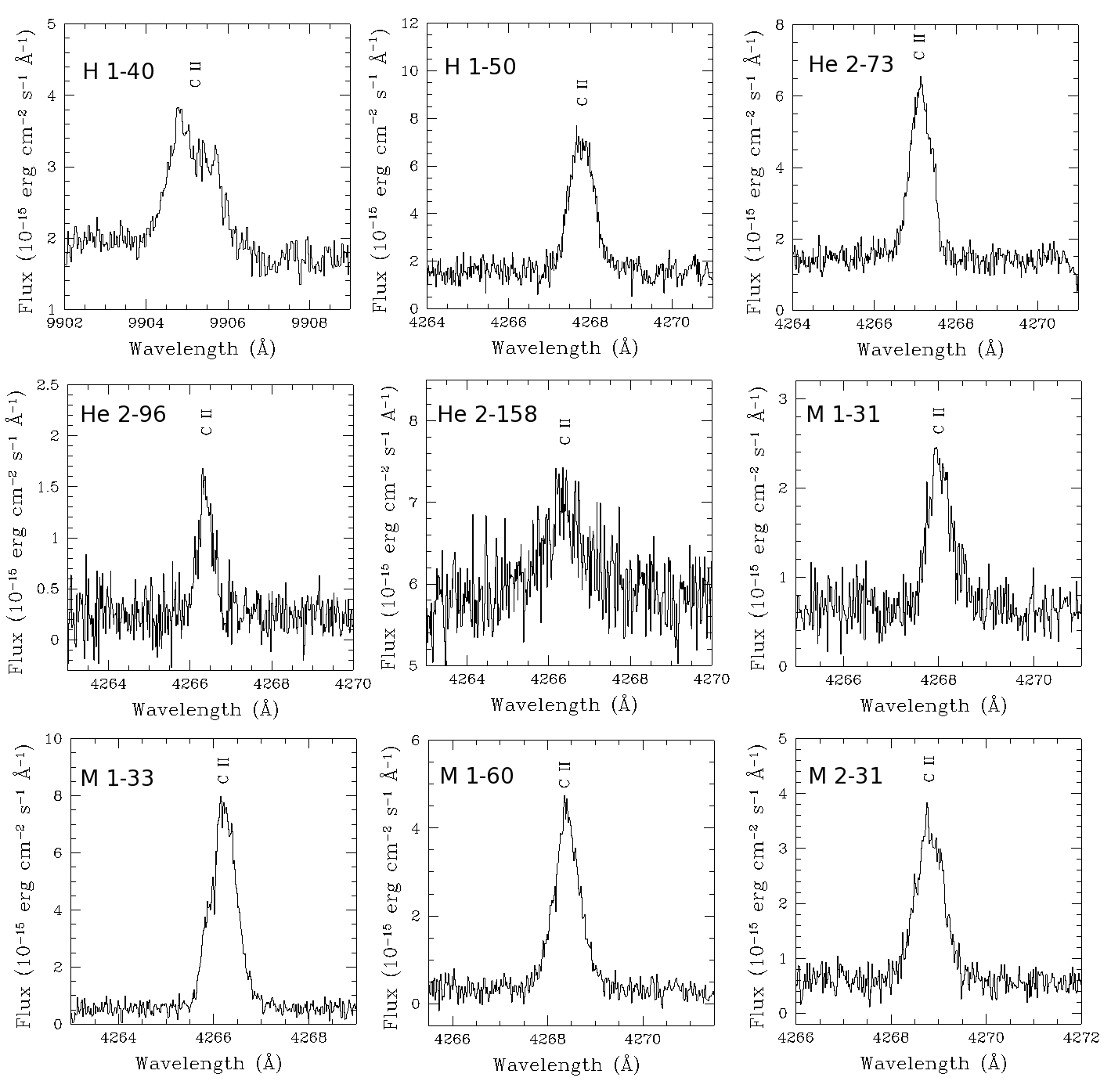}
\caption{Portion of the spectra showing the brightest {\cii} recombination line in all the planetary nebulae of our sample ({\cii} $\lambda$9903 for H\,1-40 and {\cii} $\lambda$4267 for the rest of the objects).} 
\label{cii_rec_lines}
\end{figure*}

In Fig.~\ref{cii_rec_lines} we show the brightest {\cii} recombination line for each PN in our sample. In most of the objects this line is the {\cii} $\lambda$4267 line, except for H\,1-40, where this line was not detected and the brightest one is the {\cii} $\lambda$9903 line. We compute C$^{2+}$/H$^+$ ratios using several {\cii} lines that, in principle, are excited by pure recombination. We used the recombination coefficients given by \citet{daveyetal00}. The good agreement between the abundances obtained using the different lines support the recombination origin of these lines. In Table~\ref{tab:rec_cii} we show the ionic abundances from {\cii} ORLs.

\setcounter{table}{4}
\begin{table*}
\begin{small}
\caption{Ionic abundance ratios from the {\cii} recombination lines$^{\rm a}$}
\label{tab:rec_cii}
\begin{tabular}{ccccccccccc}
\noalign{\hrule} \noalign{\vskip3pt}
Mult.& $\lambda_0$ & \mc{9}{c}{C$^{++}$/H$^+$ ($\times$10$^{-5}$)} 	\\
\noalign{\vskip3pt} \noalign{\hrule} \noalign{\vskip3pt}
&  				&          H\,1-40 &     H\,1-50 	&     He\,2-73 	&     He\,2-96 	&     He\,2-158 	&      M\,1-31     	&        M\,1-33   	& M\,1-60      	&  M\,2-31   	 \\
\noalign{\smallskip} \noalign{\hrule} \noalign{\smallskip}
6& 4267.15		& --- 		& 34$\pm$5    	& 58$\pm$7	& 34$\pm$7	& 20$\pm$4    	& 55$\pm$16    	& 98$\pm$8    	& 82$\pm$9    	& 51$\pm$11    	\\
17.06& 5342.38		& ---     		& --- 		& 88:			& --- 		& --- 		& --- 		& --- 		& --- 		& ---  		 \\
17.04& 6461.95		& ---     		& 27:		    	& --- 		& 43:			& --- 		& 60$\pm$18    	& 91$\pm$21    	& 81$\pm$26    	& ---  		 \\
17.02& 9903.46		& 20$\pm$4    	& 34$\pm$8    	& 60$\pm$5	& 32$\pm$2	&  --- 		& 47$\pm$6    	& 90$\pm$14    	& 76$\pm$7    	& 48$\pm$3    	\\
\noalign{\smallskip} \noalign{\hrule} \noalign{\smallskip}
& Adopted 		&{\bf 20$\pm$4}   &{\bf 34$\pm$4}   	&{\bf 59$\pm$4} &{\bf 32$\pm$2}&{\bf 20$\pm$4}&{\bf 49$\pm$5} &{\bf 96$\pm$7}    &{\bf 78$\pm$5}	&{\bf 48$\pm$3}  \\ 
\noalign{\smallskip} \noalign{\hrule} \noalign{\smallskip}
\end{tabular}
\begin{description}
\item[$^{\rm a}$] Colons indicate uncertainties higher than 40\%. Only lines with intensity uncertainties lower than 40\% were considered to compute averaged values (see text).
\end{description}
\end{small}
\end{table*}

\begin{landscape}
\begin{figure}
\includegraphics[width=23.5cm]{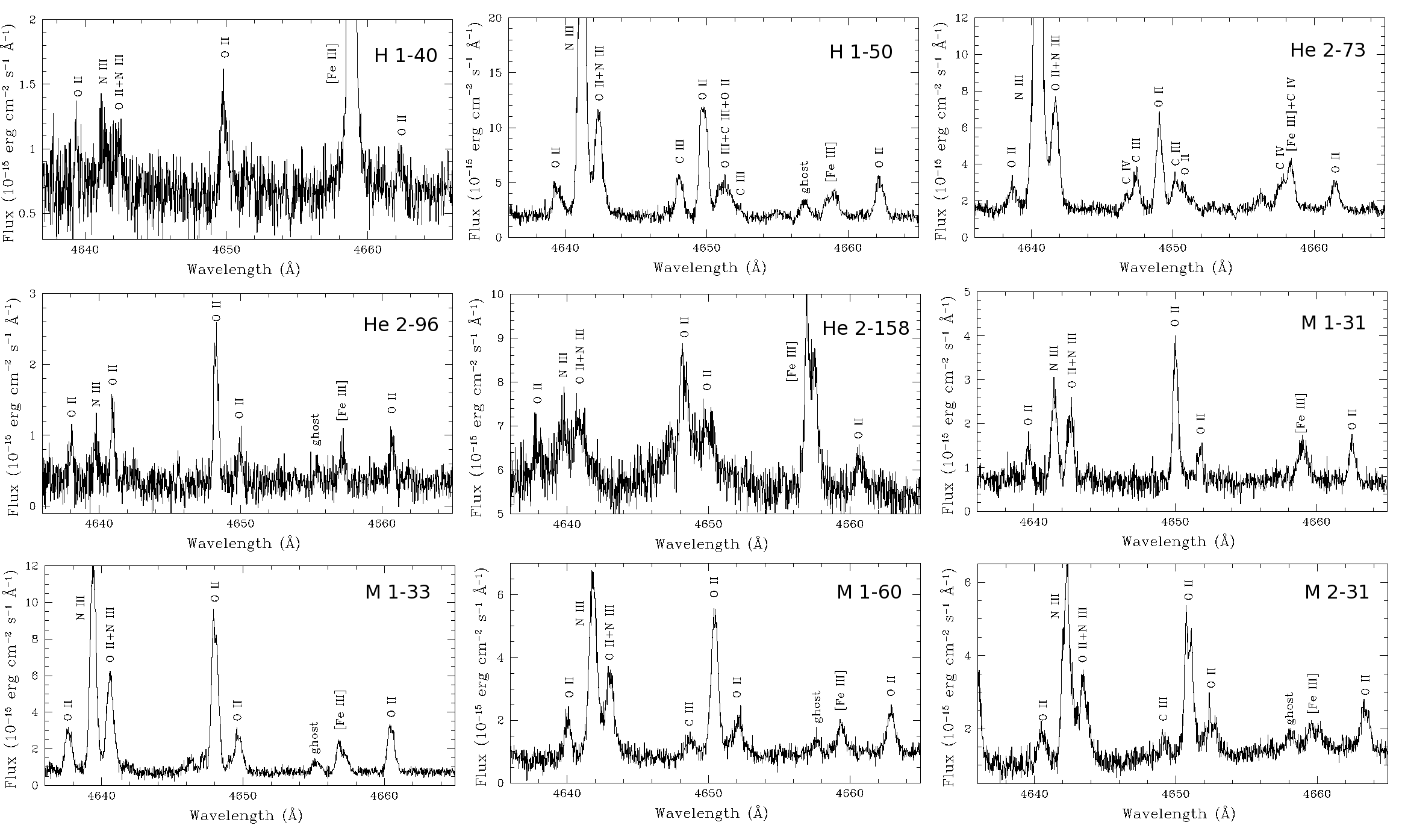}
\caption{Portion of the spectra showing the {\oii} multiplet 1 around 4650 \AA\ in all the planetary nebulae of our sample.} 
\label{oii_rec_lines}
\end{figure}
\end{landscape}

We measure {\oii} ORLs belonging to multiplet 1 in all the PN of our sample. In Fig.~\ref{oii_rec_lines} we show the quality of the spectra around this multiplet. We compute O$^{2+}$/H$^+$ ratios taking into account departures from local thermodynamic equilibrium (LTE) from the upper levels of
the transitions of this multiplet for densities {\elecd} $<$ 10$^4$ cm$^{−3}$ \citep{ruizetal03, tsamisetal03} and using the recombination coefficients computed by \citet{storey94}.  To account for such effect, we apply the non-LTE corrections estimated by \citet{apeimbertetal05}, obtaining abundances from individual lines that are in better agreement than when not considering this effect. The abundance using the sum of all lines of the multiplet \citep[see][]{estebanetal98} is not affected by non-LTE effects and generally agrees with those derived from individual lines.  In Table~\ref{tab:rec_oii} we show the ionic abundances from {\oii} ORLs. \citet{escalanteetal12} found that fluorescense may contribute significantly (up to $\sim$20\%) to the intensity of {\oii} multiplet 1 lines in low-excitation PNe, such as the case of IC\,418, but we do not have PNe in our sample with such low ionization degree. Hence, we consider the effect of fluorescense to be negligible in {\oii} multiplet 1 lines in the PNe of our sample. However, some objects of the literature sample (see Sect.~\ref{sec:lit_data}) have similar ionization degree than IC\,418 (i.e. M\,1-25 and M\,1-32) and fluorescense cannot be discarded as a possible excitation mechanism of {\oii} multiplet 1 lines. 

\setcounter{table}{5}
\begin{table*}
\begin{small}
\caption{Ionic abundance ratios from the {\oii} recombination lines$^{\rm a}$}
\label{tab:rec_oii}
\begin{tabular}{cccccccccccccc}
\noalign{\hrule} \noalign{\vskip3pt}
Mult.& $\lambda_0$ & \mc{9}{c}{O$^{++}$/H$^+$ ($\times$10$^{-5}$)} 	\\
\noalign{\vskip3pt} \noalign{\hrule} \noalign{\vskip3pt}
&  				&          H\,1-40 &     H\,1-50 	&     He\,2-73 	&     He\,2-96 	&     He\,2-158 	&      M\,1-31     &        M\,1-33   	& M\,1-60      	&  M\,2-31   	 \\
\noalign{\smallskip} \noalign{\hrule} \noalign{\smallskip}
1$^{\rm b}$& 4638.85	& 67:			& 91$\pm$28	& 64$\pm$22	& 58$\pm$18	& 31:			& 80$\pm$22	& 131$\pm$23    & 113$\pm$27	& 74$\pm$22	\\
& 4641.81			& 55:			& 91$\pm$31	& 50$\pm$15	& 50$\pm$14	& 23:			& 95$\pm$46	& 135$\pm$25    & 163$\pm$32	& 92$\pm$12	\\
& 4649.14			& 87$\pm$32	& 108$\pm$15	& 87$\pm$12	& 65$\pm$14	& 32$\pm$10	& 112$\pm$25	& 172$\pm$16    & 144$\pm$16	& 101$\pm$21	\\
& 4650.84			& ---	   	& 124$\pm$49	& 67$\pm$35	& 54$\pm$24	& 43$\pm$20	& 77:			& 136$\pm$20    & 146$\pm$27	& 67$\pm$23	\\
& 4661.64			& 112:		& 97$\pm$20	& 84$\pm$26	& 60$\pm$22	& 28:			& 123$\pm$43	& 139$\pm$36    & 119$\pm$22	& 99$\pm$23	\\
& 4673.73			& ---	   	& ---	  	& 104:		& ---		& ---		& 192:		& 130:		 & 151:		& 107:		\\
& 4676.24			& 115:		& 93$\pm$26	& 82$\pm$34	& 73$\pm$32	& ---		& 116$\pm$45 	& 158$\pm$33    & 127$\pm$34	& 85$\pm$37	\\
& 4696.14			& ---	   	& ---	  	&  ---	  	& ---		& ---		& ---		& --- 	     	& ---	    	& ---	   \\
& Sum			&{\bf 87$\pm$32}&{\bf 101$\pm$23}&{\bf 74$\pm$18}&{\bf 60$\pm$15}& {\bf 36$\pm$12}&{\bf 106$\pm$29}&{\bf 149$\pm$21} &{\bf 141$\pm$22}&{\bf 90$\pm$22}  \\
\noalign{\smallskip} \noalign{\hrule} \noalign{\smallskip}
\end{tabular}
\begin{description}
\item[$^{\rm a}$] Colons indicate uncertainties higher than 40\%. Only lines with intensity uncertainties lower than 40\% were considered to compute averaged values (see text).
\item[$^{\rm b}$] Corrected for non-LTE effects (see text).
\item[$^{\rm c}$] Blended with another line or affected by internal reflections or charge transfer in the CCD.
\end{description}
\end{small}
\end{table*}

It is to be noted here that the aim of this paper is not to address the ``abundance discrepancy problem'' \citep[see e.g.][and references therein]{garciarojasesteban07, mcnabbetal13}; some words on this topic, however, should be given here. This problem is well known in nebular physics and consist in the fact that, for a given ion, ORLs provide systematically higher abundances than those obtained using CELs; this difference is generally parametrized by the abundance discrepancy factor (ADF), which is the ratio between ORLs and CELs abundances. \citet{wangliu07} and \citet{delgadoingladarodriguez14} showed that when computed simultaneously, ADFs given by N, O and C ions show an overall agreement that translates in very similar N/O and C/O when computed from both CELs and ORLs. With this method we avoid the biases discussed by \citet{rolastasinska94} as we have computed the abundances from high-signal-to-noise optical spectra and used the same type of lines to compute C and O abundances. Additionally, \citet{rolastasinska94} pointed out that the fluxes of {\cii} faint recombination lines could be systematically overestimated if the signal to noise is low; this argument does not apply anymore as with the combination of high-sensitive detectors and large aperture telescopes, the {\cii} and {\oii} ORLs have been detected with high signal to noise in multiple PNe \citep[see Figs.~\ref{cii_rec_lines} and ~\ref{oii_rec_lines}, and ][and references therein]{garciarojas17}.

\subsection{Ionization correction factors and total abundances}\label{sec:total_ab}

To compute total abundances we have to correct for unseen ionization stages. To do this, we adopt a set of ionization correction factors (ICFs). For most elements we adopted the ICFs proposed by D-I14 from a large grid of photoionization models. We checked the validity range of the ICFs, depending on the excitation of each PN. In the majority of the cases, the ICFs can be applied. However, these ICFs are not always appropriate and we have considered alternative computations for particular cases (see below). All the elemental abundances are shown in Table~\ref{tab:total_ab}.

Following the prescriptions given by D-I14, we consider that the contribution of neutral He to the total abundance when He$^{2+}$ is present is expected to be negligible. In the case of PNe with no He$^{2+}$ we did not try any correction owing to neutral He because the available ICFs introduce a trend with the ionization degree (D-I14). However, all the objects in our sample and in the literature sample (see Sect.~\ref{sec:lit_data}) are of relatively high ionization degree, (O$^{2+}$/(O$^{+}$ + O$^{2+}$) $>$ 0.7, except M\,1-32 which shows O$^{2+}$/(O$^{+}$ + O$^{2+}$)=0.54. Hence, the correction owing to neutral He proposed by D-I14 should be very small or even negligible.

We compute the total C abundance only from ORLs. Equation 39 of D-I14 proposed an ICF based on C$^{2+}$ and O$^{2+}$ abundances derived from ORLs.

We use  the classical scheme N/O = N$^{+}$/O$^{+}$ to compute the N abundance \citep{peimberttorrespeimbert71}. The ICF proposed by D-I14 seems to introduce a trend with the degree of ionization in some nebulae (D-I15) but it is adequate for density-bounded PNe and high-excitation radiation-bounded PNe (Delgado-Inglada et al. in preparation). The fact that we observe the {\foi} $\lambda$6300 line in all but one of the PNe in our sample indicates that they are likely radiation-bounded nebulae. The low value of He$^{2+}$/(He$^{+}$ + He$^{2+}$), $< 0.2$ in all the PNe, suggests that the PNe are relatively low excitation objects. Therefore, for the subsequent analysis we will adopt the N abundances computed using the classical scheme. These arguments can be extended for the literature sample analysed in Sect.~\ref{sec:lit_data}.

To compute the total O abundance from ORLs we take into account the contribution of O$^+$ by scaling the O$^+$/O$^{2+}$ obtained from CELs and, when {\heii} lines are detected in the spectrum, the ICF provided by D-I14 to take into account the contribution of O$^{3+}$. To compute the total O/H ratio from CELs we follow the prescriptions given by D-I14.

We detect {\fneiii}, {\fneiv}, and {\fnev} lines in the spectra of H\,1-50 and He\,2-73. For these objects, we compute the abundances in two ways: i) by adding the ionic chemical abundances, and (ii) by using the ICF scheme of D-I14. We use the ICFs given by their equations 17 and 20, which give the total Ne abundance when only {\fneiii} lines are observed and when both {\fneiii} and {\fnev} lines are observed, respectively. For the other objects we correct for the unseen ionization stages by using equation 17 of D-I14. 

As we detect {\fclii}, {\fcliii} and {\fcliv} lines in all our objects we can calculate Cl abundances by simply adding up the ionic abundances of Cl$^{+}$, Cl$^{++}$ and Cl$^{+3}$ (these total abundances are labeled as Cl$^{\rm k}$ in Table~\ref{tab:total_ab}). According to the photoionization models of D-I14, the contribution of higher ionization states of Cl to the total abundance will be lower than 0.07 dex for objects with excitation characteristics like those in our sample. We compare these values with those obtained using equation 32 from D-I14 to correct for unseen Cl higher ionization states (marked as Cl$^{\rm j}$ in Table~\ref{tab:total_ab}). There is an excellent agreement between both determinations, indicating that the ICF from equation 32 is working well and that the contribution of higher ionization states is rather small or even inexistent. We also use equation 29 from D-I14, which provides an ICF when only {\fcliii} lines are observed. This equation is valid when O$^{++}$/(O$^{+}$ + O$^{++}$) $> 0.02$, which is the case for all the studied PNe. The derived values are represented as Cl$^{\rm i}$ in Table~\ref{tab:total_ab}. The Cl abundances derived with this ICF are, in general, somewhat higher (up to 0.16 dex) than the previous ones. The dispersion shown by the photoionization models around equation 29 is higher than that around equation 32 for the ranges of O$^{++}$/(O$^{+}$ + O$^{++}$) and He$^{++}$/(He$^{+}$ + He$^{++}$) displayed by the PNe studied here (see Fig.~11 and 12 from D-I14), indicating that the ICF from equation 29 provides a more uncertain Cl/H value. Therefore, we choose the Cl abundances labeled as Cl$^{\rm j, k}$ as the most reliable ones and we use the average between both determinations in the subsequent analysis.\\


H\,1-50 and He\,2-73 are the most excited PNe in our sample and we have detected {\farv} lines in their spectra, besides the {\fariii} and {\fariv} lines detected in all our sample of PNe. For these two high excitation PNe, we compute the total Ar abundance as the direct sum of the observed ionization states. As it can be seen from Table~\ref{tab:total_ab}, the abundances agree (within uncertainties) with those derived using the ICF given by equation 36 of D-I14, which is used when only {\fariii} lines are observed. For the rest of the objects in the sample, we use this equation to compute the total Ar abundance.

In H\,1-50 and He\,2-73 we measure several lines of multiple ionization stages of Fe; particularly, and thanks to the high excitation of these PNe we detect relatively bright lines of {\ffev}, {\ffevi} and, in H\,1-50, {\ffevii}. On the other hand, we measure the most commonly detected {\ffeiii} lines in all the PNe in our sample. We use the correction scheme suggested by \citet{rodriguezrubin05}, which is based on the detection of {\ffeiii} and an observational fit, given by their equation 3. An alternative approach could be to compute the total Fe abundance by simply adding up all the observed ionic species; however, this approach can only be applied to H\,1-50 and He\,2-73 and it is quite uncertain as lines from Fe$^{3+}$ were not detected in these objects. 

The total Kr abundance in He\,2-73 was computed by using equation 3 of \citet{sterlingetal15} which accounts for unobserved ions when only {\fkriv} lines are detected.

\setcounter{table}{6}
\begin{table*}
\begin{small}
\caption{Total abundances of sample A.}
\label{tab:total_ab}
\begin{center}
\begin{tabular}{lccccccccc}
\noalign{\smallskip} \noalign{\smallskip} \noalign{\hrule} \noalign{\smallskip}
 & \multicolumn{9}{c}{12 + log(X/H)} \\
\noalign{\smallskip} \noalign{\hrule} \noalign{\smallskip}
                 Ion 	&        H\,1-40 			&        H\,1-50			 &      He\,2-73 			&      He\,2-96 			&     He\,2-158 	  	&      M\,1-31 			&     M\,1-33 			&     M\,1-60  &       M\,2-31         \\
\noalign{\smallskip} \noalign{\hrule} \noalign{\smallskip}
             He  	& 11.10$\pm$0.02 		& 11.03$\pm$0.01 		& 11.05$\pm$0.01 		& 11.08$\pm$0.02 		& 11.04$\pm$0.01 	& 11.18$\pm$0.02 		& 11.12$\pm$0.01 		& 11.13$\pm$0.01 		& 11.05$\pm$0.01 		\\[3pt]
             C$^{\rm a}$&  8.38$^{+0.09}_{-0.08}$&  8.64$\pm$0.05	&  8.89$\pm$0.03	&  8.58$\pm$0.03		& 8.37$^{+0.09}_{-0.08}$& 8.76$^{+0.05}_{-0.06}$&  9.06$\pm$0.03	&  8.97$\pm$0.03	&  8.76$\pm$0.03		\\[3pt]
             N$^{\rm b}$& 8.10$^{+0.13}_{-0.18}$ 	& 8.20$\pm$0.07 	& 8.27$\pm$0.08 		& 8.09$^{+0.09}_{-0.10}$ 	& 7.91$\pm$0.05 	& 8.44$\pm$0.11 		& 8.54$\pm$0.06 		& 8.59$\pm$0.06 		& 8.19$\pm$0.06 		\\[3pt]
             O		& 8.50$\pm$0.06 		& 8.68$\pm$0.04 		& 8.61$\pm$0.04 		& 8.68$^{+0.10}_{-0.05}$ 	& 8.48$\pm$0.04 	& 8.76$^{+0.11}_{-0.06}$ 	& 8.78$\pm$0.04 		& 8.76$\pm$0.04 		& 8.60$\pm$0.04 		\\[3pt]     
            O$^{\rm d}$&  8.96$\pm$0.17&  9.05$\pm$0.10	& 8.96$^{+0.10}_{-0.11}$	& 8.82$^{+0.12}_{-0.11}$	& 8.69$^{+0.16}_{-0.15}$& 9.09$^{+0.17}_{-0.11}$&  9.21$\pm$0.07 & 9.19$\pm$0.07 &  8.99$\pm$0.11	\\[3pt]
         Ne$^{\rm e}$& 8.18$^{+0.09}_{-0.07}$ 	& 8.13$\pm$0.04		& 8.10$\pm$0.04 		& 8.38$^{+0.13}_{-0.07}$ 	& 8.16$\pm$0.05 	& 8.55$^{+0.19}_{-0.09}$ 	& 8.41$\pm$0.04 		& 8.41$\pm$0.05 		& 8.15$\pm$0.05 		\\[3pt]
         Ne$^{\rm f}$&   \nodata    			& 8.15$\pm$0.04		& 8.12$\pm$0.04 		&   \nodata    			&   \nodata        		&   \nodata    			& \nodata 				& \nodata 				&   \nodata    			\\[3pt]
         Ne$^{\rm g}$&   \nodata    			& 8.37$\pm$0.06		& 8.37$\pm$0.06 		&   \nodata    			&   \nodata        		&   \nodata    			& \nodata 				& \nodata 				&   \nodata    			\\[3pt]
            S$^{\rm h}$& 6.88$\pm$0.08 		& 6.85$\pm$0.04 		& 6.87$\pm$0.04 		& 7.08$^{+0.06}_{-0.05}$ 	& 6.68$\pm$0.03 	& 7.09$\pm$0.06 		& 7.13$\pm$0.04 		& 7.14$\pm$0.05 		& 6.92$\pm$0.04		\\[3pt]
         Cl$^{\rm i}$&5.17$^{+0.11}_{-0.09}$ 	& 5.14$\pm$0.03 		& 5.10$\pm$0.04 		& 5.28$^{+0.08}_{-0.07}$ 	& 4.94$\pm$0.06 	& 5.43$\pm$0.07 		& 5.38$\pm$0.03 		& 5.43$\pm$0.04 		& 5.17$\pm$0.04 		\\[3pt]
         Cl$^{\rm j}$&5.01$^{+0.11}_{-0.09}$ 	& 5.20$\pm$0.03 		& 5.13$\pm$0.04 		& 5.15$^{+0.08}_{-0.07}$ 	& 4.88$\pm$0.06 	& 5.31$\pm$0.07 		& 5.30$\pm$0.03 		& 5.37$\pm$0.04 		& 5.12$\pm$0.04 		\\[3pt]
         Cl$^{\rm k}$& 5.02$^{+0.11}_{-0.08}$ 	& 5.21$\pm$0.03 		& 5.13$\pm$0.03 		& 5.16$^{+0.14}_{-0.17}$ 	& 4.88$\pm$0.05 	& 5.32$^{+0.09}_{-0.07}$ 	& 5.31$\pm$0.03 		& 5.38$\pm$0.04 		& 5.13$\pm$0.04 		\\[3pt]
          Ar$^{\rm l}$& 6.61$^{+0.08}_{-0.11}$ 	& 6.32$\pm$0.03		& 6.46$\pm$0.04		& 6.73$^{+0.07}_{-0.11}$ 	& 6.21$\pm$0.05 	& 6.84$^{+0.08}_{-0.13}$ 	& 6.67$\pm$0.04		& 6.72$\pm$0.04 		& 6.39$\pm$0.03 	  \\[3pt]
          Ar$^{\rm m}$&   \nodata    			& 6.38$\pm$0.04		& 6.46$\pm$0.04		&   \nodata    			&   \nodata			&   \nodata    			& \nodata 				& \nodata 				&   \nodata    			 \\[3pt]
          Fe$^{\rm n}$& 6.32$\pm$0.07 		& 5.01$\pm$0.11 		& 5.15$^{+0.09}_{-0.11}$ 	& 4.71$^{+0.15}_{-0.16}$ 	& 5.51$\pm$0.08        & 5.46$^{+0.16}_{-0.14}$ 	& 5.43$\pm$0.08 		& 5.19$\pm$0.10	 	& 5.04$^{+0.14}_{-0.16}$ 		\\[3pt]
          Kr$^{\rm o}$&   \nodata    			& \nodata 				& 3.79$^{+0.13}_{-0.14}$ 	&   \nodata    			&   \nodata        		&   \nodata    			& \nodata 				& \nodata 				&   \nodata    			\\[3pt]
\noalign{\smallskip} \noalign{\hrule} \noalign{\smallskip}
\end{tabular}
\begin{description}
\item[$^{\rm a}$] From {\cii} ORLs and the ICF by D-I14.
\item[$^{\rm b}$] Classic ICF scheme of N/O$\sim$N$^+$/O$^+$ (see text). 
\item[$^{\rm d}$] From {\oii} ORLs, O$^+$/O$^{2+}$ from CELs, and the ICF by D-I14.
\item[$^{\rm e}$] ICF from equation 17 of D-I14.
\item[$^{\rm f}$] ICF from equation 20 of D-I14.
\item[$^{\rm g}$] Sum of Ne$^{2+}$,  Ne$^{3+}$ and  Ne$^{4+}$.
\item[$^{\rm h}$] Where {\heii} lines were not detected, ICF from equation 36 of \citet{kingsburghbarlow94}. Where {\heii} lines were detected, ICF from equation 26 of D-I14.
\item[$^{\rm i}$] ICF from equation 29 of D-I14.
\item[$^{\rm j}$] ICF from equation 32 of D-I14.
\item[$^{\rm k}$] Sum of Cl$^{+}$,  Cl$^{2+}$ and  Cl$^{3+}$.
\item[$^{\rm l}$] ICF from equation 36 of D-I14.
\item[$^{\rm m}$] Sum of Ar$^{2+}$,  Ar$^{3+}$ and  Ar$^{4+}$.
\item[$^{\rm n}$] ICF from equation 3 by \citet{rodriguezrubin05}
\item[$^{\rm o}$] ICF from equation 3 by \citet{sterlingetal15}
\end{description}
\end{center}
\end{small}
\end{table*}

\subsubsection{High-quality spectroscopic data from the literature}\label{sec:lit_data}

In order to extend our sample of DC PNe, we have computed (following exactly the same methodology) the elemental abundances for the eight DC PNe present in the sample studied by D-I15: Cn\,1-5, H\,1-50, M\,1-42, M~2-27, M~2-31, MyCn~18, NGC~6439, and NGC~7026, and for the nine DC PNe studied by \citet{garciarojasetal12} and \citet{garciarojasetal13}: Cn\,1-5, Hb\,4, He\,2-86, M\,1-25, M\,1-32, M\,1-61, M\,3-15, NGC\,2867 and PB\,8. For simplicity, we will refer to the new PNe reported here as sample A, while sample B and C will be composed by the PNe previously studied by D-I15 and Garc\'{\i}a-Rojas et al., respectively. There are two objects in sample B in common with our sample (H\,1-50 and M\,2-31) and there is also an object that is in both samples B and C (Cn\,1-5). In general, we have found a good agreement between the abundances computed for the different samples (within 0.1 dex), with the exception of Cl/H in M\,2-31, where we found a Cl abundance about 0.2 dex higher when using the data from D-I15, and for Ar/H in Cn\,1-5, which is more than 0.2 dex lower when using the data from the compilation of D-I15. As the quality of our new observational data (sample A) or the data from sample C are much better than those of sample B, we will not consider H\,1-50, M\,2-31 and Cn\,1-5 from sample B, hereinafter. All  the recomputed abundances of samples B and C, as well as some properties of the PNe, such as Galactic component they belong, Galactocentric distances, height over the Galactic plane, and dust type, are presented in Table~\ref{tab:total_ab_lit}.

\begin{landscape}
\setcounter{table}{7}
\begin{table}
\caption{Abundances recomputed for DC PNe from the literature (samples B and C).}
\label{tab:total_ab_lit}
\begin{center}
\begin{tabular}{lccccccccccccc}
\noalign{\smallskip} \noalign{\smallskip} \noalign{\hrule} \noalign{\smallskip}
 &  & & & \multicolumn{8}{c}{12 + log(X/H)} & \\
\noalign{\smallskip} \noalign{\hrule} \noalign{\smallskip}
 	&  Gal.  & $R_G$ & $z$ & Dust & 	&     	&    	&  		&     	&      &   &  & Ref.  \\
PNe 	&  Comp. & (kpc)$^{\rm a}$ & (pc)$^{\rm a}$ & type & He/H 	&     N/H	&   O/H 	&  Ne/H		&    Ar/H  	&    S/H  &  Cl/H & log(C/O) (RLs) & (Sample) \\
\noalign{\smallskip} \noalign{\hrule} \noalign{\smallskip}
Cn\,1-5 & bulge & 2.504 & $-$920	& DC$_{\rm cr}$ & 11.10$\pm$0.01  & 8.72$^{+0.04}_{-0.05}$ & 8.78$\pm$0.04 	& 8.62$^{+0.05}_{-0.04}$  & 6.82$\pm$0.05	& 7.15$\pm$0.04	& 5.47$\pm$0.03	& 0.08$\pm$0.07 & 1	(C) \\[3pt]
Hb\,4 & bulge & 2.932 &	260 & DC$_{\rm cr}$ & 11.05$\pm$0.02  & 8.93$^{+0.10}_{-0.11}$ & 8.66$\pm$0.05 	& 8.12$^{+0.06}_{-0.05}$  & 6.58$^{+0.05}_{-0.07}$ & 7.12$^{+0.09}_{-0.10}$	& 5.28$\pm$0.04	& $-$0.29$^{+0.06}_{-0.05}$ & 1	(C)\\[3pt]
He\,2-86 & disc & 7.139 & $-$217 &  DC$_{\rm am+cr}$ & 11.11$\pm$0.02  & 8.81$^{+0.09}_{-0.13}$ & 8.75$\pm$0.04  & 8.37$^{+0.06}_{-0.04}$ & 7.03$^{+0.06}_{-0.08}$ & 7.25$^{+0.08}_{-0.10}$	& 5.37$^{+0.05}_{-0.04}$ & $-$0.14$\pm$0.03 & 1 (C) \\[3pt]
M\,1-25 & bulge & 2.527 & 898 &  DC$_{\rm cr}$ & 11.11$\pm$0.02  & 8.43$^{+0.06}_{-0.04}$ & 8.78$^{+0.09}_{-0.06}$  & 7.92$^{+0.13}_{-0.08}$  & 6.70$\pm$0.05 & 7.18$^{+0.07}_{-0.06}$	& 5.42$^{+0.05}_{-0.04}$ & $-$0.21$\pm$0.08 & 1 (C) \\[3pt]
M\,1-32 & disc & 3.466  & 354 & DC$_{\rm cr}$ & 11.10$\pm$0.02  & 8.46$^{+0.06}_{-0.05}$ & 8.53$^{+0.13}_{-0.09}$  & 7.82$^{+0.15}_{-0.11}$ & 6.54$\pm$0.06 & 7.16$\pm$0.08	& 5.30$\pm$0.05 & 0.39$^{+0.09}_{-0.11}$ & 1 (C) \\[3pt]
M\,1-42 & bulge & 2.465 & $-$472 & DC$_{\rm cr}$ & 11.23$\pm$0.02  & 8.70$\pm$0.04 & 8.47$\pm$0.04 	& 8.04$^{+0.04}_{-0.05}$  & 6.46$\pm$0.04	& 6.93$\pm$0.04	& 5.23$\pm$0.04	& $-$0.24$\pm$0.06 & 2 (B) \\[3pt]
M\,1-61 & disc & 3.433  &$-$502	& DC$_{\rm am+cr}$ & 11.06$\pm$0.02  & 8.33$^{+0.11}_{-0.14}$ & 8.65$^{+0.06}_{-0.04}$  & 8.22$^{+0.07}_{-0.05}$ & 6.82$^{+0.06}_{-0.09}$ & 7.05$^{+0.10}_{-0.11}$	& 5.14$^{+0.05}_{-0.04}$ & $-$0.14$\pm$0.06 & 1 (C) \\[3pt]
M\,2-27 & bulge & 4.500 & $-$1005 & DC$_{\rm cr}$ & 11.15$^{+0.02}_{-0.03}$ & 8.91$\pm$0.06 & 8.86$\pm$0.04 	& 8.53$\pm$0.04  & 6.73$\pm$0.03 & 7.33$^{+0.06}_{-0.05}$ & 5.58$^{+0.06}_{-0.05}$	& $-$0.36$^{+0.06}_{-0.07}$	& 3 (B) \\[3pt]
M\,3-15 & bulge & 1.480 & 495 & DC$_{\rm cr}$ & 11.06$\pm$0.02  & 8.45$^{+0.11}_{-0.13}$ & 8.78$^{+0.06}_{-0.05}$  & 8.14$\pm$0.07  & 6.95$\pm$0.06 & 7.21$^{+0.11}_{-0.14}$	& 5.28$\pm$0.05 & $-$0.25$\pm$0.11 & 1 (C) \\[3pt]
MyCn\,18 & disc &  6.528 & $-$288 & DC$_{\rm am+cr}$ & 11.00$^{+0.02}_{-0.03}$  & 8.46$^{+0.06}_{-0.05}$ & 8.56$\pm$0.04 	& 8.40$\pm$0.04  & 6.49$\pm$0.04	& 7.26$\pm$0.06	& 5.49$\pm$0.04	& $-$0.50$^{+0.05}_{-0.06}$	& 4 (B) \\[3pt]
NGC\,2867 & disc &  7.996 & $-$233	& DC$_{\rm cr}$ & 11.03$\pm$0.01 & 7.94$\pm$0.05 & 8.56$\pm$0.04  & 7.88$\pm$0.04 & 6.18$\pm$0.04 & 6.67$^{+0.06}_{-0.05}$	& 4.97$\pm$0.03 & 0.38$^{+0.05}_{-0.06}$ & 5 (C) \\[3pt]
NGC\,6439 & bulge & 2.130 &	658 & DC$_{\rm cr}$ & 11.12$\pm$0.02  & 8.50$^{+0.05}_{-0.04}$ & 8.68$\pm$0.04 	& 8.20$\pm$0.04  & 6.59$\pm$0.03	& 7.08$^{+0.05}_{-0.06}$ & 5.38$\pm$0.04	& $-$0.04$^{+0.07}_{-0.06}$	& 3 (B) \\[3pt]
NGC\,7026 & disc & 8.232  & 14	& DC$_{\rm cr}$ & 11.08$\pm$0.02  & 8.60$^{+0.03}_{-0.04}$ & 8.71$\pm$0.02 	& 8.21$\pm$0.03  & 6.49$\pm$0.02	& 7.21$\pm$0.04	& 5.43$\pm$0.03	& $-$0.09$^{+0.06}_{-0.07}$	& 6 (B) \\[3pt]
PB\,8 & disc & 8.558  & 536 & DC$_{\rm cr}$ & 11.09$^{+0.01}_{-0.02}$ & 8.25$^{+0.07}_{-0.08}$ & 8.75$^{+0.05}_{-0.04}$  & 8.25$\pm$0.05 & 7.08$\pm$0.06 & 7.29$\pm$0.08	& 5.63$^{+0.07}_{-0.08}$ & $-$0.39$^{+0.05}_{-0.06}$ & 5 (C) \\[3pt]
\noalign{\smallskip} \noalign{\hrule} \noalign{\smallskip}
\end{tabular}
\begin{description}
\item[$^{\rm a}$] Galactocentric distances computed from heliocentric distances by \citet{stanghellinihaywood10} and using equation given in Table~\ref{tab:tobs}. $z$ is the height above the Galactic plane in pc and was computed following equation in Table~\ref{tab:tobs}. \\
References: (1) \citet{garciarojasetal12, garciarojasetal13}; (2) \citet{liuetal01}; (3) \citet{wangliu07}; (4) \citet{tsamisetal03}; (5) \citet{garciarojasetal09}; (6) \citet{wessonetal05}
\end{description}
\end{center}
\end{table}
\end{landscape}

\subsection{Comparison with other abundance determinations in the literature}

We have compared our abundance determinations for sample A objects with the recent study made by \citet{garciahdezgorny14} for a large sample of PNe compiled from the literature. All our sample A PNe, with the exception of M\,1-33 were studied by \citet{garciahdezgorny14}. In general, we find a very good agreement in the He and O abundances in both sets (agreement within 0.04 and 0.1 dex, respectively). The agreement between N abundances is also relatively good (within 0.2 dex); the same occurs for S, Ar, and Ne, except for some objects for which the Ar and Ne abundance differences reach higher than 0.3 dex. We ascribe these differences to the different sets of atomic data and ICFs used by \citet{garciahdezgorny14}\footnote{ \citet{garciahdezgorny14} computed Cl abundances, although they did not use them; they used an ICF scheme by \citet{liuetal00} but their reported total Cl abundances do not correspond with the use of this ICF scheme and are strongly overestimated, in most of the cases by more than 1 dex.}. In fact, the effects on the abundances of using different atomic data sets have been recently studied by \citet{juandediosrodriguez17}; these authors pointed out that for PNe with electron densities higher than 10$^4$ cm$^{-3}$, the use of different atomic data sets can produce differences in computed abundance ratios such as O/H or N/O of a factor of 4. We have avoided using the atomic data sets that reach to even higher differences, but the abundances of our high-density PNe are likely to be affected by the systematic uncertainties introduced by atomic data. On the other hand, the high quality and signal to noise of our spectra represent a substantial improvement to previous data sets; additionally we have used an up-to-date ICF scheme, which make our abundance determinations more reliable. 

C abundances have been previously computed from UV CELs only for 1 object of sample A (H\,1--50). \citet{wangliu07} reported a C/O ratio from UV/optical CELs of 0.15, which was later recomputed by \citet{delgadoingladarodriguez14} as 0.17. These values are much lower than the one obtained in this work: C/O = 0.39, which is almost coincident with the C/O ratio of 0.41 reported by \citet{delgadoingladarodriguez14} from C and O ORLs and assuming C/O$\sim$C$^{2+}$/O$^{2+}$. \citet{delgadoingladarodriguez14} compiled C/O ratios computed from both CELs and ORLs for several PNe in the literature; these authors found an overall agreement between C/O ratios computed from CELs and ORLs, however, they also pointed out some outliers, presenting significant different C/O ratios from CELs and ORLs. In the objects in common with our samples B and C, the computed C/O ratios from CELs and ORLs are somewhat different. The difference (C/O$_{CELs}$)$-$(C/O$_{ORLs}$) is 0.27, $-$0.20, $-$0.20, and $-$0.17 dex for Cn\,1-5, M\,1-42, NGC\,6439 and NGC\,7026, respectively. These differences can be attributed to several effects: i) as PNe are extended objects (at least, more extended than the slit width of optical observations), the volume of the PN covered by optical and UV observations are different and ionization structure effects could be affecting the computed C/O ratios, and ii) UV data are extremely sensitive to uncertainties in the computed extinction.

\section{Nucleosynthesis models for low- to intermediate-mass stars}\label{sec:models}

We compare the derived nebular abundances with two state-of-the-art sets of AGB nucleosynthesis models that use different approaches to the calculations. The AGB nucleosynthesis models calculated using the Monash code \citep{karakaslattanzio07} and presented by \citet[][hereinafter KL16]{karakaslugaro16} for $Z$ = 0.007, 0.014 and 0.03. The AGB nucleosynthesis models obtained by using the ATON code for stellar evolution \citep{mazzitelli89, venturaetal98, venturadantona09} presented by \citet{venturaetal13} for $Z$ = 0.008, by \citet{dicriscienzoetal16} for $Z$ = 0.018 and by \citet{garciahdezetal16b} for $Z$ = 0.04. We label this set of models as ATON16. 
The abundances represented by the models are the final AGB abundances. No extra nucleosynthetic products are expected to be carried into the nebula during the AGB-PN transition.
For the sake of simplicity, we will refer to the $Z_{\rm KL16}$=0.007 and $Z_{ATON16}$=0.008 as sub-solar metallicity models, $Z_{\rm KL16}$=0.014 and $Z_{\rm ATON16}$=0.018 as solar metallicity models and $Z_{\rm KL16}$=0.03 and $Z_{\rm ATON16}$=0.04 as super-solar metallicity models. The difference in metallicities between KL16 and ATON16 models comes from the different reference for Solar metallicity adopted by the two groups: KL16 adopt the solar abundances given by \citet{asplundetal09} while the ATON16 models assume solar abundances by \citet{grevessesauval95}. The differences in the assumed input physics between KL16 and ATON16 models come from the mass-loss rate adopted and from the treatment of convection during the AGB evolution. The KL16 models use the \citet{VassiliadisWood93} mass-loss rate whereas ATON16 models use the \citet{bloecker95} mass-loss prescription in the oxygen-rich regime with $\eta_{r}=0.02$ \citep{venturaetal2000}. In the ATON16 models during the carbon-rich phase the mass-loss rate based on the hydrodynamical models of carbon stars by \citet{wachteretal02, wachteretal08} is adopted. As for the convection treatment, the ATON16 models differ from those by KL16 in that they use the full spectrum of turbulence (FST) convective model \citep{canutomazzitelli91} instead of the mixing length theory. The net result is a more efficient HBB and a lower threshold mass for the activation of these processes in the ATON16 models. 

Finally, we note that KL16 included a partial mixing zone (PMZ) between the convective H-rich envelope and the intershell at the deepest extent of each TDU in some models; this was done to produce enough $^{13}$C nuclei in the He intershell, to release neutrons for an efficient production of the elements heavier than iron via slow neutron captures (the $s$-process), as required by the observations \citep[e.g.][]{bussoetal01}. All KL16 1.5--4.5 M$_{\odot}$ models for sub-solar metallicities, 1.5--4.75 M$_{\odot}$ models for solar metallicities and 2.5--4.5 M$_{\odot}$ models for super-solar metallicities include a PMZ. However, the presence of the PMZ has almost no effect on the abundances of interest here (it mainly affects the $s$-process elements), with at most an increase of 0.05 dex in the C abundances for models with small or no PMZ. Regarding the initial abundances, KL16 employed the solar abundances from \citet{asplundetal09} for $Z$=0.014 and scaled them by a factor of two up (for $Z$=0.03) or down ($Z$=0.007). In this way the initial elemental ratios are constant (as seen in the following plots). The $Z$ = 0.018 and $Z$ = 0.04 ATON models have a solar-scaled mixture, while an alpha-enhancement $[\alpha/{\rm Fe}]=+0.2$ is used for the $Z$=0.008 models.

\section{Comparison between observations and with AGB nucleosynthesis predictions}\label{sec:results}

The accurate abundances that we have obtained for sample A allow us to compare with the predictions of AGB stellar nucleosynthesis models to better constrain the progenitor masses of these objects. As we mentioned before, one of our aims is to understand the simultaneous presence of PAHs and silicates in DC PNe and the role of the progenitor mass in this behaviour.

\subsection{Metallicity indicators. Behaviour of chlorine and oxygen abundances in the sample}\label{sec:Cl_O}

Chlorine (Cl) is considered a better metallicity tracer for PNe than oxygen as there are observational evidences of O-enrichment in PNe at low-metallicity \citep[see e.g.][]{penaetal07} and at somewhat lower than solar-metallicity PNe (D-I15). Post-AGB stars in the Magellanic Clouds also show evidence of O-enrichment \citep[e.g.,][]{desmedtetal12}. This enrichment has been attributed to the PN progenitor stars having experienced dredge-up of material into the He-rich intershell from the C-O core, where O is present due to production via $^{12}$C($\alpha$,$\gamma$)$^{16}$O during the core He-burning \citep[see e.g.][]{pequignotetal2000}. The occurrence of such O dredge-up is supported by low-metallicity AGB theoretical models \citep{marigo01, herwig04, karakaslugaro10} and slightly subsolar-metallicity models that include diffusive overshooting \citep{garciahdezetal16a}.

\begin{figure}
\includegraphics[width=\columnwidth]{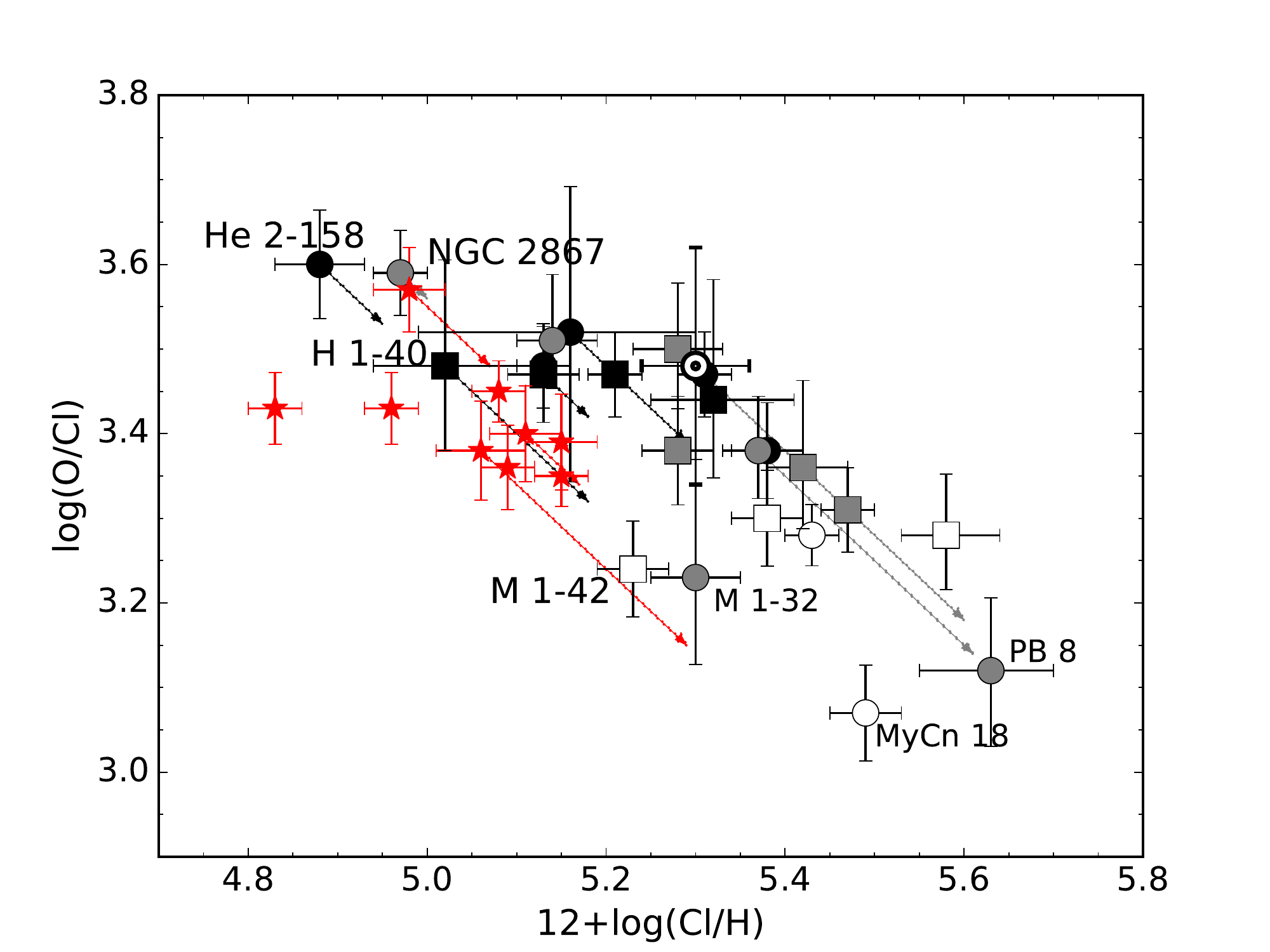}
\includegraphics[width=\columnwidth]{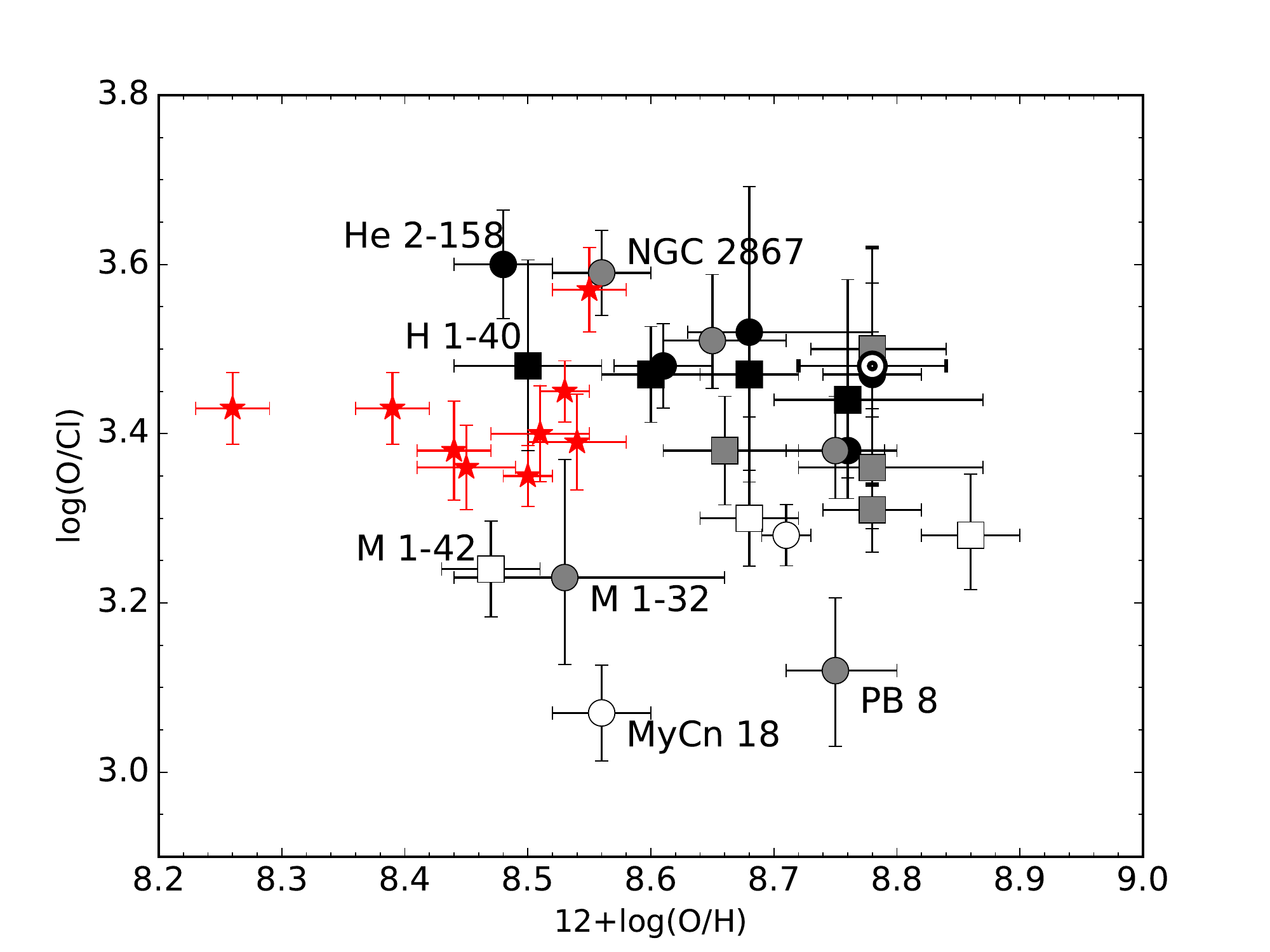}
\caption{Values of log(O/Cl) vs. 12+log(Cl/H) (upper panel) vs. 12+log(O/H) (lower panel) for our sample of objects. The Cl/H ratio has been computed by simply adding available ionic abundances or by using equation 29 of D-I14 when only {\fcliii} lines were available in the spectra (PNe of sample B and PB\,8). The effect of considering using the ICF of equation 29 of D-I14 is shown by arrows. Filled black symbols are the PNe of our sample (sample A); open symbols are DC PNe from D-I15 (sample B) and filled grey symbols are those studied by \citet[][sample C]{garciarojasetal13}. Squares are bulge PNe and circles are disc PNe. Red stars represent {\hii} region abundances from the sample of \citet{estebanetal15}. The protosolar abundances of \citet{lodders10} are overplotted with the solar symbol.} 
\label{Cl_PNHII}
\end{figure}

The values of the O/Cl abundance ratios are plotted in Fig.~\ref{Cl_PNHII} as a function of the Cl and O abundances for all our sample objects. The results are compared with the protosolar values of \citet{lodders10} and with the set of Galactic {\hii} regions (red stars in Fig.~\ref{Cl_PNHII}) studied by \citet{estebanetal15}. The most striking feature of Fig.~\ref{Cl_PNHII} is the apparent anticorrelation between the values of O/Cl and Cl/H, which would indicate that the O abundance varies more slowly than that of Cl as the metallicity increases. However, an anticorrelation between O/Cl and Cl/H might be introduced by errors in the determination of the Cl abundance. The Cl abundances of our sample objects have been calculated in an inhomogeneous way. For the PNe in sample A objects (filled black symbols in Fig.~\ref{Cl_PNHII}), the {\hii} regions, and for most of the PNe in sample C (grey symbols) the Cl abundances could be computed without applying any ICF correction, but for PB~8 and all the PNe in sample B (open symbols in Fig.~\ref{Cl_PNHII}), the Cl abundances are based on the ICF given by equation 29 of D-I14. D-I15 show in their Fig.~3 that the uncertainties of O/Cl and Cl/H, with the Cl abundances calculated with the ICF of equation 29 of D-I14, are anticorrelated, and can easily lead to an anticorrelation between these two quantities. In order to illustrate the effect of these different approaches, we show with arrows in the upper panel of Fig.~\ref{Cl_PNHII} the positions that some of the objects would occupy if the ICF of equation 29 of D-I14 had been used to derive their Cl abundances. It is clear from these arrows that most, if not all, of the anticorrelation between O/Cl and Cl/H can be explained by the effects of using an ICF for several objects. Besides, any other uncertainty in the determination of Cl/H that does not affect the calculation of O/H will have a similar effect, moving the points in the direction of the arrows.

\begin{figure}
\includegraphics[width=\columnwidth]{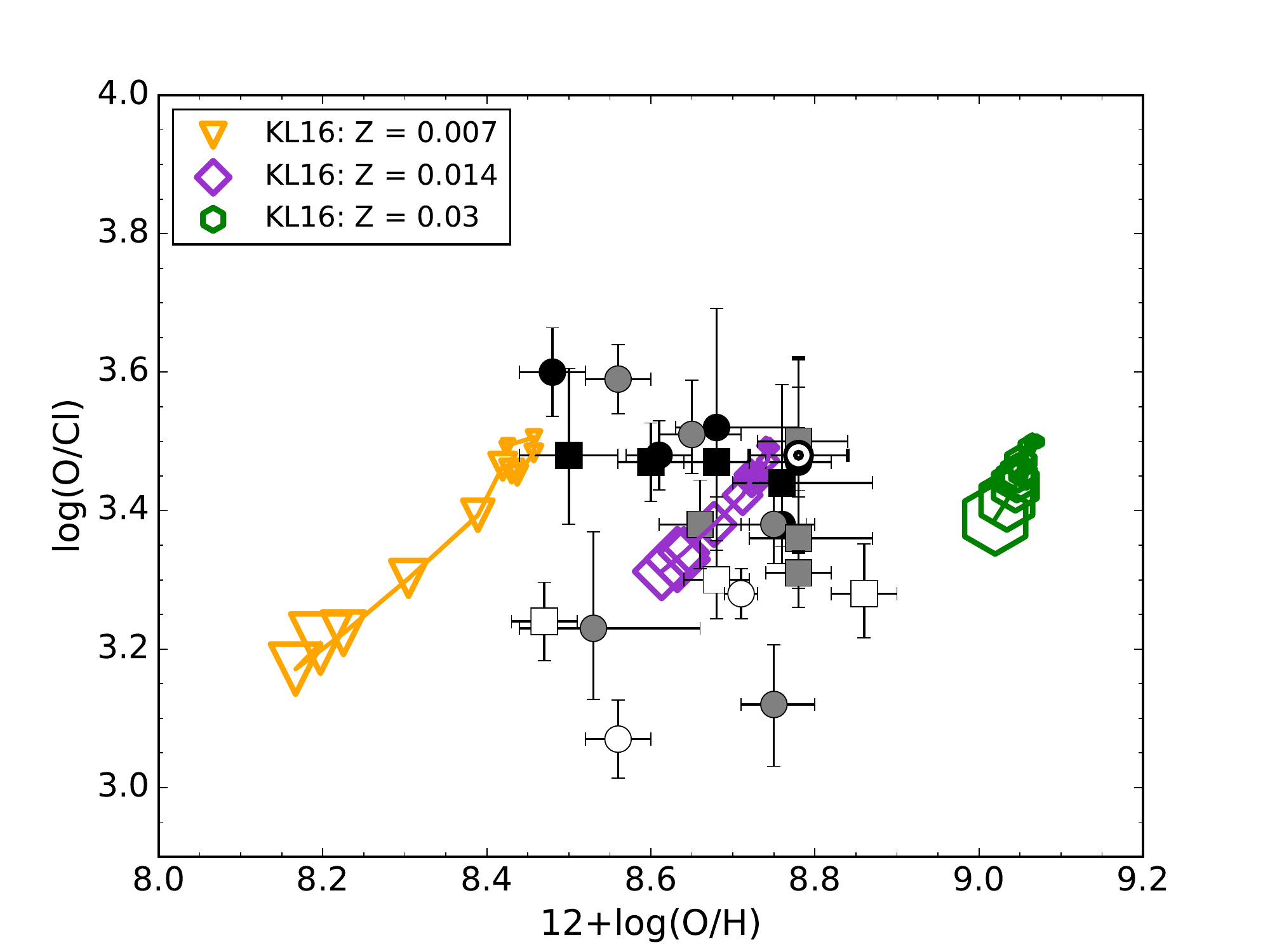}
\includegraphics[width=\columnwidth]{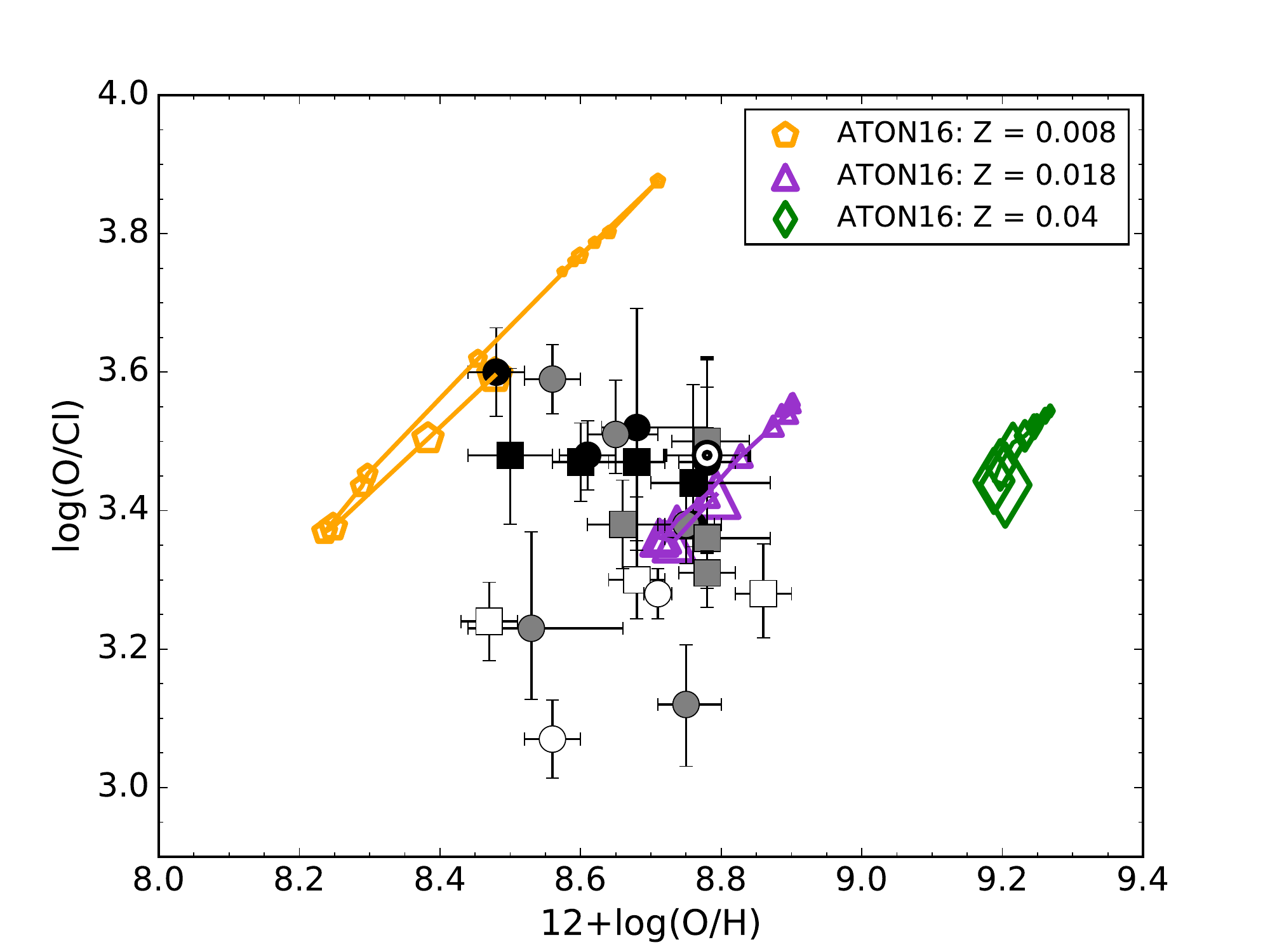}
\caption{Values of log(O/Cl) vs. 12+log(O/H) for our sample of objects compared with the predictions from the stellar nucleosynthesis models by \citet{karakaslugaro16} (K16, upper panel) and \citet{garciahdezetal16b} (ATON16, lower panel) for different metallicities. PN symbols are identified in Fig.~\ref{Cl_PNHII}. The protosolar abundances of \citet{lodders10} are overplotted with the solar symbol. The symbol sizes relate to the progenitor mass in the models.} 
\label{models_OCl}
\end{figure}

Our values of O/H are not affected by as many inhomogeneities or uncertainties as those plaguing our determinations of Cl/H, as can be seen by the lack of correlation between the values of O/Cl and O/H shown in the lower panel of Fig.~\ref{Cl_PNHII}. Furthermore, as we show in Fig.~\ref{models_OCl}, the stellar nucleosynthesis models that predict the largest variations in O/H do not seem to be representative of the PNe in our sample. Hence, for our objects O/H seems to be a better estimate than Cl/H of the initial metallicity of the progenitor stars. Taking O/H as our metallicity indicator, our PNe are the descendants of stars with metallicities from solar to half-solar.

It is also noticeable from Fig.~\ref{Cl_PNHII} that there is a significant number of PNe with higher O/H than the {\hii} regions. This would not be surprising if those PNe belong to the bulge, which has a higher metallicity population, but several of the disc PNe also show higher O/H ratios than those of {\hii} regions. D-I15 found also that several Galactic disc PNe showed higher Cl, O, Ne, and Ar abundances than Galactic {\hii} regions. Since these PNe are located between 6 and 9 kpc they can be considered solar neighborhood members, and gradient effects are excluded. D-I15 propose alternative explanations such as stellar migration of central stars, as it is well known to be required to interpret the spread in the age-metallicity relationship in the Galaxy  \citep[e.g., see][]{spitonietal15} or changes in the stellar composition arising during star formation or stellar evolution.

In Fig.~\ref{models_OCl} we show the O/Cl ratios as a function of the O/H abundance ratio. We have overplotted the two sets of stellar evolution models with different metallicities. In the KL16 models, the oxygen abundance decreases with progenitor mass due to oxygen destruction via HBB, whereas the ATON16 models show both oxygen production and destruction.
The metallicities of around solar and half-solar from the models of KL16 and ATON16 seem appropriate to cover all the metallicity range covered by our PNe, but we also include some higher metallicity models for completeness.

\subsection{The alpha-elements: Ne, S and Ar abundances}\label{sec:alpha}

\begin{figure}
\includegraphics[width=\columnwidth]{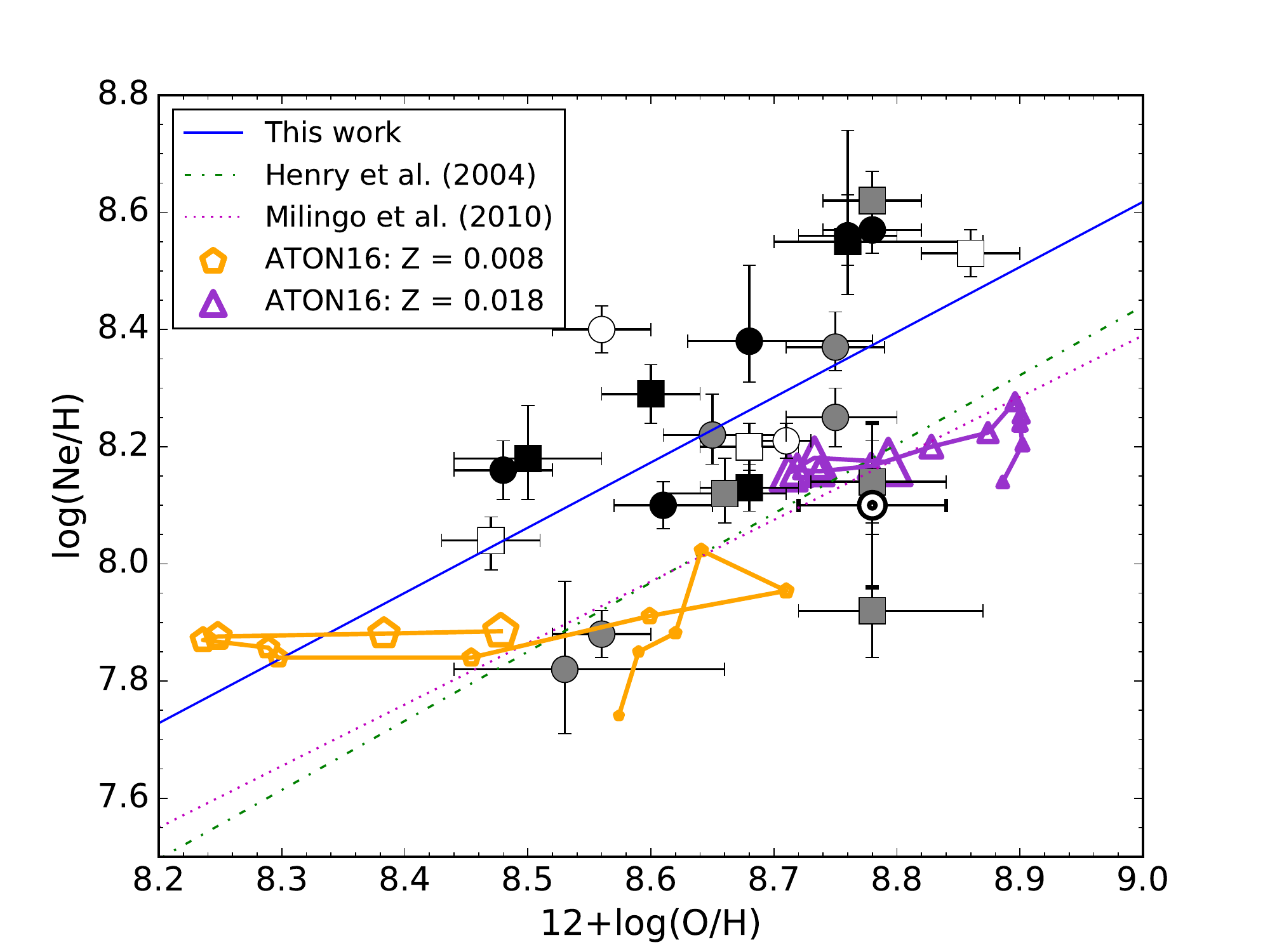}
\includegraphics[width=\columnwidth]{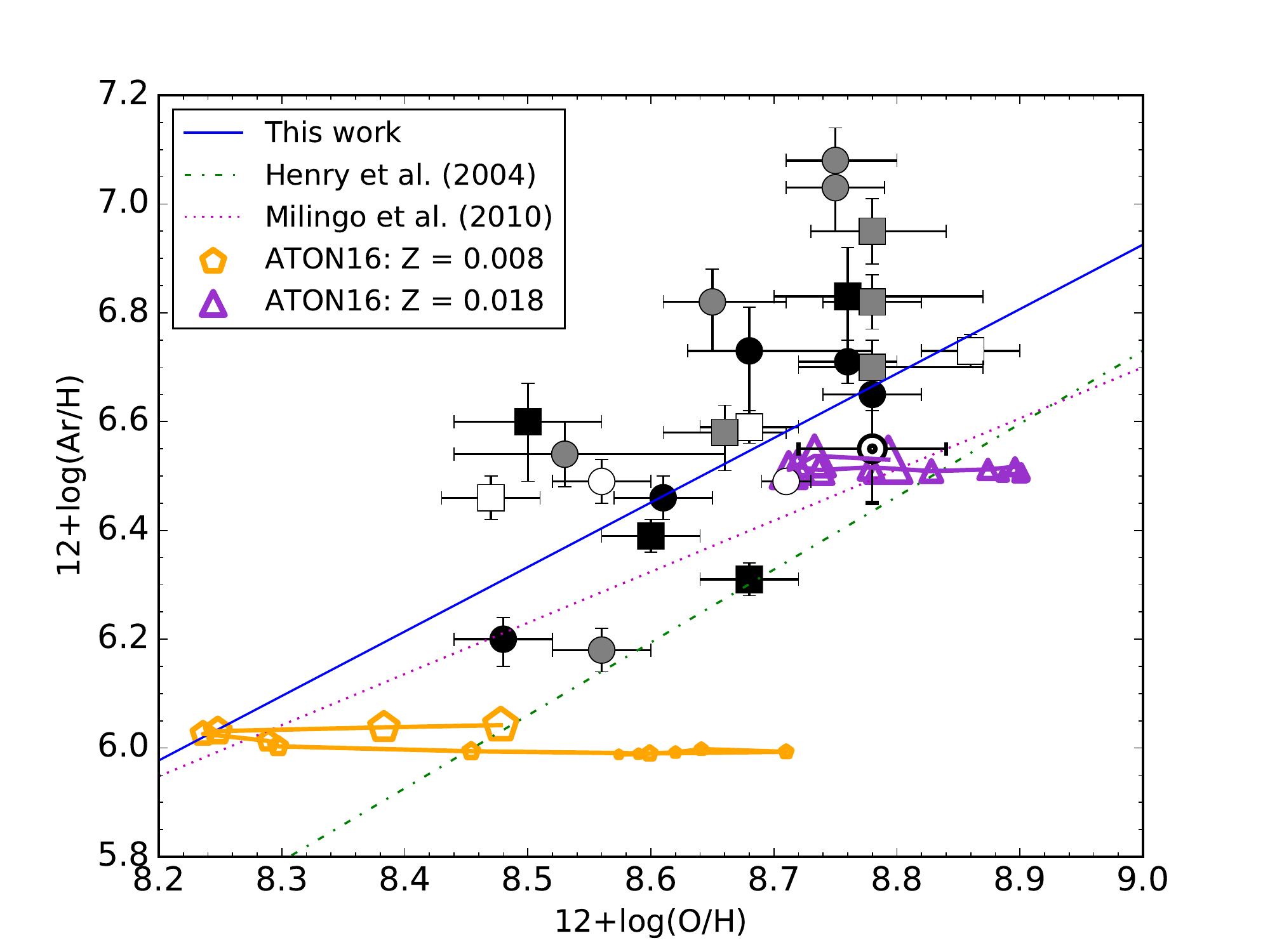}
\includegraphics[width=\columnwidth]{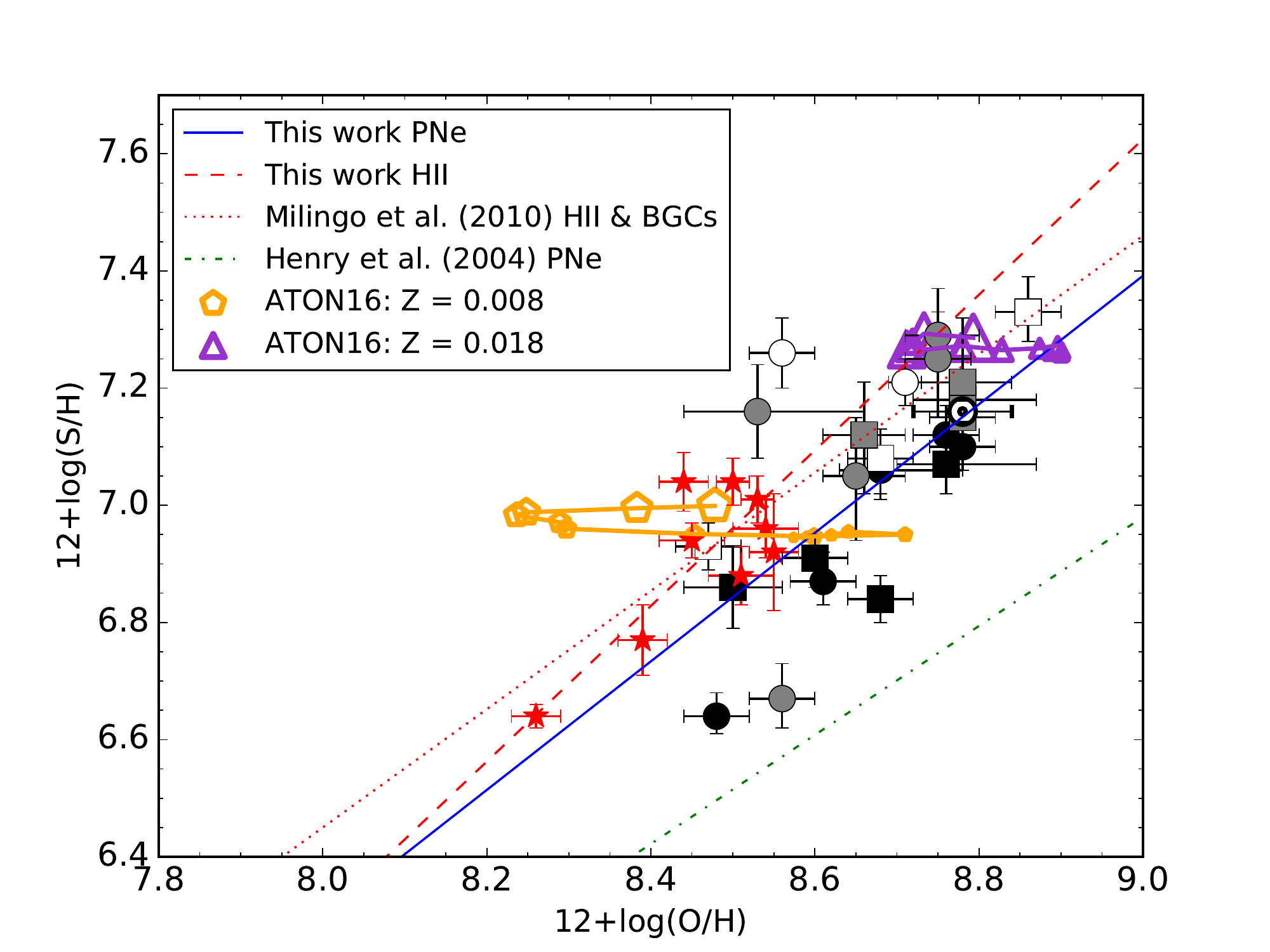}
\caption{Values of 12+log(Ne/H) (upper panel) and 12+log(Ar/H) (middle panel) and 12+log(S/H) (lower panel) vs. 12+log(O/H) for our sample of objects. Symbols for observations are the same than in Fig.~\ref{models_OCl}. The protosolar abundances of \citet{lodders10} are overplotted with the solar symbol. Fits to the observed relations in PNe are shown as continuous blue lines in each panel. Nucleosynthesis ATON16 models for Z= 0.008, 0.018 are also included for comparison (see text). } 
\label{Ne_Ar_S}
\end{figure}

We have also computed the abundance of other $\alpha$-elements, such as Ne, S and Ar, that are supposed to evolve in lock-step during the chemical evolution of galaxies and hence, a correlation is expected between their abundances and O abundance. In Fig.~\ref{Ne_Ar_S} we show the behaviour of Ne/H vs. O/H (upper panel), Ar/H vs. O/H (middle panel) and S/H vs. O/H (lower panel). A correlation between (Ne, Ar, S) abundances and O abundances is found, as expected. The fits to our data are shown as a continuous blue line on each panel. However, there is a large dispersion in all the relations, which is mainly due to uncertainties in the adopted ICFs. In the upper and middle panel of Fig.~\ref{Ne_Ar_S} we include the fits obtained to the Ne/H vs. O/H and Ar/H vs. O/H relations by \citet[][green dotted-dashed line]{henryetal04} and \citet[][dotted magenta line]{milingoetal10} from large samples of Galactic PNe. It is clear that the slopes are very similar although the new ICFs by D-I14 used here seem to provide somewhat higher abundances in average ($\sim$0.2 dex) for both Ne and Ar.

It is interesting to note that when comparing to nucleosynthesis ATON16 models Ne/H and Ar/H seem to be overestimated. However, the ICFs for these two elements from optical data have large uncertainties (see discussion in D-I14). KL16 models behave very similarly to ATON16 models in these plots; however, we avoid to show them to not overload Fig.~\ref{Ne_Ar_S}.  

Furthermore, we can also check the puzzling behaviour of S abundances in our DC PNe sample. \citet{henryetal04} coined the expression ``sulphur anomaly'' to define the phenomenon consisting on the fact that, for a given O/H value, the S/H ratios computed for PNe are systematically lower than those computed for {\hii} regions in a given galaxy; PNe with C-rich dust features seem to be, on average, more depleted in S than other types of PNe \citep[see e.g.][]{garciahdezgorny14}. This behaviour has been found for PNe in the Milky Way \citep[see e.g.][]{henryetal12, milingoetal10} and in nearby galaxies \citep{bernardsalasetal08, shawetal10, garciarojasetal16}. Sulphur depletion into dust and sulphur destruction by nucleosynthetic processes during stellar evolution have been proposed to explain this anomaly, but the theoretical models have failed to reproduce the amount of missing sulphur \citep[see][and references therein]{henryetal12, shingleskarakas13}. \citet{henryetal04} proposed a failure in the proposed ionization correction schemes to account for the highly ionized S$^{3+}$ state; infrared observations of the {\fsiv} 10.5 $\mu$m line significantly reduce the sulphur deficit but they cannot completely resolve the anomaly \citep{bernardsalasetal08, henryetal12}. This seems to indicate that the contribution from S$^{3+}$  and probably from other ionization stages are commonly underestimated in ICFs computed from photoionization models grids. On the other hand, \citet{badnelletal15} computed revised dielectronic recombination rates that lead to a higher fraction of S$^+$ in the gas with respect to S$^{2+}$; unfortunately, these authors claim that this new computations are not enough to solve the ``sulphur anomaly''. 

In the S/H vs. O/H plot (lower panel of Fig.~\ref{Ne_Ar_S}) we plot the fit to our PNe data (solid blue line) together with the fit to our {\hii} regions data (dashed red line). For comparison, we include also the fit to a large sample of Galactic PNe data by \citet{henryetal04} (dotted-dashed green line) and the fit computed by \citet{milingoetal10} to a large sample of {\hii} regions and Blue Compact Dwarf (BCD) galaxies (dotted red line). Besides the relatively low metallicity coverage of our sample, it seems that the use of the newly computed ICFs by D-I14 improves the situation but does not solve the problem. However, \citet{garciahdezgorny14} found from their analysis of a large sample of PNe with different dust-types, that DC PNe showed the highest S abundances, and that C-rich dust disc PNe seem to show a higher degree of S depletion. This agrees with the hypothesis that S is more depleted into dust grains in the C-rich dust objects, which is supported by the detection of the broad 30 $\mu$m feature in C-rich AGB, post-AGBs and PNe and that is commonly attributed to sulphur-based dust like MgS \citep[e.g.][]{honyetal02}; although a definitive identification is still under debate \citep[see e.g.][and references therein]{zhangetal09, garciahdez12}. A recomputation of the ICF of sulphur using the approach of D-I14 and the new dielectronic recombination rates by \citet{badnelletal15}, as well as a detailed comparison with high-quality observations of a large sample of PNe covering the different dust-types, would be interesting to see if we are getting closer to resolve the ``sulphur anomaly''.  Additionally, from  Fig.~\ref{Ne_Ar_S} it is clear that ATON16 nucleosynthesis models (and also KL16 ones) reproduce much better the S abundances obtained for our PNe than abundances obtained by \citet{henryetal04}. 

\subsection{Nucleosynthesis indicators. Behaviour of nitrogen, helium and carbon abundances}\label{mass}

\begin{figure*}
\includegraphics[width=\columnwidth]{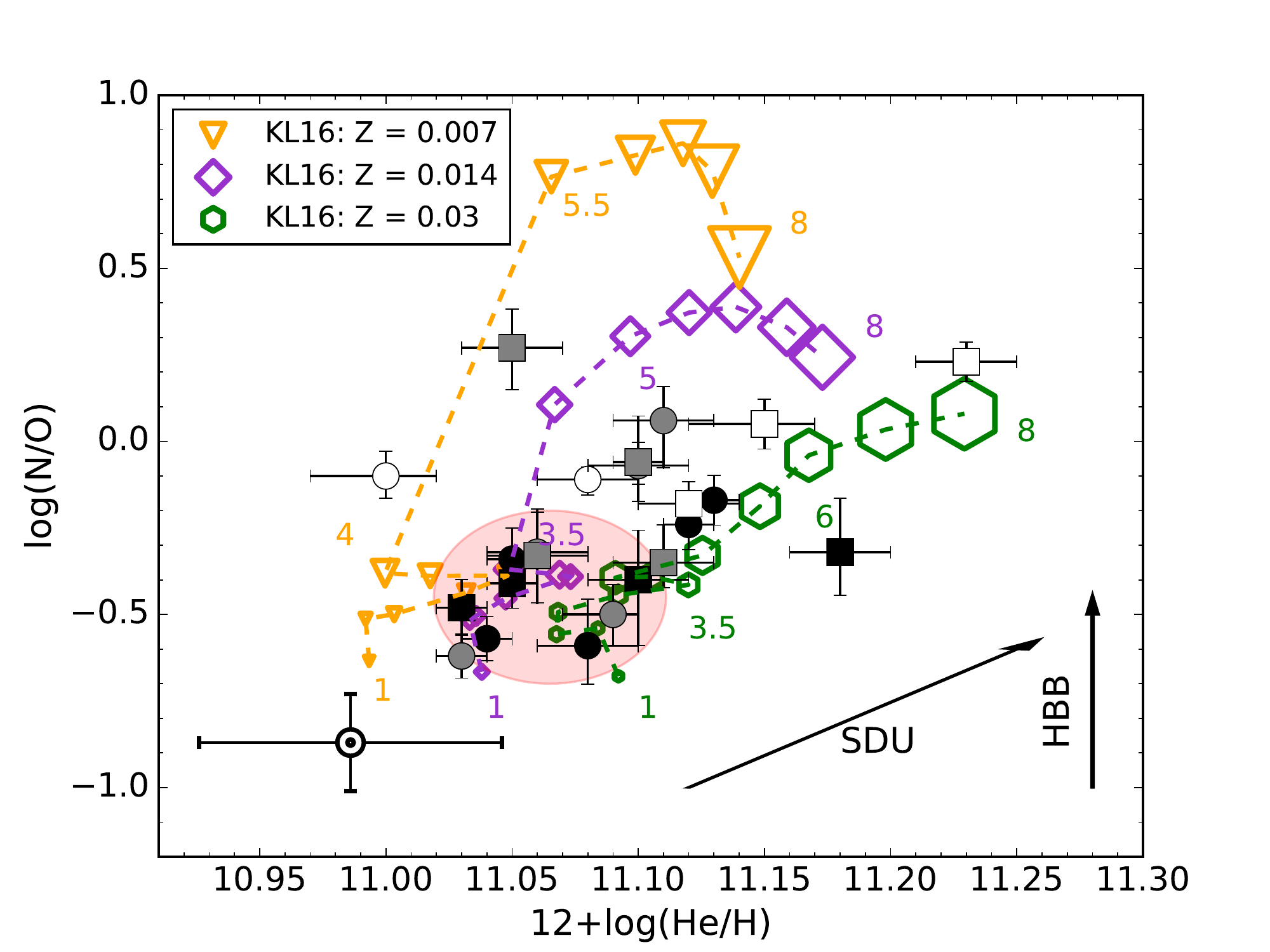}
\includegraphics[width=\columnwidth]{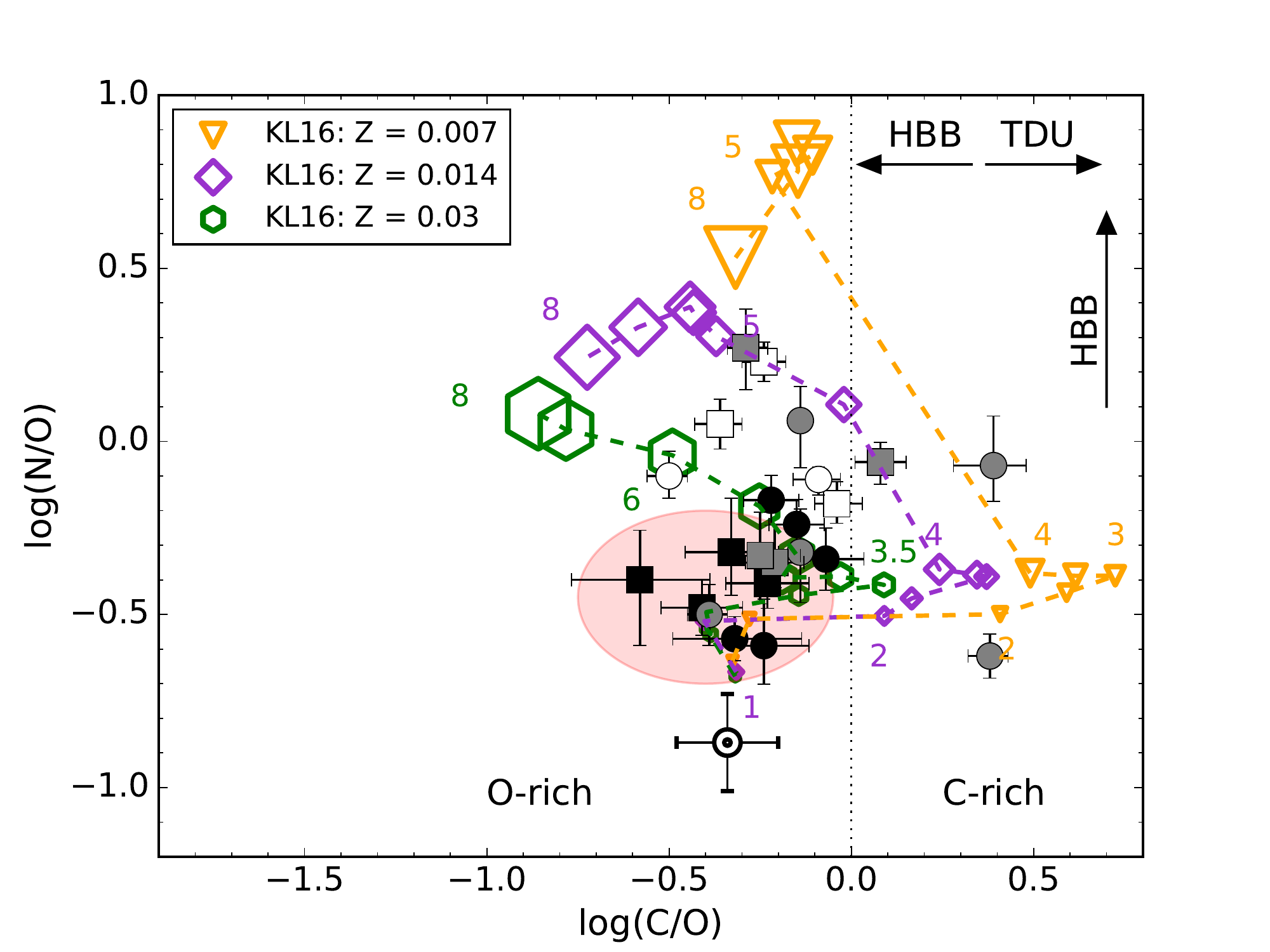}
\includegraphics[width=\columnwidth]{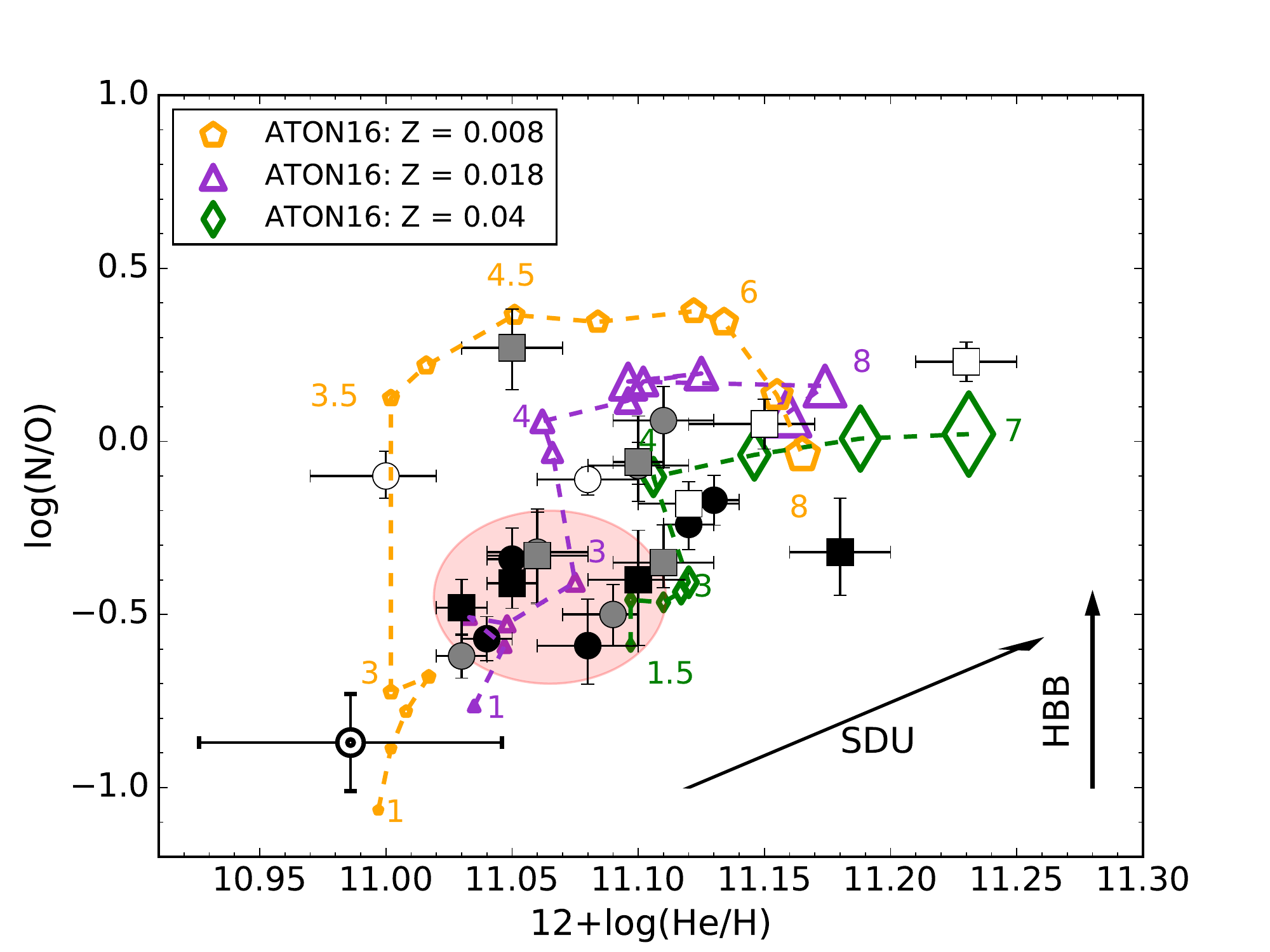}
\includegraphics[width=\columnwidth]{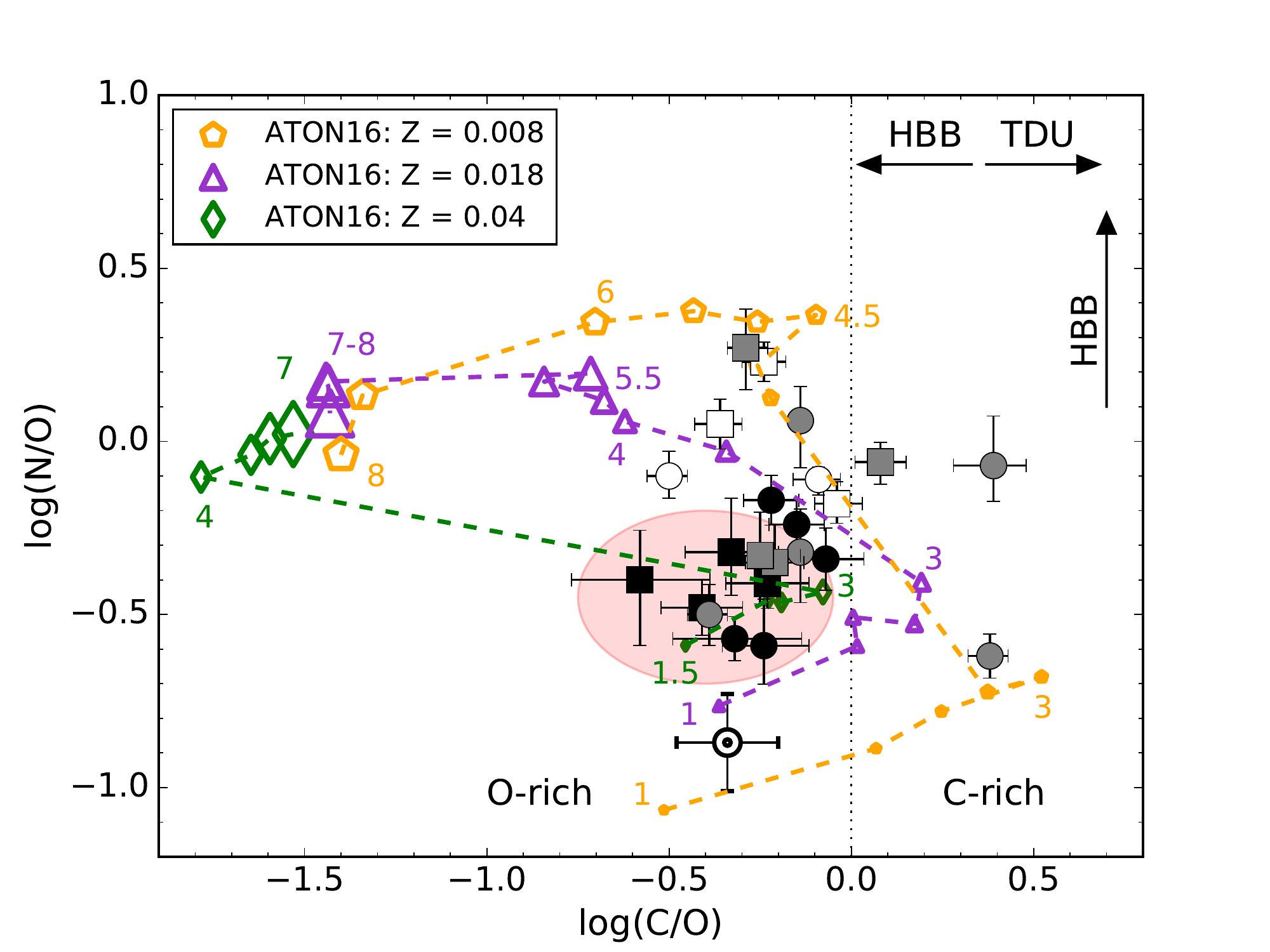}
\caption{Values of log(N/O) vs. 12+log(He/H) (upper panel) and log(N/O) vs. log(C/O) (lower panel) for our sample of objects compared with the predictions from the stellar nucleosynthesis models by \citet{karakaslugaro16} (KL16, left panels) and from the ATON  stellar nucleosynthesis models (ATON16, right panels) for different metallicities (see text). The size of the model symbols increases with the mass. We have included some mass labels and connected consecutive models with a dashed line to ease interpretation. Symbols for observations are the same than in Fig.~\ref{models_OCl}. Red-shadowed region points to the PNe compatible with having a low-mass (M$<$1.5 M$_{\odot}$) progenitor (see text). Arrows indicate what would be the action of the second dredge up (SDU), third dredge up (TDU) and hot bottom burning (HBB) in the represented abundances. The protosolar abundances of \citet{lodders10} are overplotted with the solar symbol.} 
\label{modelsNO_He_CO}
\end{figure*}

Fig.~\ref{modelsNO_He_CO} shows the values of N/O as a function of He/H (upper panel) and C/O (lower panel) for the full sample of DC PNe. The predictions from the nucleosynthesis models by KL16 (left panels) and from ATON models (right panels) are also plotted for sub-solar, solar and super-solar metallicities. As we mentioned in Section~\ref{Intro}, the observed abundances of N, He and C in PNe may be significantly affected by the previous occurrence of nucleosynthesis processes, depending on the initial mass and metallicity. We briefly describe here the main behaviour of the AGB models as a function of the progenitor mass and metallicity, in terms of the N/O, C/O and He/H ratios. We also compare the predictions from the nucleosynthesis models with the abundance ratios derived for the three samples (A, B and C) of DC PNe.

The enhancement in the N/O ratio is, in principle, a clear sign of hydrogen burning. The source of such hydrogen burning is unknown but is inferred to be HBB nucleosynthesis that is active in the more massive AGB stars ($>$3--4 M$_{\odot}$, depending on the description adopted for the convective instabilities). The HBB efficiency increases with decreasing metallicity. In the KL16 models, the combined effect of the TDU that increases the surface carbon and the HBB, which converts the dredge-up carbon into nitrogen, produces higher N/O ratios compared to the ATON16 models, because the latter ones experience a less extended TDU. In the low mass regime ($\leq$3 M$_{\odot}$), the N/O ratio slightly increases as a function of the initial mass. This is a result of the first dredge-up that brings to the surface the products of the CNO nucleosynthesis, thus N-rich material. Moreover, extra-mixing in low-mass stars (M$<$2 M$_{\odot}$) and rotationally induced mixing on the main sequence can also raise the N/O ratio \citep[][]{karakaslattanzio14}.

According to the N/O predictions from KL16 and ATON16 models, the progenitor stars of most of the PNe in sample A (filled black symbols) have not gone through any significant HBB, indicating initial masses below $\sim$3--4 M$_\odot$. The PNe from sample B (empty symbols) exhibit, on average, larger N/O ratios and He abundances, which suggests more massive progenitors \citep[see also][]{garciahdezetal16b}. As for sample C, while some PNe show large N/O ratios, none of them exhibit He-enrichment. The observed behaviour of the whole sample in the N/O vs. He/H diagram agrees better with the high metallicity models (those with solar or super-solar metallicities).

The TDU is responsible for a large enhancement in the surface carbon abundance of some AGB stars. Low-mass stars, with M $ < 1.5$ M$_\odot$ (at Solar metallicity, but at slightly lower mass for lower metallicities), do not suffer the TDU and thus preserve C/O$<1$ during their entire life. The strength and the number of dredge-up episodes, together with the initial metallicity, define the final C/O ratio in the stellar surface. For example, solar and super-solar metallicity stars, which formed with a higher O, hardly become carbon stars. The ignition of HBB is evident in the most massive AGB stars where carbon is strongly depleted, in particular in the ATON models. Eventually, final episodes of TDU can be responsible for a final increase in the carbon abundances during the last thermal pulses. Nucleosynthesis models predict that there is a relatively narrow mass range, starting in the limit mass for HBB activation, where the HBB and TDU mechanisms can co-exist and the final stellar surface abundances will depend on which mechanism dominates. 

From the log(N/O) vs. log(C/O) panels in Fig.~\ref{modelsNO_He_CO}, it is clear that most of the PNe have C/O$<1$, which according to model predictions suggest progenitor stars with: 1) M $ < 1.5$ M$_\odot$, that have not experienced the TDU and HBB or 2) M $>$ 3--4 M$_\odot$, that have undergone both the TDU and the HBB, showing the lowest C/O values. Taking into account that Initial Mass Function predicts a larger portion of progenitor stars being low-mass stars, we consider the former channel is more probable, although with the available data, we cannot discard the high-mass channel. Only three PNe from sample C have C/O$>1$: Cn\,1--5, M\,1-32 and NGC\,2867, being the last two PNe extremely C enriched (C/O$>2$). 

As for helium abundances, all the PNe present a clear He enrichment relative to the Solar value. He is enhanced by the second dredge-up experienced by stars with M $\gtrsim$ 4M$_{\odot}$. The efficiency of this process depends on the H-exhausted core mass of the star and on treatment of convection. The helium abundances predicted by the two set of stellar nucleosynthesis models are similar. The observed values of He/H run over the whole range of He/H predictions, making a difficult task to constrain the progenitor masses of the PNe. However, the highest He abundances (12+log(He/H) $\geq$ 11.15) found for M\,2-27 and M\,1-42, can be explained by a massive progenitor star formed in a super-solar metallicity environment ($Z$ = 0.03); although only M\,2-27 has such a large metallicity (see Table~\ref{tab:total_ab_lit}).

In summary, we conclude that it is very difficult to reconcile consistently the interpretations on a one-by-one object basis (in terms of progenitor mass and metallicity) from the observations$-$predictions comparison on the N/O. vs. He/H panel with those in the N/O vs. C/O panel, and we must seek alternative explanations (see Sect.~\ref{sec:preHe}).


\section{Discussion}\label{sec:discuss}

Regarding the N/O vs. C/O behaviour, our sample show most PNe having C/O $< 1$ and hence, their ionized gas is ``now'' oxygen-rich. This behaviour can be explained via two different scenarios: i) the progenitor stars are low-mass stars (M $\lesssim$ 1.5 M$_{\odot}$ at solar metallicity but slightly lower for sub-solar metallicity) that have not gone through the TDU, and they remain O-rich during their whole life. In this case the PNe should show 
, in principle, no enrichment of C, N and He. This applies for some of the PNe of our sample, which simultaneously show low C/O, N/O and He/H. Curiously, most of these PNe, red-shadowed in Fig.~\ref{modelsNO_He_CO}, display also low Cl/H (or sub-solar metallicities) suggestive of relatively old and low-mass stars. The abundances for three PNe of sample A (M\,1-31, M\,1-33 and M\,1-60, the most metal-rich PNe in sample A) and the whole sample by D-I15 (also more metal-rich objects, on average) seem to disagree with this scenario, mainly due to their high He/H and/or N/O ratios (see below); or  
ii) the progenitor stars are intermediate-mass stars (M$>$3--4 M$_{\odot}$) that have gone through HBB and possibly some TDU events. They end up their evolution as O-rich stars, because HBB destroys the surface C. Their final chemical composition is expected to show up a significant enrichment of N and, depending on the initial mass, some enrichment in He. As it can be seen from Fig.~\ref{modelsNO_He_CO} both scenarios fail to explain simultaneously the N/O, He/H and C/O ratios observed. However, this second scenario would be plausible if the progenitor stars have experienced some He pre-enrichment, as \citet{karakas14} proposed for bulge PNe in order to explain the paucity of C stars in high-metallicity environments. In Sect.~\ref{sec:preHe} we will go further on this. Interestingly, the addition of our accurate C/O ratios into the observations$-$models comparison for a larger sample of DC PNe reveal for the first time that DC PNe may be formed via different channels (e.g., very low-mass objects or more massive HBB stars). \citet{garciahdezgorny14} only compared the median He, N, O and Ar nebular abundances (as obtained from low-resolution optical spectroscopy) observed in a larger number of compact Galactic (both bulge and disc) DC PNe from \citet{stanghellinietal12} (i.e., similar to our sample A) with the older Karakas (2010) AGB nucleosynthesis models, finding that they were consistent with relatively massive ($\sim$3$-$5 M$_{\odot}$) and high-metallicity (solar/supersolar) HBB-AGB stars as progenitors; sub-solar metallicity DC PNe are also present in their sample but they do not dominate their median chemical abundance pattern of Galactic DC PNe. On the other hand, \citet{garciahdezetal16b} reached a similar conclusion (i.e., DC PNe mostly descend from high-mass solar/supersolar metallicity HBB-AGBs) by comparing the D-I15 He, N, O and Cl nebular abundances from sample B DC PNe with the ATON16 model predictions; they could not discard, however, another possible formation channels in lower mass AGB stars (like extra mixing, stellar rotation, binary interaction, or even He pre-enrichment) until more accurate C/O ratios would be obtained. 

The new DC PNe (sample A) presented here thus suggest the low-mass channel for the first time. This interpretation is similar for both sets of the most recent nucleosynthesis models (KL16 and ATON16); although the exact progenitor masses are somewhat model-dependent. As we already mentioned before, our sample A DC PNe is biased towards compact ($<$4$''$ in diameter, with the exception of H\,1-50 and M\,1-60 which are 10$''$ in diameter) and (presumably) young PNe but, accidentally, it is mostly composed (6/9) by sub-solar metallicity objects with no significant He and N enrichment. This kind of objects do not dominate the bulk of compact and presumably young DC PNe previously studied by \citet{garciahdezgorny14} neither the DC PNe samples B and C previously studied in the literature \citep{garciarojasetal13,delgadoingladaetal15,garciahdezetal16b}. 
In principle, both sample B and C objects are also relatively compact. However they are generally more extended than the ones in sample A. All of them with the exception of NGC\,2867 (sample B, 14$''$ diameter) and M\,1-42 and NGC\,7026 (sample C, 9$''$ and 14$''$ diameter respectively) have optical diameters below 7.6$''$ \citep{ackeretal92}. On the other hand, sample B objects belong to a sample of relatively early-[WC] type stars, which have, in general, $T_{eff}$ around 50--60 kK \citep[see][and references therein]{garciarojasetal09, garciarojasetal12}, with the exception of Hb\,4 and NGC\,2867 which have central stars with $T_{eff}$ of 86 kK \citep[][]{tylendastasinska94} and 150 kK \citep[][]{kelleretal14} respectively . The slower evolution of low-mass stars (most of sample A DC PNe) could favor the detection of these objects in very early PN phases (young), while the more massive sources (the ones with the hottest central stars of sample B), evolving faster and even more obscured in the optical, should be more easily detected at later PN phases.

Regarding the dual-dust chemistry phenomenon, the lack of pure C-rich PNe (those with C/O$>$1 and carbon-rich dust) in our sample seem to discard the hypothesis of a late thermal pulse as a main channel to produce the dual-dust features by converting the central stars from O-rich to C-rich \citep[see][]{pereacalderonetal09}, but it still work for the C-rich DC PNe (C/O$>$1 and mixed chemistry dust) because with the late thermal pulse channel we really would expect such behaviour (i. e, C/O slightly above 1). Therefore, it seems that in both the two hypotheses stated above (low-mass progenitor star or soft HBB stars) the PAHs should have been formed through the dissociation of CO molecule, as proposed by \citet{guzmanramirezetal11} \citep[already discussed by][]{garciahdezgorny14}. However, the scenario proposed by \citet{guzmanramirezetal11} assume the presence of a dense toroidal structure, which is prevalent in PNe with binary central stars \citep{miszalskietal09} and has been also proposed as a result of the ejection of a common-envelope \citep{passyetal12}. Moreover, very recently, \citet{sowickaetal17} have reported the discovery of a binary central star in the DC PN IC\,4776 and therefore, this scenario should be taken into account in the future.

\subsection{Helium pre-enriched models}\label{sec:preHe}

\begin{figure}
\includegraphics[width=\columnwidth]{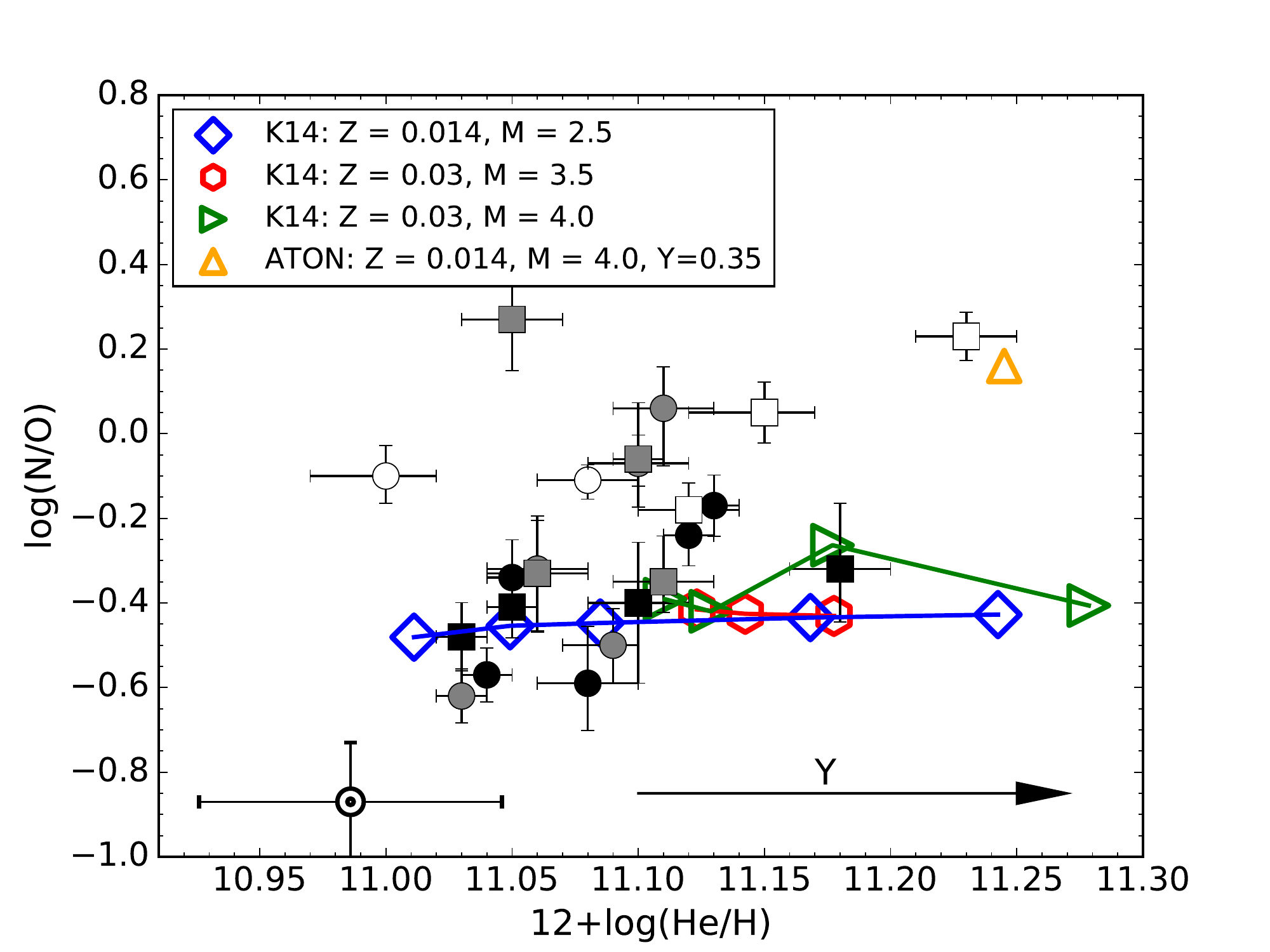}
\includegraphics[width=\columnwidth]{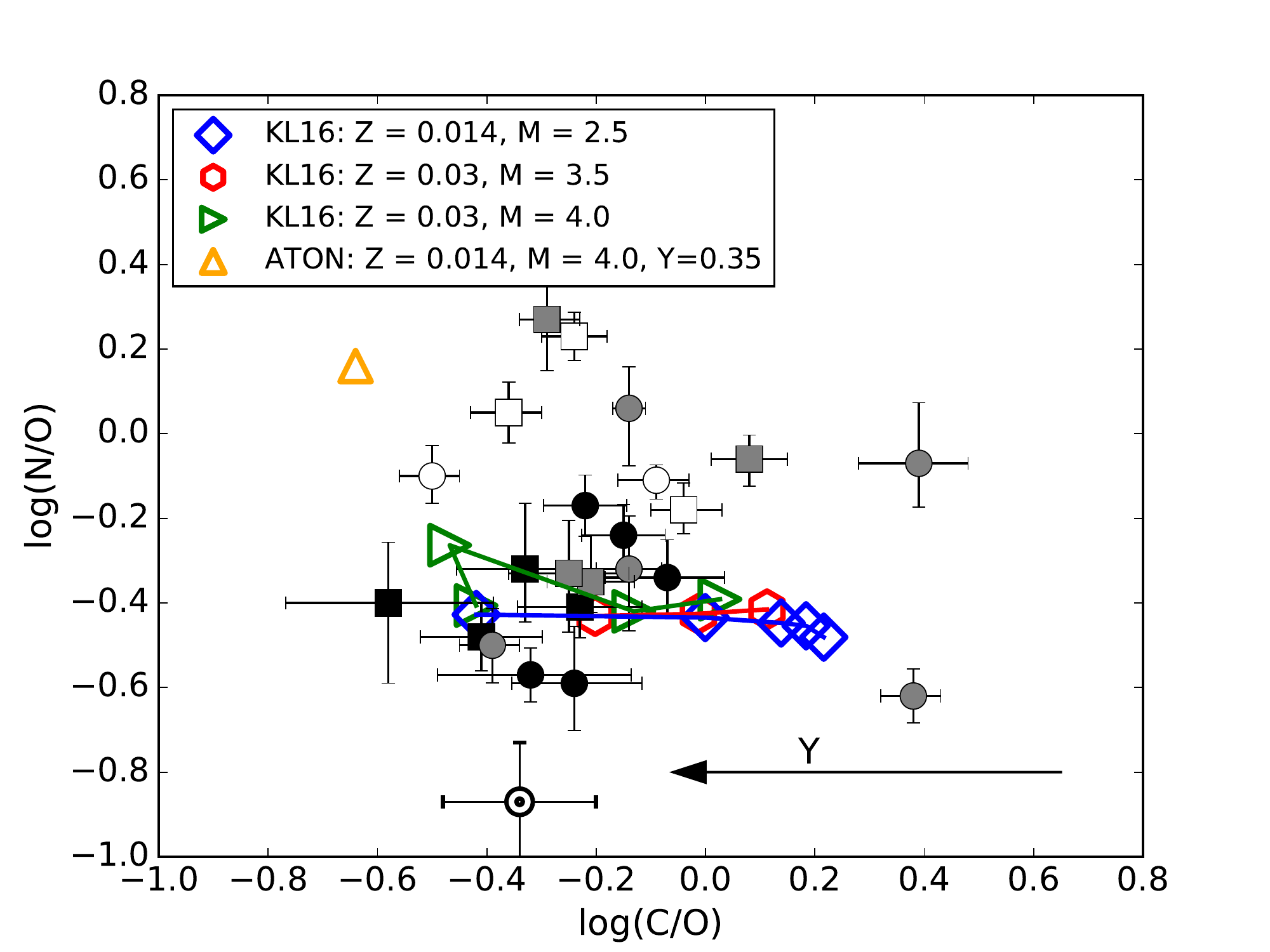}
\caption{Comparison with helium enriched stellar nucleosynthesis models by \citet{karakas14}. The arrow show the increasing direction of Y. For solar metallicity models Y=0.26, 0.28, 0.30, 0,35 and 0.4; for super-solar metallicity models Y=0.30, 0.32, 0.32 and 0.35 for M = 3.5M$_{\odot}$, and Y=0.30, 0.32, 0.35 and 0.49 for M = 4.0M$_{\odot}$.} 
\label{K14_models}
\end{figure}

As we have commented in the previous section, nucleosynthesis models can not reproduce the observed behaviour of N/O, He/H and C/O simultaneously for the entire sample. Some of these objects belong to the bulge of the Galaxy with a higher metallicity component according to the Cl measurements (M\,1-42, M\,2-27 from sample B and M\,1-31 from sample A). The models presented by \citet{karakas14} consider the AGB evolution of stars from the canonical value to helium-enriched compositions at solar and super-solar metallicities. Higher-helium stars evolve at higher luminosities, thus loose the envelope faster and experience a smaller number of TDU events. Furthermore, a higher helium favors a higher entropy barrier at the hydrogen/helium interface, which diminishes the efficiency of TDU. These results in the inhibition of carbon star production which can explain the paucity of C-rich stars in high-metallicity environments.
For the moment a grid of models that cover a wide range of masses is not available yet, but for some of them the detailed stellar nucleosynthesis have been calculated. In Figure~\ref{K14_models} we present the final abundances for M = 2.5 M$_{\odot}$ at solar metallicity and for M = 3.5 and 4.0 M$_{\odot}$ for super-solar metallicity. For solar metallicities the initial He abundances adopted were:  Y=0.26, 0.28, 0.30, 0,35 and 0.4; for super-solar metallicity models, the initial He abundances were: Y=0.30, 0.32, 0.32 and 0.35 for M = 3.5M$_{\odot}$, and Y=0.30, 0.32, 0.35 and 0.4 for M = 4.0 M$_{\odot}$.  These models have not experienced the HBB and limited episodes of TDU. This is in nice agreement with the objects that have C/O$<1$ and with log(N/O)$<-0.2$ dex, simultaneously, indicating 2.5--4 M$_{\odot}$ as possible progenitors of these PNe.
The large helium abundances detected would be compatible with a very massive progenitor, with M$\sim$7--8 M$_{\odot}$, that experienced deep SDU. This interpretation is however at odds with the measured C/O, which is far in excess than expected. This evidence shifts the attention towards He-rich, lower-mass progenitors, with M$\sim$4--5 M$_{\odot}$, which are exposed to a weaker nucleosynthesis at the base of the envelope, compared to their higher mass counterparts. To test this possibility, we calculated a helium-enriched (initial helium Y=0.35) model, with initial mass 4.5 M$_{\odot}$. It has been calculated with the ATON code until the very final AGB phases and the correspondent final abundances are shown with orange triangle in the two panels of Figure~\ref{K14_models}. The results indicate that after the latest thermal pulses, when HBB is turned off, a few TDU events can raise the surface carbon (hence, the C/O ratio significantly) to the observed values. Note that because the mass of the envelope at this stage is significantly reduced, the increase in the surface carbon associated to these late events does not require deep TDU. A wider grid of models would be extremely helpful to better explore this interpretation.

\section{Conclusions}\label{sec:conclu}

We present deep high-resolution spectrophotometry of nine compact PNe obtained with UVES at the VLT. PNe were selected from the sample of Galactic disc and bulge PNe with dual dust features (cristalline or amorphous silicates + PAHs) detected in their infrared spectra presented by \citet{stanghellinietal12}.

We computed physical conditions from a large number of diagnostics, and hence, taking advantage of the large number of emission lines measured in the spectra, computed ionic chemical abundances for a large number of ions from both collisionally excited lines (CELs) and optical recombination lines (ORLs). The total abundances were computed mainly with the ICFs derived by D-I14 but alternative ICFs were considered in a few cases (N and Fe). In particular we take advantage of the quality of our spectra to compute C/O ratios using ORLs.

We extend the initial sample (Sample A) with seven PNe studied by D-I15 and nine by \citet{garciarojasetal12, garciarojasetal13}. These two new samples are refered to as Sample B and Sample C, respectively. The abundances of these PNe were re-computed following the same procedure as for Sample A.

Although Cl has been recently invoked as a better proxy for metallicity for C-rich dust PNe, from the comparison of O/Cl abundance ratios $vs.$ Cl/H and O/H ratios, we reach to the conclusion that O is a better metallicity tracer than Cl for our sample, owing to the inhomogeneities or uncertainties involved in our determinations of Cl/H. From the comparison with nucleosynthesis models, it seems that metallicities around solar and half-solar covers all the metallicity range in the PNe in our sample.

We studied the abundances of Ne, S, and Ar because these elements are expected to evolve in lockstep and we compared them with the O abundances, that can be  slightly affected by nucleosynthesis. The abundances derived here define a similar trend between Ne/H, S/H, Ar/H and O/H than the ones previously obtained in the literature \citep{henryetal04, milingoetal10} but our abundances are somewhat higher (likely due to the use of different ICFs). 

The abundances of N, C, and He derived for the PNe were used to study nucleosynthesis occurring in the progenitor stars. Many of the PNe from sample B have higher N/O values than the other PNe, suggesting that many of the progenitors of Sample B PNe have masses above 3--4 M$_{\odot}$. Most of the PNe (with the exception of Cn~1.5, M~1-32, and NGC~2867) have $-$0.5 $<$ C/O $<1$ which can be explained with progenitors with masses i) below 1.5 M$_{\odot}$, that have not gone through the TDU, or ii) above 3--4 M$_{\odot}$, that have experienced TDU and HBB. All the PNe show some He enrichment, where M~2-27 and M~1-42 are the most extreme in our sample and therefore, likely arise from more massive progenitors.

The values of C/O in the PNe reflect the present dominant chemistry (either oxygen- or carbon-rich) in the gas. As we mentioned before, since all but three PNe have  C/O $<1$, the hypothesis of the last thermal pulse that turns O-rich PNe into C-rich PNe is discarded (with the exception of these three PNe). The alternatives are low- (M $<$ 1.5 M$_{\odot}$) or high-mass (M $>$ 3--4 M$_{\odot}$) progenitors or massive stars (that have suffered a soft HBB) in which the PAHs could have been formed through the dissociation of the CO molecule.

The new data (sample A) suggest a low-mass channel for the DC PNe phenomenon. The few He-pre-enriched models that we have are encouraging to explain some outliers in the N/O vs. He/H diagram. However, we still cannot discard another scenarios  like extra mixing, stellar rotation or binary interactions.

Further precise determinations of the C/O ratios (i.e., based on ORLs as we have done in this work from high-resolution and high-quality spectra) in a larger sample of DC PNe (i.e., covering the different types of {\it Spitzer} spectra and central stars) are needed to learn about the dominant mechanism for PAH formation (HBB deactivation and/or hydrocarbon chemistry within O-rich shells).

\section*{Acknowledgements}

This work is based on observations collected at the European Southern Observatory, Chile, proposal number ESO 095.D-0067(A). This work has been funded by the Spanish Ministry of Economy and Competitiveness (MINECO) under the grant AYA2015-65205-P. J.~G-R acknowledges fruitful discussions with D. Jones. J.~G-R acknowledges support from Severo Ochoa excellence program (SEV-2011-0187) postdoctoral fellowship and Severo Ochoa Excellence Program (SEV-2015-0548) Advanced Postdoctoral Fellowship. G.~D-I gratefully acknowledges support from: CONACYT grant no. CB-2014/241732 and PAPIIT-UNAM grant no. IA-101517. D.~A.~G-H. was funded by the Ram\'on y Cajal fellowship number
RYC-2013-14182. D.~A.~G-H. and F.~D.~A acknowledge support
provided by the MINECO under grant AYA-2014-58082-P. M.~L. is a Momentum (``Lend\"ulet-2014'') project leader of the Hungarian Academy of Sciences.  M.~L. acknowledges financial support from the MINECO under
the 2015 Severo Ochoa Program MINECO SEV-2015-0548. M.~R. ackowledges support from Mexican CONACYT grant CB-2014-240562.








\appendix



\onecolumn
\setcounter{table}{8}
\begin{longtable}{ccccccccc} 
\caption{Observed and reddening corrected line ratios (F(H$\beta$) = 100) and line identifications in H\,1-40.\label{tab:lineid1}}\tabularnewline
\hline\hline 
$\lambda_0$ (\AA\ ) & Ion & Mult. & $\lambda_{\rm obs}$ & $V_{\rm rad}$ (km s$^{-1}$) & $F(\lambda)/F({\rm H}\beta)$$^{\rm a}$ & $I(\lambda)/I({\rm H}\beta)$$^{\rm b}$ & Err(\%) & notes \tabularnewline
\hline
\endfirsthead
\caption{continued.}\tabularnewline
\hline\hline 
$\lambda_0$ (\AA\ ) & Ion & Mult. & $\lambda_{\rm obs}$ & $V_{\rm rad}$ (km s$^{-1}$) & $F(\lambda)/F({\rm H}\beta)$$^{\rm a}$ & $I(\lambda)/I({\rm H}\beta)$$^{\rm b}$ & Err(\%) & notes \tabularnewline
\hline
\endhead
\hline
\endfoot
\endlastfoot
 3721.93 &            {\hi} &          H14 & 3722.39 &  37.04 &   0.4202 &   2.5239 &   : &     \\
 3726.03 &          {\foii} &           1F & 3726.57 &  43.44 &   5.3795 &  32.158 &   10 &     \\
 3728.82 &          {\foii} &           1F & 3729.31 &  39.39 &   2.3745 &  14.150 &   17 &     \\
 3734.37 &            {\hi} &          H13 & 3734.92 &  44.13 &   0.5717 &   3.3852 &   : &     \\
 3750.15 &            {\hi} &          H12 & 3750.69 &  43.17 &   0.6887 &   4.0051 &   : &     \\
 3770.63 &            {\hi} &          H11 & 3771.24 &  48.50 &   0.8942 &   5.0775 &   19 &     \\
 3797.90 &            {\hi} &          H10 & 3798.50 &  47.36 &   1.2692 &   6.9764 &   15 &     \\
 3819.61 &           {\hei} &           22 & 3820.20 &  46.29 &   0.2727 &   1.4598 &   38 &     \\
\end{longtable}

\clearpage

\onecolumn
\setcounter{table}{9}
\begin{longtable}{ccccccccc} 
\caption{Observed and reddening corrected line ratios (F(H$\beta$) = 100) and line identifications in H\,1-50.\label{tab:lineid2}}\tabularnewline
\hline\hline 
$\lambda_0$ (\AA\ ) & Ion & Mult. & $\lambda_{\rm obs}$ & $V_{\rm rad}$ (km s$^{-1}$) & $F(\lambda)/F({\rm H}\beta)$$^{\rm a}$ & $I(\lambda)/I({\rm H}\beta)$$^{\rm b}$ & Err(\%) & notes \tabularnewline
\hline
\endfirsthead
\caption{continued.}\tabularnewline
\hline\hline 
$\lambda_0$ (\AA\ ) & Ion & Mult. & $\lambda_{\rm obs}$ & $V_{\rm rad}$ (km s$^{-1}$) & $F(\lambda)/F({\rm H}\beta)$$^{\rm a}$ & $I(\lambda)/I({\rm H}\beta)$$^{\rm b}$ & Err(\%) & notes \tabularnewline
\hline
\endhead
\hline
\endfoot
\endlastfoot
 3121.63 &          {\oiii} &    3S-3P$_0$ & 3122.20 &  54.74 &   0.5880 &   1.6297 &   14 &     \\
 3132.79 &          {\oiii} &    3S-3P$_0$ & 3133.27 &  45.92 &   9.8532 &  26.995 &    6 &     \\
 3187.84 &           {\hei} &            3 & 3188.23 &  36.66 &   1.0326 &   2.6751 &   11 &     \\
 3203.10 &          {\heii} &          3.5 & 3203.62 &  48.66 &   1.9755 &   5.0427 &    8 &     \\
 3299.39 &          {\oiii} &            3 & 3299.90 &  46.33 &   0.5171 &   1.2173 &   15 &     \\
 3312.33 &          {\oiii} &            3 & 3312.84 &  46.15 &   1.3494 &   3.1467 &    9 &     \\
 3340.77 &          {\oiii} &            3 & 3341.29 &  46.66 &   1.9604 &   4.4843 &    8 &     \\
 3345.83 &          {\fnev} &        3P-1D & 3346.32 &  43.90 &   0.4038 &   0.9207 &   18 &     \\
\end{longtable}

\clearpage

\onecolumn
\setcounter{table}{10}
\begin{longtable}{ccccccccc} 
\caption{Observed and reddening corrected line ratios (F(H$\beta$) = 100) and line identifications in Hen\,2-73.\label{tab:lineid3}}\tabularnewline
\hline\hline 
$\lambda_0$ (\AA\ ) & Ion & Mult. & $\lambda_{\rm obs}$ & $V_{\rm rad}$ (km s$^{-1}$) & $F(\lambda)/F({\rm H}\beta)$$^{\rm a}$ & $I(\lambda)/I({\rm H}\beta)$$^{\rm b}$ & Err(\%) & notes \tabularnewline
\hline
\endfirsthead
\caption{continued.}\tabularnewline
\hline\hline 
$\lambda_0$ (\AA\ ) & Ion & Mult. & $\lambda_{\rm obs}$ & $V_{\rm rad}$ (km s$^{-1}$) & $F(\lambda)/F({\rm H}\beta)$$^{\rm a}$ & $I(\lambda)/I({\rm H}\beta)$$^{\rm b}$ & Err(\%) & notes \tabularnewline
\hline
\endhead
\hline
\endfoot
\endlastfoot
 3132.79 &          {\oiii} &    3S-3P$_0$ & 3132.77 &  -1.92 &  11.490 &  56.505 &    6 &     \\
 3203.10 &          {\heii} &          3.5 & 3203.10 &   0.00 &   2.3576 &  10.368 &   10 &     \\
 3299.39 &          {\oiii} &            3 & 3299.36 &  -2.71 &   0.6618 &   2.5600 &   21 &     \\
 3312.33 &          {\oiii} &            3 & 3312.31 &  -1.81 &   1.6445 &   6.2678 &   12 &     \\
 3340.77 &          {\oiii} &            3 & 3340.76 &  -0.90 &   2.2261 &   8.2296 &   10 &     \\
 3345.83 &          {\fnev} &        3P-1D & 3345.77 &  -5.38 &   1.0622 &   3.9067 &   16 &     \\
 3425.87 &          {\fnev} &        3P-1D & 3425.83 &  -3.50 &   3.1781 &  10.875 &    9 &     \\
 3428.65 &          {\oiii} &           15 & 3428.59 &  -5.23 &   0.8922 &   3.0460 &   17 &     \\
\end{longtable}

\clearpage

\onecolumn
\setcounter{table}{11}
\begin{longtable}{ccccccccc} 
\caption{Observed and reddening corrected line ratios (F(H$\beta$) = 100) and line identifications in Hen\,2-96.\label{tab:lineid4}}\tabularnewline
\hline\hline 
$\lambda_0$ (\AA\ ) & Ion & Mult. & $\lambda_{\rm obs}$ & $V_{\rm rad}$ (km s$^{-1}$) & $F(\lambda)/F({\rm H}\beta)$$^{\rm a}$ & $I(\lambda)/I({\rm H}\beta)$$^{\rm b}$ & Err(\%) & notes \tabularnewline
\hline
\endfirsthead
\caption{continued.}\tabularnewline
\hline\hline 
$\lambda_0$ (\AA\ ) & Ion & Mult. & $\lambda_{\rm obs}$ & $V_{\rm rad}$ (km s$^{-1}$) & $F(\lambda)/F({\rm H}\beta)$$^{\rm a}$ & $I(\lambda)/I({\rm H}\beta)$$^{\rm b}$ & Err(\%) & notes \tabularnewline
\hline
\endhead
\hline
\endfoot
\endlastfoot
 3697.15 &            {\hi} &          H17 & 3696.50 & -52.71 &   0.3357 &   1.5335 &   : &     \\
 3703.86 &            {\hi} &          H16 & 3703.16 & -56.68 &   0.3659 &   1.6612 &   38 &     \\
 3705.04 &           {\hei} &           25 & 3704.31 & -59.08 &   0.1704 &   0.7729 &   : &     \\
 3711.97 &            {\hi} &          H15 & 3711.29 & -54.92 &   0.4179 &   1.8834 &   35 &     \\
 3721.83 &         {\fsiii} &           2F & 3721.11 & -58.00 &   0.8835 &   3.9452 &   22 &     \\
 3721.93 &            {\hi} &          H14 &       * &       * &       * &       * &    * &     \\
 3726.03 &          {\foii} &           1F & 3725.36 & -53.91 &   7.8821 &  35.058 &    7 &     \\
 3728.82 &          {\foii} &           1F & 3728.11 & -57.09 &   3.2708 &  14.510 &   10 &     \\
\end{longtable}

\clearpage

\onecolumn
\setcounter{table}{12}
\begin{longtable}{ccccccccc} 
\caption{Observed and reddening corrected line ratios (F(H$\beta$) = 100) and line identifications in Hen\,2-158.\label{tab:lineid5}}\tabularnewline
\hline\hline 
$\lambda_0$ (\AA\ ) & Ion & Mult. & $\lambda_{\rm obs}$ & $V_{\rm rad}$ (km s$^{-1}$) & $F(\lambda)/F({\rm H}\beta)$$^{\rm a}$ & $I(\lambda)/I({\rm H}\beta)$$^{\rm b}$ & Err(\%) & notes \tabularnewline
\hline
\endfirsthead
\caption{continued.}\tabularnewline
\hline\hline 
$\lambda_0$ (\AA\ ) & Ion & Mult. & $\lambda_{\rm obs}$ & $V_{\rm rad}$ (km s$^{-1}$) & $F(\lambda)/F({\rm H}\beta)$$^{\rm a}$ & $I(\lambda)/I({\rm H}\beta)$$^{\rm b}$ & Err(\%) & notes \tabularnewline
\hline
\endhead
\hline
\endfoot
\endlastfoot
3187.84 &           {\hei} &            3 & 3187.23 & -57.39 &   1.4925 &   2.7369 &   18 &     \\
 3587.28 &           {\hei} &           31 & 3586.58 & -58.51 &   0.2142 &   0.3362 &   : &     \\
 3613.64 &           {\hei} &            6 & 3612.96 & -56.42 &   0.2407 &   0.3753 &   : &     \\
 3634.25 &           {\hei} &           28 & 3633.58 & -55.27 &   0.3659 &   0.5677 &   : &     \\
 3667.68 &            {\hi} &          H26 & 3666.99 & -56.41 &   0.2930 &   0.4508 &   : &     \\
 3669.47 &            {\hi} &          H25 & 3668.74 & -59.65 &   0.3809 &   0.5857 &   : &     \\
 3671.48 &            {\hi} &          H24 & 3670.80 & -55.53 &   0.3903 &   0.5998 &   : &     \\
 3673.76 &            {\hi} &          H23 & 3673.08 & -55.50 &   0.4547 &   0.6985 &   : &     \\
\end{longtable}

\clearpage

\onecolumn
\setcounter{table}{13}
\begin{longtable}{ccccccccc} 
\caption{Observed and reddening corrected line ratios (F(H$\beta$) = 100) and line identifications in M\,1-31.\label{tab:lineid6}}\tabularnewline
\hline\hline 
$\lambda_0$ (\AA\ ) & Ion & Mult. & $\lambda_{\rm obs}$ & $V_{\rm rad}$ (km s$^{-1}$) & $F(\lambda)/F({\rm H}\beta)$$^{\rm a}$ & $I(\lambda)/I({\rm H}\beta)$$^{\rm b}$ & Err(\%) & notes \tabularnewline
\hline
\endfirsthead
\caption{continued.}\tabularnewline
\hline\hline 
$\lambda_0$ (\AA\ ) & Ion & Mult. & $\lambda_{\rm obs}$ & $V_{\rm rad}$ (km s$^{-1}$) & $F(\lambda)/F({\rm H}\beta)$$^{\rm a}$ & $I(\lambda)/I({\rm H}\beta)$$^{\rm b}$ & Err(\%) & notes \tabularnewline
\hline
\endhead
\hline
\endfoot
\endlastfoot
 3691.56 &            {\hi} &          H18 & 3692.26 &  56.83 &   0.2869 &   1.3876 &   : &     \\
 3697.15 &            {\hi} &          H17 & 3697.92 &  62.43 &   0.3053 &   1.4683 &   : &     \\
 3703.86 &            {\hi} &          H16 & 3704.62 &  61.50 &   0.3621 &   1.7309 &   : &     \\
 3705.04 &           {\hei} &           25 & 3705.78 &  59.86 &   0.2340 &   1.1172 &   : &     \\
 3711.97 &            {\hi} &          H15 & 3712.77 &  64.60 &   0.4319 &   2.0484 &   : &     \\
 3721.83 &         {\fsiii} &           2F & 3722.59 &  61.21 &   0.8108 &   3.8087 &   30 &     \\
 3721.93 &            {\hi} &          H14 &       * &       * &       * &       * &    * &     \\
 3726.03 &          {\foii} &           1F & 3726.76 &  58.72 &   7.8908 &  36.919 &    6 &     \\
\end{longtable}

\clearpage

\onecolumn
\setcounter{table}{14}
\begin{longtable}{ccccccccc} 
\caption{Observed and reddening corrected line ratios (F(H$\beta$) = 100) and line identifications in M\,1-33.\label{tab:lineid7}}\tabularnewline
\hline\hline 
$\lambda_0$ (\AA\ ) & Ion & Mult. & $\lambda_{\rm obs}$ & $V_{\rm rad}$ (km s$^{-1}$) & $F(\lambda)/F({\rm H}\beta)$$^{\rm a}$ & $I(\lambda)/I({\rm H}\beta)$$^{\rm b}$ & Err(\%) & notes \tabularnewline
\hline
\endfirsthead
\caption{continued.}\tabularnewline
\hline\hline 
$\lambda_0$ (\AA\ ) & Ion & Mult. & $\lambda_{\rm obs}$ & $V_{\rm rad}$ (km s$^{-1}$) & $F(\lambda)/F({\rm H}\beta)$$^{\rm a}$ & $I(\lambda)/I({\rm H}\beta)$$^{\rm b}$ & Err(\%) & notes \tabularnewline
\hline
\endhead
\hline
\endfoot
\endlastfoot
 3634.25 &           {\hei} &           28 & 3633.44 & -66.84 &   0.2885 &   0.9804 &   31 &     \\
 3667.68 &            {\hi} &          H26 & 3666.73 & -77.67 &   0.0930 &   0.3089 &   : &     \\
 3669.47 &            {\hi} &          H25 & 3668.56 & -74.36 &   0.2001 &   0.6636 &   39 &     \\
 3671.48 &            {\hi} &          H24 & 3670.61 & -71.05 &   0.1652 &   0.5471 &   : &     \\
 3673.76 &            {\hi} &          H23 & 3672.85 & -74.27 &   0.1942 &   0.6419 &   : &     \\
 3676.37 &            {\hi} &          H22 & 3675.49 & -71.79 &   0.2382 &   0.7861 &   35 &     \\
 3679.36 &            {\hi} &          H21 & 3678.53 & -67.65 &   0.2345 &   0.7723 &   35 &     \\
 3682.81 &            {\hi} &          H20 & 3681.95 & -70.03 &   0.2811 &   0.9234 &   31 &     \\
\end{longtable}

\clearpage

\onecolumn
\setcounter{table}{15}
\begin{longtable}{ccccccccc} 
\caption{Observed and reddening corrected line ratios (F(H$\beta$) = 100) and line identifications in M\,1-60.\label{tab:lineid8}}\tabularnewline
\hline\hline 
$\lambda_0$ (\AA\ ) & Ion & Mult. & $\lambda_{\rm obs}$ & $V_{\rm rad}$ (km s$^{-1}$) & $F(\lambda)/F({\rm H}\beta)$$^{\rm a}$ & $I(\lambda)/I({\rm H}\beta)$$^{\rm b}$ & Err(\%) & notes \tabularnewline
\hline
\endfirsthead
\caption{continued.}\tabularnewline
\hline\hline 
$\lambda_0$ (\AA\ ) & Ion & Mult. & $\lambda_{\rm obs}$ & $V_{\rm rad}$ (km s$^{-1}$) & $F(\lambda)/F({\rm H}\beta)$$^{\rm a}$ & $I(\lambda)/I({\rm H}\beta)$$^{\rm b}$ & Err(\%) & notes \tabularnewline
\hline
\endhead
\hline
\endfoot
\endlastfoot
 3634.25 &           {\hei} &           28 & 3635.31 &  87.42 &   0.2187 &   0.8154 &   36 &     \\
 3664.68 &            {\hi} &          H28 & 3665.63 &  77.69 &   0.0853 &   0.3110 &   : &     \\
 3666.10 &            {\hi} &          H27 & 3667.07 &  79.30 &   0.0951 &   0.3463 &   : &     \\
 3667.68 &            {\hi} &          H26 & 3668.72 &  84.99 &   0.1192 &   0.4333 &   : &     \\
 3669.47 &            {\hi} &          H25 & 3670.49 &  83.31 &   0.1010 &   0.3666 &   : &     \\
 3671.48 &            {\hi} &          H24 & 3672.52 &  84.90 &   0.1892 &   0.6860 &   39 &     \\
 3673.76 &            {\hi} &          H23 & 3674.77 &  82.40 &   0.1821 &   0.6589 &   : &     \\
 3676.37 &            {\hi} &          H22 & 3677.44 &  87.21 &   0.2514 &   0.9082 &   33 &     \\
\end{longtable}

\clearpage

\onecolumn
\setcounter{table}{16}
\begin{longtable}{ccccccccc} 
\caption{Observed and reddening corrected line ratios (F(H$\beta$) = 100) and line identifications in M\,2-31.\label{tab:lineid9}}\tabularnewline
\hline\hline 
$\lambda_0$ (\AA\ ) & Ion & Mult. & $\lambda_{\rm obs}$ & $V_{\rm rad}$ (km s$^{-1}$) & $F(\lambda)/F({\rm H}\beta)$$^{\rm a}$ & $I(\lambda)/I({\rm H}\beta)$$^{\rm b}$ & Err(\%) & notes \tabularnewline
\hline
\endfirsthead
\caption{continued.}\tabularnewline
\hline\hline 
$\lambda_0$ (\AA\ ) & Ion & Mult. & $\lambda_{\rm obs}$ & $V_{\rm rad}$ (km s$^{-1}$) & $F(\lambda)/F({\rm H}\beta)$$^{\rm a}$ & $I(\lambda)/I({\rm H}\beta)$$^{\rm b}$ & Err(\%) & notes \tabularnewline
\hline
\endhead
\hline
\endfoot
\endlastfoot
  3187.84 &           {\hei} &            3 & 3189.01 & 109.98 &   0.8723 &   4.0921 &   20 &     \\
 3554.42 &           {\hei} &           34 & 3555.84 & 119.73 &   0.1069 &   0.3448 &   : &     \\
 3669.47 &            {\hi} &          H25 & 3670.80 & 108.63 &   0.1692 &   0.5071 &   : &     \\
 3671.48 &            {\hi} &          H24 & 3672.93 & 118.35 &   0.1505 &   0.4504 &   : &     \\
 3673.76 &            {\hi} &          H23 & 3675.20 & 117.46 &   0.2197 &   0.6564 &   : &     \\
 3676.37 &            {\hi} &          H22 & 3677.69 & 107.59 &   0.2309 &   0.6889 &   : &     \\
 3679.36 &            {\hi} &          H21 & 3680.81 & 118.09 &   0.3249 &   0.9673 &   37 &     \\
 3682.81 &            {\hi} &          H20 & 3684.26 & 117.98 &   0.3355 &   0.9966 &   37 &     \\
\end{longtable}



\bsp	
\label{lastpage}
\end{document}